\documentclass[12pt]{article}
\usepackage{latexsym}
\usepackage{amsmath}
\usepackage{amssymb}
\catcode`\@=11
\newcount\hour
\newcount\minute
\newtoks\amorpm \hour=\time\divide\hour by 60\minute
=\time{\multiply\hour by 60 \global\advance\minute by-\hour}
\edef\standardtime{{\ifnum\hour<12 \global\amorpm={am}%
        \else\global\amorpm={pm}\advance\hour by-12 \fi
        \ifnum\hour=0 \hour=12 \fi
        \number\hour:\ifnum\minute<10
        0\fi\number\minute\the\amorpm}}
\edef\militarytime{\number\hour:\ifnum\minute<10
0\fi\number\minute}
\def\draftlabel#1{{\@bsphack\if@filesw {\let\thepage\relax
   \xdef\@gtempa{\write\@auxout{\string
      \newlabel{#1}{{\@currentlabel}{\thepage}}}}}\@gtempa
   \if@nobreak \ifvmode\nobreak\fi\fi\fi\@esphack}
        \gdef\@eqnlabel{#1}}
\def\@eqnlabel{}
\def\@vacuum{}
\def\marginnote#1{}
\def\draftmarginnote#1{\marginpar{\raggedright\scriptsize\tt#1}}
\def\draft{
        \pagestyle{plain}
        \overfullrule=2pt
        \oddsidemargin -.1truein
        \def\@oddhead{\sl \phantom{\today\quad\militarytime} \hfil
        \smash{\Large\sl DRAFT} \hfil \today\quad\militarytime}
        \let\@evenhead\@oddhead
        \let\label=\draftlabel
        \let\marginnote=\draftmarginnote
        \def\ps@empty{\let\@mkboth\@gobbletwo
        \def\@oddfoot{\hfil \smash{\Large\sl DRAFT} \hfil}
        \let\@evenfoot\@oddhead}
        \def\@eqnnum{(\theequation)\rlap{\kern\marginparsep\tt\@eqnlabel}%
        \global\let\@eqnlabel\@vacuum}  }

\renewcommand{\theequation}{\thesection.\arabic{equation}}
\renewcommand{\thefootnote}{\fnsymbol{footnote}}
\newcommand{\newsection}{    
\setcounter{equation}{0}\section}
\def\appendix#1{\addtocounter{section}{1}\setcounter{equation}{0}
\renewcommand{\thesection}{\Alph{section}}
\section*{Appendix \thesection\protect\indent \parbox[t]{11.15cm}{#1}}
\addcontentsline{toc}{section}{Appendix \thesection\ \ \ #1}}

\def \lc {{light-cone}}

\def \bi{\bibitem}
\def \la {\label}

\def \b {\beta}

\def \t {\theta}
\def \s{\sigma}
\def \d {\partial}

\def\det{\hbox{det}}
\def\be{\begin{equation}}
\def\ee{\end{equation}}

\newcommand{\cont}[1]{{}_{#1}{}^{#1}}

\catcode`\@=11

\def\bea{\begin{eqnarray}}
\def\eea{\end{eqnarray}}
\def\beann{\begin{eqnarray*}}
\def\eeann{\end{eqnarray*}}
\def\beq{\begin{equation}}
\def\eeq{\end{equation}}
\def\ba{\begin{array}}
\def\ea{\end{array}}
\def\ben{\begin{enumerate}}
\def\een{\end{enumerate}}

\def\s {\sigma }
 \def \la {\label}
 \def\be{\begin{equation}}
\def\ee{\end{equation}}

\def \la {\label}

\def \r {\rho}

\font\mybb=msbm10 at 11pt

\def\bb#1{\hbox{\mybb#1}}

\def\bR {\bb{R}}

\def\bC {\bb{C}}

\def\e  {\epsilon}

\def \k {\kappa}

 \def\ep {\epsilon}

\def \t {\theta}

\def \ee {\epsilon}

\def \g {\gamma}
\def \bi{\bibitem}
\def\a{\alpha }

\def \ep {\epsilon}
\def \s {\sigma}

\def \r {\rho}
\def \d {\delta}
\def \G {\Gamma}

\def \g {\gamma}

\def \b {\beta}

\def\lc{\lrcorner}

\def \E {{E}}
\def \LF {{LF}}
\def \LG {{LG}}
\def \LP {{LP}}
\def \BG {{BG}}
\def \BP {{BP}}
\def \BF {{BF}}

\def \xone {G_{-+p}}
\def \xtwo {G_{-1p}}
\def \xthree {G_{- \bar{1} p}}
\def \xfour {G_{+1p}}
\def \xfive {G_{+ \bar{1} p}}

\def \yone {F_{-+1 \bar{1} p}}
\def \ytwo {F_{-+pq}{}^q}
\def \ythree {F_{-1pq}{}^q}
\def \yfour {F_{- \bar{1} pq}{}^q}
\def \yfive {F_{+1pq}{}^q}

\def \xone {G_{-+{\bar{\alpha}}}}
\def \xtwo {G_{\bar{\alpha} \beta}{}^\beta}
\def \xthree {\epsilon_{\bar{\alpha}}{}^{\beta_1 \beta_2 \beta_3} G_{\beta_1 \beta_2 \beta_3}}
\def \xfour {\epsilon_{\bar{\alpha}}{}^{\beta_1 \beta_2 \beta_3} F_{-+\beta_1 \beta_2 \beta_3}}
\def \xfive {F_{-+{\bar{\alpha}} \beta}{}^\beta}

\def \yone {\Omega_{\beta,}{}^\beta{}_{\bar{\alpha}}}
\def \ytwo {\epsilon_{\bar{\alpha}}{}^{\beta_1 \beta_2 \beta_3} \Omega_{\beta_1, \beta_2 \beta_3}}
\def \ythree {\Omega_{-,+\bar{\alpha}}}
\def \yfour {\Omega_{\bar{\alpha},-+}}
\def \yfive {\Omega_{\bar{\alpha}, \beta}{}^\beta}

\def \xonet {G_{-+\alpha}}
\def \xtwot {G_{\alpha \beta}{}^\beta}
\def \xthreet {\epsilon_\alpha{}^{{\bar{\beta}}_1  {\bar{\beta}}_2 {\bar{\beta}}_3}
G_{{\bar{\beta}}_1  {\bar{\beta}}_2 {\bar{\beta}}_3}}
\def \xfourt {\epsilon_\alpha{}^{{\bar{\beta}}_1  {\bar{\beta}}_2 {\bar{\beta}}_3}
 F_{-+{\bar{\beta}}_1  {\bar{\beta}}_2 {\bar{\beta}}_3}}
\def \xfivet {F_{-+\alpha \beta}{}^\beta}

\def \yonet {\Omega_{{\bar{\beta}},}{}^{{\bar{\beta}}}{}_\alpha}
\def \ytwot {\epsilon_\alpha{}^{{\bar{\beta}}_1  {\bar{\beta}}_2 {\bar{\beta}}_3}
\Omega_{{\bar{\beta}}_1,  {\bar{\beta}}_2 {\bar{\beta}}_3}}
\def \ythreet {\Omega_{-,+\alpha}}
\def \yfourt {\Omega_{\alpha,-+}}
\def \yfivet {\Omega_{\alpha, \beta}{}^\beta}

\def\be{\begin{equation}}
\def\ee{\end{equation}}

\def \bi {\bibitem}
\def \la{\label}

\def \t {\tau}

\def \nn {\nonumber}

\begin{document}
\date{November 2002}
\begin{titlepage}
\begin{center}
\hfill hep-th/0507087\\

\vspace{3.0cm}
{\Large \bf Systematics  of IIB spinorial geometry}\\[.2cm]

\vspace{1.5cm}
 {\large  U. Gran$^1$, J. Gutowski$^2$,  G. Papadopoulos$^1$ and D. Roest$^1$}

 \vspace{0.5cm}
${}^1$ Department of Mathematics\\
King's College London\\
Strand\\
London WC2R 2LS, UK\\
\vspace{0.5cm}
${}^2$ DAMTP, Centre for Mathematical Sciences\\
University of Cambridge\\
Wilberforce Road, Cambridge, CB3 0WA, UK
\end{center}

\vskip 1.5 cm
\begin{abstract}

We reduce the classification of all supersymmetric backgrounds of IIB supergravity to the evaluation
of the Killing spinor equations and their integrability conditions, which contain the field equations,
on five types of spinors. This extends the work of [hep-th/0503046] to IIB supergravity.
 We give the expressions of the Killing spinor equations on
  all five types of spinors. In this way,  the  Killing spinor equations
become  a linear system for the fluxes, geometry and spacetime derivatives
of the functions that determine the Killing spinors. This system can be solved to express the fluxes
in terms of the geometry and determine the conditions on the geometry of any supersymmetric
background. Similarly, the integrability conditions of the Killing spinor equations
 are turned into a linear system.  This can be used to determine  the field equations
  that are implied by the Killing spinor equations for any supersymmetric background.
We show that these linear systems simplify for generic backgrounds with maximal and half-maximal number of
$H$-invariant Killing spinors, $H\subset Spin(9,1)$. In the maximal case, the Killing spinor equations
factorize, whereas in the half-maximal case they do not.
As an example, we solve the Killing spinor equations of backgrounds with two $SU(4)\ltimes \bR^8$-invariant
Killing spinors. We also solve the linear systems associated with the integrability
conditions of  maximally supersymmetric $Spin(7)\ltimes\bR^8$- and $SU(4)\ltimes\bR^8$-backgrounds
and determine the field equations that are not implied by the Killing spinor equations.

\end{abstract}
\end{titlepage}
\newpage
\setcounter{page}{1}
\renewcommand{\thefootnote}{\arabic{footnote}}
\setcounter{footnote}{0}

\setcounter{section}{0}
\setcounter{subsection}{0}
\newsection{Introduction}

Supersymmetric solutions of IIB supergravity  have found widespread applications in string theory and gauge theories.
Some of these solutions have been discovered in the context of branes, see e.g.
\cite{gibbons, duff, strominger} and  in the context of AdS/CFT correspondence \cite{maldacena}, see e.g. \cite{schwarz,
 klebanov, chamseddine, nunez, georgea}.
Most of these results rely on Ans\"{a}tze appropriately adapted to the requirements of the
physical problems.
Progress has also been made towards a systematic understanding of the supersymmetric
solutions of IIB supergravity.
The maximally supersymmetric solutions of IIB supergravity have been classified in \cite{jfgpa} and they have been
found to be locally isometric to Minkowski space, $AdS_5\times S^5$ \cite{schwarz} and a maximally supersymmetric
plane wave \cite{georgea}.  In addition, these backgrounds are  related by Penrose limits \cite{georgeb}.
More recently, the Killing spinor equations of IIB have been solved
for one Killing spinor \cite{ugjggpa, ugjggpb}, and for all supersymmetric
backgrounds with two $Spin(7)\ltimes \bR^8$-invariant spinors, and  four $SU(4)\ltimes \bR^8$- and $G_2$-invariant
spinors \cite{ugjggpb}.

In the spinorial geometry approach to supersymmetric backgrounds \cite{uggp}, the Killing
spinor equations of M-theory and their integrability
conditions for any number of supersymmetries turn into linear systems \cite{ugdrgp}. The linear system of the Killing spinor
equations can be solved to express
the fluxes of the theory in terms of the geometry and to find the conditions on the geometry imposed by supersymmetry
for any number of Killing spinors. The linear system associated with the integrability conditions
determines the  field equations that are implied by the Killing spinor equations.
The main purpose of this paper is to adapt the above results to the Killing spinor equations
of IIB supergravity
and their integrability conditions.
The construction relies on the linearity of the Killing spinor equations  and the observation that
the IIB Killing spinor equations of any spinor can be determined from those of five types of spinors.
These five types of spinors are
\bea
1~,~~~e_{ij}~,~~~~e_{1234}~,~~~e_{i5}~,~~~e_{ijk5}~,~~~~~~~i,j,k=1,\dots,4~,
\eea
which we denote collectively by $\sigma_I$,
where we have expressed the spinors in terms of forms. For IIB supergravity, this has been explained in
\cite{ugjggpa}, see also appendix A.
We evaluate the Killing spinor equations of IIB supergravity on all five types of spinors and express
the result in terms of an oscillator basis in the space of spinors. In this way, we can construct
a linear system associated with the Killing spinor equations of backgrounds with {\it any
number} of Killing spinors. This linear system can be used to determine the fluxes
in terms of the geometry and the conditions on the geometry imposed by supersymmetry.
In IIB supergravity, it is first convenient to solve for the complex fluxes, i.e.~the
three-form field strength $G$ and the one-form field strength $P$ associated with the two scalars.
Then the remaining equations determine some of the components of the five-form flux $F$ and
constrain the geometry of spacetime.

The Killing spinor equations of supergravity theories imply some of
the field equations. In IIB supergravity, this is related to the computation of the field equations from the
commutator of supersymmetry transformations  \cite{schwarz}, see also \cite{tomas}. We identify the
integrability conditions ${\cal I}$ and ${\cal I}_A$
that contain the field equations and the Bianchi identities\footnote{The $\G$-matrix algebra has been carried out
using the computer programme GAMMA \cite{gamma}.}. Then, we show that the integrability
conditions of a Killing spinor can be expressed in terms of those on five types of spinors $\sigma_I$.
We evaluate ${\cal I}\sigma_I$ and ${\cal I}_A\sigma_I$ in terms of an oscillator basis in the space of spinors
and thus derive a linear system. This linear system can be used to determine which
field equations and Bianchi identities are implied by the Killing spinor equations for backgrounds
with any number of Killing spinors.

The main purpose of this paper is to be used as a manual to solve the Killing spinor equations
of IIB supergravity for backgrounds with any number of Killing spinors, and to determine
the field equations that are implied from the Killing spinor equations for such backgrounds.
Because of this, apart from giving the general formulae of the Killing spinor equations
acting on all the spinors $\sigma_I$, we also list the explicit results in the appendices. From these
results, one can construct the linear system associated with the Killing spinor equations
of any supersymmetric background. The same applies for the linear system associated
with the field equations.

There are several ways to characterize IIB supersymmetric backgrounds. One way
is to count the number
of Killing spinors\footnote{The number of Killing spinors is counted over the real numbers because
the Killing spinor equations of IIB supergravity are $\bR$-linear.} $N$ and their stability  subgroup $H$
 in $Spin(9,1)\times U(1)$. The role of the stability subgroup of the Killing spinors in the classification
 of supersymmetric backgrounds has been stressed in
\cite{jose}. Backgrounds for which $H$ contains a Berger holonomy group,
i.e. $H$ contains $SU(n)$, $G_2$, $Sp(2)$ and $Spin(7)$, are of particular interest. The Killing spinors
of most of the known solutions have stability subgroups in $Spin(9,1)\times U(1)$ which are
of Berger type.
 It has been demonstrated
in \cite{ugjggpb} that for any  subgroup $H$ in $Spin(9,1)$,  there
is a basis in the space of $H$-invariant spinors $\Delta^H$ which
can be written as $(\eta_j, i\eta_j), j=1,\dots,{1\over2} {\rm
dim}_{\bR}\Delta^H$, where $\eta_j$ are Majorana-Weyl spinors.
Moreover it was shown that  the Killing spinor equations factorize
for some backgrounds which admit $N={\rm dim}_{\bR}\Delta^H$ Killing
spinors. Here, we shall show that this is the case for {\it all}
backgrounds with $N={\rm dim}_{\bR}\Delta^H$ $H$-invariant Killing
spinors, i.e. the {\it maximally supersymmetric $H$-backgrounds} or {\it maximal $H$-backgrounds}.

In addition, we shall examine the backgrounds that admit $N={1\over 2} {\rm dim}_{\bR}\Delta^H$
H-invariant Killing spinors, i.e. they admit half of the maximal possible number of $H$-invariant
Killing spinors. We refer to these backgrounds either as {\it half-maximally supersymmetric}
$H$-backgrounds or as {\it half-maximal} $H$-backgrounds.
 We show that the Killing spinors of such backgrounds can be written as
\bea
\epsilon= z \,\eta
\eea
where $z$ is a $N\times N$ complex matrix and $\eta$ is a basis of
$N$ Majorana-Weyl $H$-invariant spinors. There are two classes of
half-maximally supersymmetric  $H$-backgrounds. One class consists
of those backgrounds for which the Killing spinors are linearly
independent over the complex numbers, and so over the real numbers.
Such backgrounds are associated with a complex $N\times N$
invertible matrix $z$, ${\rm det} \, z\not=0$. We refer to these
models as {\it generic half-maximal} $H$-backgrounds. We shall show
that although the Killing spinor equations do not factorize in this
case, they simplify. In particular, the gravitino Killing spinor
equations can be rewritten so that the only contributions that
include terms with more than two gamma matrices are those of the $F$
flux. The dependence on the functions of the Killing spinors is also
restricted in the terms that contain up to two  gamma matrices. In
addition, the solution to the Killing spinor equations gives rise to
a parallel transport equation
\bea
z^{-1} dz+ C=0~,
\eea
where $C$ can be interpreted as the restriction of the supercovariant connection
on the bundle of Killing spinors ${\cal K}$.
This is similar to the parallel transport equation\footnote{For maximally supersymmetric backgrounds, $H=1$ and $C$ is the supercovariant connection.} that arises in the maximally supersymmetric
$H$-backgrounds \cite{ugjggpb} but in this case $C$  may depend on $z$ and so the resulting first order system
is {\it non-linear}.
The other class consists of those backgrounds for which the Killing spinors are
linearly independent over the real numbers but linearly dependent over the complex numbers, so
${\rm det}\, z=0$. We refer to these models as {\it degenerate half-maximal} $H$-backgrounds.
Clearly this subclass can be further characterized by the rank of $z$. If the rank of $z$ is $r$, then
the space of Killing spinors of such backgrounds is of   co-dimension $2(N-r)$ in the space of Killing spinors.
In particular if the rank of $z$ is $N-1$, then one of the Killing spinors will be linearly dependent
over the complex numbers on the remaining $N-1$ Killing spinors but linearly independent over the reals.

As an example of our construction, we consider backgrounds
with two $SU(4)\ltimes \bR^8$-invariant spinors. The dimension of  $SU(4)\ltimes \bR^8$-invariant spinors
in the (complex) chiral representation of $Spin(9,1)$ is four. So backgrounds with
two $SU(4)\ltimes \bR^8$-invariant Killing spinors are half-maximally supersymmetric $SU(4)\ltimes \bR^8$-backgrounds.
We solve the Killing spinor equations for both generic and degenerate backgrounds.
For generic half-maximally supersymmetric $SU(4)\ltimes \bR^8$-backgrounds, we find that there are two cases to consider. In both
cases, we compute the non-linear parallel transport equation $z^{-1} dz+ C=0$ but
we do not give the most general solution. Instead, we analyze two examples. In one of the examples
$z$ is diagonal with complex entries and in the other $z$ is a real matrix.
In both examples, we determine the geometry of the supergravity backgrounds. In particular, we find
that the spacetime admits a null Killing vector field and compute all spacetime form bi-linears.
In the degenerate half-maximally supersymmetric $SU(4)\ltimes \bR^8$-backgrounds,
the second Killing spinor  is proportional to the first Killing spinor, $\epsilon_2=w\, \epsilon_1$,
where $w$ is a complex function with {\it non-vanishing}
imaginary part, and $\epsilon_1=(f-g_2+i g_1) 1+(f+g_2+i g_1) e_{1234}$ as in \cite{ugjggpa}, $f, g_2\not=0$.
The geometry of these backgrounds is similar to those with one $SU(4)\ltimes\bR^8$-invariant spinor
investigated in \cite{ugjggpa}.

We also solve the linear systems associated with the integrability conditions of the Killing spinor
equations of maximally supersymmetric $Spin(7)\ltimes\bR^8$- and $SU(4)\ltimes\bR^8$-backgrounds
of \cite{ugjggpb}. We find that in both cases, if the Bianchi identities are imposed, the only field equations
that are not implied by the Killing spinor equations are the $E_{--}$ components of the Einstein
equations. We explicitly give these equations  in terms of the connection and fluxes of the
backgrounds.

This paper is organized as follows. In section two, we review the construction of the
bosonic sector of IIB supergravity together with the Killing spinor equations
and their integrability conditions. In section three,  we show that the
Killing spinor equations and the integrability conditions of any spinor
can be determined from those on five types of spinors. We also introduce the
maximal and half-maximal supersymmetric $H$-backgrounds and investigate their
Killing spinor equations and  integrability  conditions. In section four,
 we construct the linear systems of the Killing spinor equations and the
 integrability conditions for any supersymmetric background. In section five,
 we present two examples of generic half-maximally supersymmetric
 $SU(4)\ltimes\bR^8$-backgrounds. In section six,
 we solve the Killing spinor equations of  degenerate half-maximally supersymmetric
 $SU(4)\ltimes\bR^8$-backgrounds. In section seven, we solve the linear systems
 of the integrability conditions of maximally supersymmetric
 $Spin(7)\ltimes\bR^8$- and $SU(4)\ltimes\bR^8$-backgrounds, and in section eight
 we give our conclusions. In appendix A, we summarize the construction of spinors
 in terms of forms and compute various spinor bi-linears. In appendix B, we explicitly give the
Killing spinor equations on all five types of spinors. In appendix C, we explicitly give the
 integrability  conditions of the Killing spinor equations on all five types of spinors.
 In appendix D, we present the linear system of the  Killing spinor equations of
generic half-maximally supersymmetric
  $SU(4)\ltimes\bR^8$-backgrounds and give its solution, and in appendix E, we present
 the linear system of degenerate half-maximally supersymmetric
  $SU(4)\ltimes\bR^8$-backgrounds.


\newsection{Killing spinor equations and integrability conditions}

\subsection{Killing spinor equations}

The bosonic fields of IIB supergravity \cite{west,schwarz, howe} are the spacetime metric $g$, two real scalars,
 the axion $\sigma$
and the dilaton $\phi$, which are combined into a complex one-form field strength $P$,
two three-form field strengths $G^1$ and $G^2$ which are combined to a (complex) three-form
field strength $G$, and a self-dual
five-form field strength $F$. To describe these\footnote{For a recent account of the
description of the field strengths in terms of gauge potentials see \cite{eric}.}, we introduce
a $SU(1,1)$ matrix $U=(V_+^\a, V_-^\a)$, $\a=1,2$ such that
\bea
V_-^\a V_+^\b-V_-^\b V_+^\a=\epsilon^{\a\b}~,
\eea
where $\epsilon^{12}=1=\epsilon_{12}$,  $(V_-^1)^*=V_+^2$ and $(V_-^2)^*=V_+^1$.  The signs denote
$U(1)\subset SU(1,1)$ charge. Then
\bea
P_M&=&-\epsilon_{\a\b} V_+^\a \partial_M  V_+^\b
\cr
Q_M&=&-i \epsilon_{\a\b} V_-^\a\partial_M V_+^\b~.
\eea
The three-form  field strengths $G_{MNR}^\a=3\partial_{[M} A^\a_{NR]}$, with $(A^1_{MN})^*= A^2_{MN}$ combine into the complex
field strength
\bea
G_{MNR}=-\epsilon_{\a\b} V^\a_+ G_{MNR}^\b~.
\eea
The five-form self-dual field strength is
\bea
F_{M_1M_2M_3M_4M_5}=5\partial_{[M_1} A_{M_2M_3M_4M_5]}+{5i\over8} \epsilon_{\a\b} A^\a_{[M_1M_2} G^\b_{M_3M_4M_5]}~,
\eea
where
$F_{M_1\dots M_5}=- {1\over 5!}
\epsilon_{M_1\dots M_5}{}^{N_1\dots N_5}
F_{N_1\dots N_5}$ and $\epsilon_{01\dots9}=1$.
To identify the scalars define the variables $\rho=V_-^2/V_-^1$ and
\bea
\rho={1+i\tau\over 1-i\tau}~.
\eea
In turn $\tau=\sigma+i e^{-\phi}$, where $\sigma$ is the axion (RR scalar) and $\phi$ is the dilaton.

The  Killing spinor
equations of IIB supergravity are the parallel transport equations of the
supercovariant derivative ${\cal D}$
\be
{\cal D}_M\epsilon=\tilde\nabla_M \epsilon+{i\over 48} \Gamma^{N_1\dots N_4 } \epsilon
 F_{N_1\dots N_4 M}
 -{1\over 96} (\Gamma_{M}{}^{N_1N_2N_3}
G_{N_1N_2N_3}-9 \Gamma^{N_1N_2} G_{MN_1N_2}) (C\epsilon)^*=0
\la{kseqna}
\ee
and the algebraic condition
\be
{\cal A}\epsilon=P_M \Gamma^M (C\epsilon)^*+ {1\over 24} G_{N_1N_2N_3} \Gamma^{N_1N_2N_3} \epsilon=0~,
\la{kseqnb}
\ee
where
$$
\tilde\nabla_M=D_M+{1\over4} \Omega_{M,AB} \Gamma^{AB}~,~~~~~~D_M=\partial_M-{i\over2}Q_M
$$
is the
spin connection, $\nabla_M=\partial_M+{1\over4} \Omega_{M,AB} \Gamma^{AB}$,
twisted with $U(1)$ connection $Q_M$, $Q_M^*=Q_M$, $\epsilon$
is a (complex) Weyl spinor,
$\Gamma^{0\dots 9}\epsilon=-\epsilon$, and  $C$ is a charge conjugation
 matrix. For our spinor conventions\footnote{We use a mostly plus
convention for the metric.
To relate this  to the conventions of  \cite{schwarz}, one
takes  $\Gamma^A\rightarrow
i\Gamma^A$ and every time a index is lowered there is also
an additional minus sign.} see appendix A.
 The Killing spinor equations are the  vanishing conditions of the supersymmetry
transformations of the gravitino, and  the supersymmetric partners of the
dilaton and axion  restricted to the bosonic sector of IIB supergravity, respectively.
For a superspace formulation of IIB supergravity
see \cite{howe}. The recent modification of IIB supergravity with ten-form
potentials \cite{eric} does not change our analysis below because the Killing spinor equations
remain the same.

\subsection{Integrability conditions}

To determine the field equations which are implied by the Killing spinor equations, one has to investigate
the integrability conditions of the Killing spinor equations. This calculation is essentially the same as that
of \cite{schwarz} where the field equations of the IIB supergravity were found from the commutator
of the supersymmetry transformations. However, we cast the results in such a way that is more suitable to our purpose.
The integrability conditions are
\be
[{\cal D}_A,{\cal
D}_B]\ep={\cal R}_{AB}\ep=0~,
\la{inta}
\ee
and
 \be
 [{\cal
D}_A,{\cal A}]\ep=0~,
\la{intb}
\ee
where ${\cal R}$ has been given in \cite{tsimpis} and so the expression will not be repeated here.
It turns out that some field equations and Bianchi identities of IIB supergravity
are contained in the  ${\cal I}_A= \tfrac{1}{2} \Gamma_A{}^{BC}{\cal R}_{BC}$ and ${\cal I}=\Gamma^A[{\cal D}_A,{\cal A}]$ components
of the integrability conditions\footnote{The integrability conditions of the Killing spinor equations (\ref{inta}) and (\ref{intb})
  may impose further conditions on the Bianchi identities and field equations
 than those implied by the vanishing of  these two components.}. In particular, we have\footnote{
 In terms of $\BF$, we have

${\cal I}_A\ep=\big[\tfrac{1}{2}\Gamma^B
E_{AB}+ i \G_{A}{}^{B_1\ldots B_6}BF_{B_1\ldots B_6}\big]\ep
-\big[\G^B LG_{AB} -\G_A{}^{B_1\ldots B_4}BG_{B_1\ldots B_4}
\big](C\ep)^*$.}
\bea
{\cal I}_A\ep=\big[\tfrac{1}{2}\Gamma^B
E_{AB}- i \G^{B_1B_2 B_3}\LF_{AB_1B_2 B_3}\big]\ep
-\big[\G^B LG_{AB} -\G_A{}^{B_1\ldots B_4}BG_{B_1\ldots B_4}
\big](C\ep)^*
\la{IAint}
 \eea
and similarly, $\Gamma^A[{\cal D}_A,{\cal A}]\ep$ can be written
as
\bea {\cal
I}\epsilon=\big[\tfrac{1}{2}\G^{AB}LG_{AB}+\G^{A_1\ldots
A_4}BG_{A_1\ldots
A_4}\big]\ep+\big[LP+\G^{AB}BP_{AB}\big](C\ep)^*~,
\la{Iint}
\eea
where
\bea
 E_{AB}&:=& R_{AB}-\tfrac{1}{2}g_{AB}R -\tfrac{1}{6}F_{AC_1\ldots C_4}F_{B}{}^{C_1\ldots C_4}
 -\tfrac{1}{4}G_{(A}{}^{C_1C_2}G^*_{B)C_1C_2}\nn\\
 &&+\tfrac{1}{24}g_{AB}G^{C_1C_2C_3}G^*_{C_1C_2C_3}  -2P_{(A}P^*_{B)}+g_{AB}P^C P^*_C\,,\nn\\
 \LG_{AB}&:=& \tfrac{1}{4}(\tilde\nabla^C G_{ABC}-P^C G^*_{ABC}+\tfrac{2i}{3}F_{ABC_1C_2C_3}G^{C_1C_2C_3})
 \,,\nn\\
 \LP&:=& \tilde\nabla^A P_A+\tfrac{1}{24}G_{A_1A_2A_3}G^{A_1A_2A_3}\,,\nn\\
 \LF_{A_1\ldots A_4}&:=& \tfrac{1}{3!}(\nabla^B F_{A_1\ldots
 A_4 B}+\tfrac{i}{288} \e_{A_1 \ldots A_4}{}^{B_1 \ldots B_6} G_{B_1B_2B_3}G^*_{B_4B_5B_6})\,,\nn\\
 \BF_{A_1\ldots A_6}&:=& \tfrac{1}{5!}(\partial_{[A_1}F_{A_2\ldots
 A_6]}-\tfrac{5i}{12}G_{[A_1A_2A_3}G^*_{A_4A_5A_6]})\,,\nn\\
 BG_{A_1\ldots
 A_4}&:=&\tfrac{1}{4!}(D_{[A_1}G_{A_2A_3A_4]}+P_{[A_1}G^*_{A_2A_3A_4]})\,,\nn\\
 BP_{AB}&:=& D_{[A}P_{B]}\,.
\eea
One can show that $\LF$ and $\BF$ are not independent but are related by the self duality condition on $F$.
The field strengths $P$ and $G$ have different $U(1)\subset SU(1,1)$ charges. In particular, one has
\bea
D_A P_B&=&\partial_A P_B-2i Q_A P_B
\cr
D_A G_{BCD}&=&\partial_A G_{BCD}-i Q_A G_{BCD}~.
\eea
To derive the  above expressions  some very painful Dirac algebra is required but we have been
 assisted  by GAMMA
\cite{gamma} to perform most of the computation.   The algebraic Killing spinor
equation (\ref{kseqnb}) has also been used to convert expressions containing  $G$ and
$P$ fluxes. The above choice of components of integrability conditions  that contain
the field equations and the Bianchi identities is not unique. For example, the component $\Gamma^{B}{\cal R}_{AB}$ may also
be used giving identical results. However, it turns out that the computation is more involved.

\newsection{The five types of spinors}

\subsection{General case}

The spinors that appear in type IIB supergravity are complex Weyl spinors of positive chirality.
A direct consequence of this is that
the most general Killing spinor of IIB supergravity can be written as
\bea
\epsilon= p 1+ q e_{1234}+ u^i e_{i5}+{1\over2} v^{ij} e_{ij}+{1\over6} w^{ijk} e_{ijk5}~,
\eea
where $p,q,u,v$ and $w$ are complex functions on the spacetime, and $i,j,k=1,2,3,4$.
The supercovariant derivative acting on $\epsilon$ gives
\bea
{\cal D}_A\epsilon&=&\partial_A p 1+ \partial_A q e_{1234}+ \partial_A u^i e_{i5}
+{1\over2} \partial_A v^{ij} e_{ij}+{1\over6} \partial_A w^{ijk} e_{ijk5}
\cr
&+&  p_0 {\cal D}_A 1+q_0 {\cal D}_A e_{1234}+  u_0^i {\cal D}_A e_{i5}
+{1\over2}v_0^{ij} {\cal D}_A e_{ij}
+{1\over6}w_0^{ijk} {\cal D}_A e_{ijk5}
\cr
&+& p_1 {\cal D}_A (i1)+q_1 {\cal D}_A(i e_{1234})+u_1^i {\cal D}_A (ie_{i5})
+{1\over2}v_1^{ij} {\cal D}_A(i e_{ij})
+{1\over6}w_1^{ijk} {\cal D}_A(i e_{ijk5})~~~~
\eea
and the algebraic Killing spinor equation becomes
\bea
{\cal A}\epsilon&=&p_0 {\cal A}1+ q_0 {\cal A}e_{1234}+ u_0^i {\cal A}e_{i5}
+{1\over2} v_0^{ij} {\cal A}e_{ij}+{1\over6} w_0^{ijk}
{\cal A}e_{ijk5}
\cr
&&+p_1 {\cal A}(i1)+ q_1 {\cal A}(ie_{1234})+ u_1^i {\cal A}(ie_{i5})
+{1\over2} v_1^{ij} {\cal A}(ie_{ij})+{1\over6} w_1^{ijk}
{\cal A}(ie_{ijk5})
~,
\eea
where $p=p_0+ip_1$ and similarly for the rest of the components.
Therefore to compute the Killing spinor equations of the most general spinor in IIB supergravity,
it suffices
to compute the supercovariant derivative and ${\cal A}$ on the ten types of spinors $1$,
$e_{1234}$, $e_{i5}$, $e_{ij}$ and $e_{ijk5}$, and $i1$,
$ie_{1234}$, $ie_{i5}$, $ie_{ij}$ and $ie_{ijk5}$. However it is straightforward to see that
 ${\cal D}_A(i 1)$ and ${\cal A}(i1)$ can be easily read from the expressions of ${\cal D}_A 1$ and ${\cal A}1$,
  respectively, and similarly for the rest of the pairs.  The only effect is a sign which appears
  in those terms of the Killing spinor equation which contain the charge conjugation matrix.
  Of course the  Killing spinor equations acting on $1$ should be in addition multiplied by the
  complex unit $i$ to recover those acting on $i 1$, and similarly for the rest of the pairs.
  Therefore to construct the linear system associated with
  any number of Killing spinors, it suffices
to compute
\bea
&&{\cal D}_A1~,~~~{\cal D}_Ae_{1234}~,~~~{\cal D}_Ae_{i5}~,~~~{\cal D}_Ae_{ij}~,~~~{\cal D}_Ae_{ijk5}~,
\cr
&&{\cal A}1~,~~~{\cal A}e_{1234}~,~~~{\cal A}e_{i5}~,~~~{\cal A}e_{ij}~,~~~{\cal A}e_{ijk5}~,
\eea
i.e. the Killing spinor equations on five types of spinors.

It remains to show that the integrability conditions ${\cal I}_A, {\cal I}$ on a Killing spinor $\epsilon$ can also be
determined in terms of those on the above five types of spinors.  Since these integrability conditions are algebraic,
one finds that
\bea
{\cal I}_A\epsilon&=&p_0 {\cal I}_A1+ q_0 {\cal I}_Ae_{1234}+ u_0^i {\cal I}_Ae_{i5}
+{1\over2} v_0^{ij} {\cal I}_Ae_{ij}+{1\over6}
w_0^{ijk} {\cal I}_Ae_{ijk5}
\cr
&&+p_1 {\cal I}_A(i1)+ q_1 {\cal I}_A(ie_{1234})+ u_1^i {\cal I}_A(ie_{i5})+{1\over2} v_1^{ij} {\cal I}_A(ie_{ij})+
{1\over6} w_1^{ijk} {\cal I}_A(ie_{ijk5})~~
\eea
and similarly for the ${\cal I}$ integrability condition.
Because the expressions for ${\cal I}_A(i1)$ can be easily recovered from that of ${\cal I}_A1$, and similarly
for the rest, one has to compute
\bea
&& {\cal I}_A1~,~~~{\cal I}_Ae_{1234}~,~~~{\cal I}_Ae_{i5}~,~~~{\cal I}_Ae_{ij}~,~~~{\cal I}_Ae_{ijk5}~,
\cr
&& {\cal I}1~,~~~{\cal I}e_{1234}~,~~~{\cal I}e_{i5}~,~~~{\cal I}e_{ij}~,~~~{\cal I}e_{ijk5}~,
\eea
i.e. the integrability conditions on  five types of spinors.
In what follows, we shall give the general formulae of the Killing spinor equations and their integrability
conditions acting on all five types of spinors. In the appendices, we shall list the various components
of these equations in the basis (\ref{hbasis}).

\subsection{Invariant spinors} \la{invspin}

\subsubsection{Maximally supersymmetric $H$-backgrounds}

In many cases of interest, the Killing spinors are invariant under some proper subgroup
$H$ of $Spin(9,1)$. In such cases, it has been shown in
\cite{ugjggpb} that  the space of $H$-invariant spinors, $\Delta^H$, is even-dimensional,
 ${\rm dim} \Delta^H=k=2m$, and there is a basis
  $(\eta_i, i=1, \dots, k)=(\eta_p=\eta_p, \eta_{m+p}=i\eta_p, p=1, \dots, m)$, where $\eta_p$
are $H$-invariant Majorana-Weyl spinors. The most general $H$-invariant Killing spinors in this case are
\bea
\epsilon_r=\sum^k_{i=1}f_{ri} \eta_i~,~~~~r=1, \dots, N
\eea
where $(f_{ri})$ is a $N\times k$ matrix of real functions and $N$ is the number of Killing spinors of the background.
 It has also been shown in \cite{ugjggpb} that the Killing spinor equations of backgrounds
with $H$-invariant Killing spinors whose number
of Killing spinors is equal to the real dimension of $\Delta^H$, i.e. of maximally
supersymmetric $H$-backgrounds, dramatically simplify. In
particular the terms in Killing spinor equations that contain the $P$ and $G$ fluxes
factorize from those that contain the
$F$ fluxes and geometry. This was shown in some special cases, here we shall
present the proof of the general case.

In the maximally-supersymmetric $H$-backgrounds, $f=(f_{ri})$ is invertible. Because of this, the Killing spinor
equations can be written as
\bea
\sum_{j=1}^N(f^{-1}D_M f)_{ij} \eta_j+ {\cal D}_M \eta_i=0~,
\cr
{\cal A}\eta_i=0~.
\la{magicf}
\eea
First consider the algebraic Killing spinor equation for $i=1$ and $i=m+1$. In this case $\eta_{m+1}=i\eta_1$ and so
\bea
{\cal A}\eta_1= P_A \Gamma^A\eta_1+{1\over24} \Gamma^{ABC} G_{ABC}\eta_1&=&0~,
\cr
{\cal A}\eta_{m+1}=-i P_A \Gamma^A \eta_1+{i\over24} \Gamma^{ABC} G_{ABC} \eta_1&=&0~.
\eea
Therefore, we conclude that $P_A \Gamma^A\eta_1=0$ and $\Gamma^{ABC} G_{ABC}\eta_1=0$. Applying this
for all pairs, we get
\bea
P_A \Gamma^A~\eta_p&=&0~,~~~p=1, \dots,m
\cr
\Gamma^{ABC} G_{ABC}~\eta_p&=&0~,~~~p=1, \dots,m~.
\la{facalg}
\eea
Similarly, for $i=p$ and $i=m+p$ in the first equation in (\ref{magicf}), we find
\bea
\sum_{j=1}^N(f^{-1}D_M f)_{pj}~ \eta_j+ \nabla_M \eta_p+{i\over 48} \Gamma^{N_1\dots N_4 } \eta_p
 F_{N_1\dots N_4 M}
 \cr
 -{1\over 96} (\Gamma_{M}{}^{N_1N_2N_3}
G_{N_1N_2N_3}-9 \Gamma^{N_1N_2} G_{MN_1N_2}) \eta_p=0~,
\cr
-i\sum_{j=1}^N(f^{-1}D_M f)_{m+pj}~\eta_j+\nabla_M \eta_p+{i\over 48} \Gamma^{N_1\dots N_4 } \eta_p
 F_{N_1\dots N_4 M}
 \cr
 +{1\over 96} (\Gamma_{M}{}^{N_1N_2N_3}
G_{N_1N_2N_3}-9 \Gamma^{N_1N_2} G_{MN_1N_2}) \eta_p=0~.
\eea
Adding and subtracting these equations, we get
\bea
{1\over2}[ \sum_{j=1}^N(f^{-1}D_M f)_{pj}~ \eta_j-i\sum_{j=1}^N(f^{-1}D_M f)_{m+pj}~\eta_j]+\nabla_M \eta_p
+{i\over 48} \Gamma^{N_1\dots N_4 } \eta_p
 F_{N_1\dots N_4 M}=0
 \cr
 \sum_{j=1}^N(f^{-1}D_M f)_{pj}~ \eta_j+i\sum_{j=1}^N(f^{-1}D_M f)_{m+pj}~\eta_j+{1\over 4} G_{MBC} \Gamma^{BC}\eta_p=0~~~~
 \la{facpar}
\eea
where we have also used the second equation in (\ref{facalg}). It is easy to see from (\ref{facalg}) and (\ref{facpar})
that, as we have mentioned, the Killing spinor equations factorize.

It has been observed in \cite{ugjggpb} that the solution to the Killing spinor equation in this case gives rise
to a parallel transport equation
\bea
f^{-1} df+ C=0~.
\eea
The connection $C$ can be thought of as the restriction of the supercovariant connection on the bundle
of Killing spinors
\bea
0\rightarrow {\cal K}\rightarrow {\cal S}\rightarrow {\cal S}/{\cal K}\rightarrow 0~,
\eea
where ${\cal S}$ is the spin bundle of the theory. A necessary condition for the existence of a solution
to the parallel transport equation is the vanishing of the curvature $F(C)=dC-C\wedge C=0$. It is worth pointing out
that for maximally supersymmetric backgrounds, $H=1$, ${\cal K}={\cal S}$ and $C$ is the supercovariant connection.
The curvature $F(C)$ is the supercovariant curvature ${\cal R}=[{\cal D}, {\cal D}]$. The vanishing
of ${\cal R}$ was precisely the condition analyzed in \cite{jfgpa} to classify the supersymmetric
solutions of ten- and eleven-dimensional supergravities.

The integrability conditions of the Killing spinor equations of maximally supersymmetric $H$-backgrounds factorize as well. In
particular since ${\rm det} f\not=0$, it is easy to see that ${\cal I}_A\epsilon_i=0$ and ${\cal I}\epsilon_i=0$ imply that
${\cal I}_A\eta_i=0$ and ${\cal I}\eta_i=0$, $i=1, \dots, 2m$. In turn these two equations give
\bea
\big[\tfrac{1}{2}\Gamma^B
E_{AB}- i \G^{B_1B_2 B_3}\LF_{AB_1B_2B_3}\big]\eta_j&=&0\,,
\cr
\big[\G^B LG_{AB} -\G_A{}^{B_1\ldots B_4}BG_{B_1\ldots B_4}
\big]\eta_j&=&0\,,
\cr
\big[\tfrac{1}{2}\G^{AB}LG_{AB}+\G^{A_1\ldots
A_4}BG_{A_1\ldots
A_4}\big]\eta_j&=&0\,,
\cr
\big[LP+\G^{AB}BP_{AB}\big]\eta_j&=&0
~,~~~~j=1,\dots,m~.
\eea
It is clear that if one assumes that the Bianchi identities are satisfied, then the above conditions impose
strong conditions on the field equations. We analyze these in detail for the maximally supersymmetric
$SU(4)\ltimes \bR^8$- and $Spin(7)\ltimes \bR^8$- backgrounds.

\subsubsection{Half-maximally supersymmetric $H$-backgrounds}

Apart from the  maximally supersymmetric backgrounds above, there is also another class of backgrounds with $H$-invariant
spinors for which the the Killing spinor equations simplify. These are the backgrounds for which the number of Killing spinors
is $N={1\over2}{\rm dim}\Delta^H=m$. We refer to such backgrounds as half-maximal $H$-backgrounds. It turns out
that if the Killing spinors are generic\footnote{We shall make this precise later.}, then the Killing spinor equations
of half-maximal $H$ backgrounds simplify
but they do not necessarily factorize as in the maximal case.

The Killing spinors for half-maximal $H$-backgrounds can be written as
\bea
\epsilon_i=\sum_{j=1}^m z_{ij} \eta_j~,~~~~i,j=1,\dots, m~,
\eea
where $z=(z_{ij})$ is an $m\times m$  matrix of complex functions on the spacetime.
This can be easily seen by expressing  $m$  $H$-invariant spinors in the basis $(\eta_j, i\eta_j)$ of $\Delta^H$.

Next suppose that $\epsilon_i$ are generic, i.e. that ${\rm det} z\not=0$.
In such case, the
algebraic Killing spinor equation can be written as
\bea
P_A \Gamma^A z^*\eta +{1\over24} G_{ABC}\Gamma^{ABC} z \eta=0~,
\eea
where we have used matrix notation for $z$ and $\eta$. This can be then rewritten as
\bea
P_A \Gamma^A z^{-1}z^*\eta+{1\over24} G_{ABC} \Gamma^{ABC} \eta=0~.
\eea
Acting on this equation with $\Gamma_A$, we find that
\bea
(P^B \Gamma_{AB}+ P_A)z^{-1}z^*\eta + {1\over24} G^{BCD} \Gamma_{ABCD} \eta+ {1\over8} G_{ABC} \Gamma^{BC}\eta=0~.
\eea
Solving for the fourth order term in the gamma matrices and substituting the result into   (\ref{kseqna}), we get
\bea
D_A z \eta+ z \nabla_A\eta +{i\over 48} \Gamma^{B_1\dots B_4 } z \eta
 F_{B_1\dots B_4 A}+ {1\over4} z^* z^{-1} z^* [P^B \Gamma_{AB}+ P_A]\eta
 \cr
 + {1\over 8} G_{ABC} \Gamma^{BC} z^*\eta=0
 \eea
 or equivalently,
\bea
z^{-1}D_A z \eta+ \nabla_A\eta +{i\over 48} \Gamma^{B_1\dots B_4 }  \eta
 F_{B_1\dots B_4 A}+ {1\over4}z^{-1} z^* z^{-1} z^* [P^B \Gamma_{AB}+ P_A]\eta
 \cr
 + {1\over 8} G_{ABC} \Gamma^{BC} z^{-1} z^*\eta=0~.
 \la{hmaxkse}
 \eea
 There is no factorization of the Killing spinor equations in this case,
 in contrast to the maximal supersymmetric $H$-backgrounds.
 Nevertheless, the Killing spinor equations simplify because the contribution of the $G$ and the $P$ fluxes
 in the (\ref{kseqna}) is contained in the up to gamma square terms and the $F$ flux term is independent
 from the spacetime functions $z$. Therefore the effect of the $G$ and $P$ fluxes is to modify the spin connection
 $\Omega$ and the $U(1)$ connection $Q$ with terms that depend on the $P$ and $G$ fluxes and the functions
 that determine the Killing spinors.

 The solution to the Killing spinor equations of generic half-maximally supersymmetric $H$-backgrounds gives rise
 to a parallel transport equation
 \bea
 z^{-1} dz+ C=0~.
 \eea
 The connection $C$ can again be thought of as the restriction of the supercovariant connection
 on the bundle of Killing spinors ${\cal K}$.
However unlike the case of maximally supersymmetric $H$-backgrounds, $C$ depends on $z$, $C=C(z)$.
To see this  observe that some fluxes in (\ref{hmaxkse}) depend on the functions $z$. We have also
confirmed this in an example. Because of this, although it is always possible to solve the linear system
associated with the Killing spinor equations, the resulting parallel transport equation may be rather
involved.

 Next let us consider the special or degenerate cases ${\rm det} ~z=0$. These cases arise whenever
 the Killing spinors $\epsilon_i$ are linearly dependent over the complex numbers but linearly
 independent over the real numbers. These cases are characterized by the rank of $z$. If the rank of $z$ is $m-1$,
 then it can be arranged such that the first $m-1$ Killing spinors are linearly independent over the complex numbers but the
 last one is linearly dependent. In such a case, we can write
 \bea
 \epsilon_m= w_1 \epsilon_1+\dots +w_{m-1} \epsilon_{m-1}~,
 \eea
where at least one of $w_1, \dots, w_{m-1}$ has a non-vanishing imaginary part. If all the imaginary parts vanish, then
$\epsilon_m$ is linearly dependent on $\epsilon_1, \dots, \epsilon_{m-1}$ over the reals and the background
has $m-1$ supersymmetries. One can modify the above argument in the cases for which $z$ has rank $r<m$ for $r=1, \dots, m-1$.
It appears that the solution of the Killing spinor equations in the degenerate cases
requires information on the solutions of the Killing spinor equations for $N<m$ $H$-invariant Killing spinors.
We shall see that this is the case in the special case of ($N=2$) half-maximal $SU(4)\ltimes \bR^8$-backgrounds.

The integrability conditions ${\cal I}_A\epsilon_i={\cal I}\epsilon_i=0$ also simplify for half-maximally $H$-supersymmetric
backgrounds. We shall focus on the case where ${\rm det}\,z\not=0$. In this case, these integrability conditions
can be rewritten as
\begin{align}
& \big[\tfrac{1}{2}\Gamma^B
\E_{AB}
- i \G^{B_1\ldots B_3} \LF_{A B_1\ldots B_3}\big]\eta
-\big[\G^B \LG_{AB} -\G_A{}^{B_1\ldots B_4} \BG_{B_1\ldots B_4}
\big] z^{-1} z^*\eta=0\,,
\cr
& \big[\tfrac{1}{2}\G^{AB} \LG_{AB}+\G^{A_1\ldots
A_4}\BG_{A_1\ldots
A_4}\big]\eta+\big[\LP+\G^{AB} \BP_{AB}\big]z^{-1} z^*\eta=0~.
\end{align}
It is clear that unlike the maximal $H$-backgrounds, the integrability conditions do not factorize in this case.

 Combining the maximal and half-maximal cases we have mentioned above,  and considering those $H\subset Spin(9,1)$ that contain
 a Berger type
 of group, one can investigate several cases of supersymmetric backgrounds. These
 include backgrounds with four, eight and sixteen supersymmetries. These cases are summarized in
the conclusions in table 1.

\newsection{Linear systems}

\subsection{The linear system of Killing spinor equations}

We have shown that the Killing spinor equations of an arbitrary spinor can be expressed in terms
of those on  five types of spinors given by
 \be
 \sigma_I= e_{i_1 \cdots i_I} = \frac{1}{\sqrt{{2}^I}} \Gamma^{\bar i_1 \cdots \bar i_I} 1 \,,
 \ee
where the index $i = (1, \ldots, 5)$ contains four holomorphic and one null
indices\footnote{In this section we will not use the notation $(-,+)$ for the null
indices but rather $(5, \bar 5)$. Thus $\Gamma^{\bar 5} = \Gamma^+$ and $\Gamma^5 = \Gamma^-$.}.
Note that $\Gamma^a$ acts as an annihilation or a creation operator on the above spinor, depending
on whether the label $i_1 \cdots i_I$ does or does not contain $a$. For this reason it is convenient
 to reshuffle $(\a, 5)$ and $(\bar \a, \bar 5)$ into indices $p,q,r$ defined by
 \bea
  p = (\bar i_1, \ldots, \bar i_I, i_{I+1}, \ldots, i_5) \,, \qquad
  \bar p = (i_1, \ldots, i_I, \bar i_{I+1}, \ldots, \bar i_5) \,,
 \eea
where the indices $(i_1, \ldots, i_5)$ are some permutation of $(1, \ldots, 5)$,
i.e.~$\e_{i_1 \cdots i_5} = \pm 1$. Thus, $\Gamma^p$ act as annihilation
operators on this  spinor while $\Gamma^{\bar p}$ are the creation operators.

With this notation at hand, the expression for the algebraic Killing spinor equation (\ref{kseqnb}) on an
arbitrary  spinor\footnote{We have used, however, that $I$ is even for all IIB spinors.} can be written as
 \bea
  \mathcal{A} e_{i_1 \cdots i_I} & =  & [\tfrac{1}{4} G_{\bar q} \cont{r}] \Gamma^{\bar q} e_{i_1 \cdots i_I} + \nonumber \\
  & & [\tfrac{1}{24} G_{\bar q_1 \cdots \bar q_3} -
  \tfrac{s}{12} \e_{\bar q_1 \cdots \bar q_3}{}^r P_r] \Gamma^{\bar q_1 \cdots \bar q_3} e_{i_1 \cdots i_I} + \nonumber \\
  & &  [\tfrac{s}{96} \e_{\bar q_1 \cdots \bar q_4} P_{\bar q_5}] \Gamma^{\bar q_1 \cdots \bar q_5} e_{i_1 \cdots i_I} \,,
  \la{oscalg}
 \eea
where the Levi-Civita symbols are defined by $\e_{\bar 1 \bar 2 \bar 3 \bar 4}
= +1$ and $\e_{\bar q_1 \cdots \bar q_3 \bar 5} = 0$, i.e.~these are only non-zero
when all its four indices are (anti-)holomorphic.
The sign $s$ depends on whether there is a null
index in $e_{i_1 \cdots i_I}$:
 \be
  s = \begin{cases}
   + 1 \,, \quad \text{for~~} e_{\a_1 \cdots \a_I} \,, \\
   - 1 \,, \quad \text{for~~} e_{\a_1 \cdots \a_{I-1} 5} \,.
   \end{cases}
 \ee
Observe that
\bea
  \{ e_{i_1 \cdots i_I}, \Gamma^{\bar p} e_{i_1 \cdots i_I}, \ldots, \Gamma^{\bar p_1 \cdots \bar p_5}
  e_{i_1 \cdots i_I}\} \, \label{oscbasis}
\eea
is a basis in the space of Dirac spinors and so $e_{i_1 \cdots i_I}$ can be thought of as another Clifford vacuum.
Therefore the terms in the square brackets in (\ref{oscalg}) are linearly independent.

We can apply the same procedure to the parallel transport Killing spinor equation (\ref{kseqna}).
In particular, we find  for $M=p$ that
 \bea
  \mathcal{D}_p e_{i_1 \cdots i_I} & = & [D_p + \tfrac{1}{2} \Omega_{p,} \cont{r}
  + \tfrac{i}{4} F_{p} \cont{r_1} \cont{r_2}] e_{i_1 \cdots i_I} +
  \nonumber \\
  && [\tfrac{1}{4} \Omega_{p, \bar q_1 \bar q_2} + \tfrac{i}{4} F_{p \bar q_1 \bar q_2} \cont{r}
   - \tfrac{s}{48} g_{p \bar q_1} \e_{\bar q_2 }{}^{r_1 \cdots r_3} G_{r_1 \cdots r_3} -
   \tfrac{s}{16} \e_{\bar q_1 \bar q_2}{}^{r_1 r_2} G_{p r_1 r_2}] \Gamma^{\bar q_1 \bar q_2} e_{i_1 \cdots i_I} + \nonumber \\
  && [ \tfrac{i}{48} F_{p \bar q_1 \cdots \bar q_4}
  + \tfrac{s}{96} \e_{\bar q_1 \cdots \bar q_3}{}^r G_{p \bar q_4 r}
  - \tfrac{s}{192} g_{p \bar q_1} \e_{\bar q_2 \cdots \bar q_4}{}^{r_1} G_{r_1} \cont{r_2} ]
  \Gamma^{\bar q_1 \cdots \bar q_4} e_{i_1 \cdots i_I} \,.
  \la{oscdiffa}
 \eea
Similarly, the resulting expression for (\ref{kseqna}) with $M = \bar p$ is
 \bea
  \mathcal{D}_{\bar p} e_{i_1 \cdots i_I} & = & [D_{\bar p} + \tfrac{1}{2} \Omega_{\bar p,} \cont{r}
   + \tfrac{i}{4} F_{\bar p} \cont{r_1} \cont{r_2}
   - \tfrac{s}{24} \e_{\bar p}{}^{r_1 \cdots r_3} G_{r_1 \cdots r_3}] e_{i_1 \cdots i_I} +
   \nonumber \\
  && [\tfrac{1}{4} \Omega_{\bar p, \bar q_1 \bar q_2}
  + \tfrac{i}{4} F_{\bar p \bar q_1 \bar q_2} \cont{r}
  + \tfrac{s}{32} \e_{\bar p \bar q_1 \bar q_2}{}^{r_1} G_{r_1} \cont{r_2}
  \nonumber
  \\
  &&
  - \tfrac{s}{32} \e_{\bar q_1 \bar q_2}{}^{r_1 r_2} G_{\bar p r_1 r_2}
  + \tfrac{s}{16} \e_{\bar p \bar q_1}{}^{r_1 r_2} G_{\bar q_2 r_1 r_2} ] \Gamma^{\bar q_1 \bar q_2} e_{i_1 \cdots i_I}
   +
   \nonumber \\
  && [ \tfrac{i}{48} F_{\bar p \bar q_1 \cdots \bar q_4}
  + \tfrac{s}{256} \e_{\bar q_1 \cdots \bar q_4} G_{\bar p} \cont r
  + \tfrac{s}{192} \e_{\bar p \bar q_1 \cdots \bar q_3} G_{\bar q_4} \cont{r}
  \nonumber
  \\
  &&
  + \tfrac{s}{48} \e_{\bar q_1 \cdots \bar q_3}{}^{r} G_{\bar p \bar q_4 r}  ]
  \Gamma^{\bar q_1 \cdots \bar q_4} e_{i_1 \cdots i_I} \,.
  \la{oscdiffb}
 \eea

To derive the linear system associated with the Killing spinor equations of a background with any number
of supersymmetries,  (\ref{oscalg}), (\ref{oscdiffa}) and (\ref{oscdiffb})  must be converted from
 the oscillator basis (\ref{oscbasis}) to the ``canonical basis''
 \be
  \{ 1, \Gamma^{\bar i} 1, \ldots, \Gamma^{\bar i_1 \cdots \bar i_5}
  1 \} \,. \label{canonical-basis}
 \ee
{}To achieve this, we expand the products of $\Gamma^{\bar p}$ matrices,
which are creation operators on $e_{i_1 \cdots i_I}$, into a sum of
products of $\Gamma^{j}$ and $\Gamma^{\bar j}$ matrices, which are
annihilation and creation operators, respectively, on $1$. Then we
act on $e_{i_1 \cdots i_I}$ with the annihilation operators. In
particular, we have
 \bea
  {\cal A} e_{i_1 \cdots i_I} &=& \sum_l [{\cal A} e_{i_1 \cdots
i_I}]_{\bar p_1 \cdots \bar p_l}
  \Gamma^{\bar p_1\cdots \bar p_l} e_{i_1 \cdots i_I} \nonumber \\ &=&
\sum_l \sum_{m+n=l}
 \frac{l!}{m! n!} [{\cal A} e_{i_1
 \cdots i_I}]_{j_1 \cdots j_m \bar k_1   \cdots \bar k_n}
 \Gamma^{j_1 \cdots j_m} \Gamma^{\bar k_1 \cdots
 \bar k_n} e_{i_1\dots i_I}
 \nonumber \\
 &=& \sum_l \sum_{m+n=l}
 \frac{l!}{m! n!}   \frac{(-1)^{[m/2]+nI}}{{2}^{{I / 2}-m} (I-m)!}
  \epsilon^{j_1 \cdots j_m}{}_{\bar j_{m+1} \cdots \bar j_I}
 \nonumber \\
  &&
  [{\cal A} e_{i_1 \cdots i_I}]_{j_1 \cdots j_m \bar k_1 \cdots \bar k_n}
   \Gamma^{\bar j_{m+1} \cdots \bar j_I \bar k_1 \cdots
  \bar k_n} 1\,,
  \la{genform}
 \eea
with the obvious restrictions $m \leq I$ and $n \leq 5-I$ and the
convention that $\epsilon_{\bar i_1 \cdots \bar i_I} = 1$. A similar
formula holds for
all components of $\cal{D}_M$. Using these
expressions one can easily compute the components of ${\cal
A} e_{i_1 \cdots i_I}$ and ${\cal
D}_M e_{i_1 \cdots i_I}$ in the canonical basis
(\ref{canonical-basis}). For convenience we give the explicit
expressions for ${\cal A}\sigma_I$ and ${\cal D}\sigma_I$ in the appendices.

\subsection{The linear system of integrability conditions}

The integrability conditions of the Killing spinor equations  (\ref{Iint}) and (\ref{IAint}) of
a IIB background with any number of supersymmetries can also be expressed in terms of those
on five types of spinors $\sigma_I$. To expand ${\cal I}\sigma_I$ and ${\cal I}_A\sigma_I$
in the canonical basis (\ref{canonical-basis}), we follow the same procedure as that
for the Killing spinor equations in the previous section. In particular, we first compute the integrability
 condition ${\cal I}$ in the oscillator basis
(\ref{oscbasis}) to find
 \bea
  \mathcal{I} e_{i_1 \cdots i_I} &=& [\LG \cont{r} + 12 \BG \cont{r_1} \cont{r_2}]
  e_{i_1 \cdots i_I} +  [ \tfrac{1}{2} \LG_{\bar q_1 \bar q_2}
  \nonumber \\
  &&
  + 12 \BG_{\bar q_1 \bar q_2} \cont{r} - \tfrac{s}{2}
   \e_{\bar q_1 \bar q_2}{}^{r_1 r_2} \BP_{r_1 r_2}] \G^{\bar q_1 \bar q_2} e_{i_1 \cdots i_I} + [ \BG_{\bar q_1 \cdots \bar q_4}
   \nonumber \\
  &&
  + \tfrac{s}{96}
  \e_{\bar q_1 \cdots \bar q_4} \LP + \tfrac{s}{48}
  \e_{\bar q_1 \cdots \bar q_4} \BP \cont{r} + \tfrac{s}{6} \e_{\bar q_1 \cdots \bar q_3}{}^r
  \BP_{\bar q_4 r} ] \G^{\bar q_1 \cdots \bar q_4} e_{i_1 \cdots i_I} \,.
 \eea
Similarly for the integrability condition $\mathcal{I}_A$,  we get
 \bea
  \mathcal{I}_p e_{i_1 \cdots i_I} &=& [\tfrac{1}{2} \E_{p \bar q}
  - 6 i \LF_{p \bar q} \cont{r} + 4 s g_{p \bar q} \e^{r_1 \cdots r_4}
  \BG_{r_1 \cdots r_4} - 8 s \e_{\bar q}{}^{r_1 \cdots r_3} \BG_{p r_1 \cdots r_3} ] \G^{\bar q} e_{i_1 \cdots i_I} + \nonumber \\
  && [-i \LF_{p \bar q_1 \cdots \bar q_3} + \tfrac{s}{12} \e_{\bar q_1
  \cdots \bar q_3}{}^r \LG_{pr} - 4 s g_{p \bar q_1} \e_{\bar q_2}{}^{r_1 \cdots r_3}
  \BG_{\bar q_3 r_1 \cdots r_3} + \nonumber \\
  && \; \; - 2 s \e_{\bar q_1 \cdots \bar q_3}{}^{r_1} \BG_{p r_1} \cont{r_2} ] \G^{\bar q_1 \cdots \bar q_3} e_{i_1 \cdots i_I} +
  [ - \tfrac{s}{96} \e_{\bar q_1 \cdots \bar q_4} \LG_{p \bar q_5}
   \nonumber \\
  &&
  + \tfrac{1}{8} s g_{p \bar q_1} \e_{\bar q_2 \cdots \bar q_5} \BG \cont{r_1} \cont{r_2}
  - \tfrac{1}{4} s \e_{\bar q_1 \cdots \bar q_4} \BG_{p \bar q_5} \cont{r} ]
  \G^{\bar q_1 \cdots \bar q_5} e_{i_1 \cdots i_I} \,,
 \eea
and
  \bea
  \mathcal{I}_{\bar p} e_{i_1 \cdots i_I} &=& [\tfrac{1}{2} \E_{\bar p \bar q}
  - 6 i \LF_{\bar p \bar q} \cont{r} - 16 s \e_{\bar p}{}^{r_1 \cdots r_3} \BG_{\bar q r_1 \cdots r_3}
  + 8 s \e_{\bar q}{}^{r_1 \cdots r_3} \BG_{\bar p r_1 \cdots r_3} + \nonumber \\
  && \; \; - 24 s \e_{\bar p \bar q}{}^{r_1 r_2} \BG_{r_1 r_2} \cont{r_3}]
  \G^{\bar q} e_{i_1 \cdots i_I} + \nonumber \\
   && [ -i \LF_{\bar p \bar q_1 \cdots \bar q_3} + \tfrac{s}{12} \e_{\bar q_1 \cdots \bar q_3}{}^r \LG_{\bar p r}
   - 9 s \e_{\bar p \bar q_1}{}^{r_1 r_2} \BG_{\bar q_2 \bar q_3 r_1 r_2}
   - 3 s \e_{\bar q_1 \bar q_2}{}^{r_1 r_2} \BG_{\bar p \bar q_3 r_1 r_2} + \nonumber \\
  && \; \; + 6 s \e_{\bar p \bar q_1 \bar q_2}{}^{r_1} \BG_{\bar q_3 r_1} \cont{r_2} ]
  \G^{\bar q_1 \cdots \bar q_3} e_{i_1 \cdots i_I} + \nonumber \\
  && [ - \tfrac{s}{96} \e_{\bar q_1 \cdots \bar q_4} \LG_{\bar p \bar q_5}
  + \tfrac{s}{4} \e_{\bar q_1 \cdots \bar q_4} \BG_{\bar p \bar q_5} \cont{r} ]
  \G^{\bar q_1 \cdots \bar q_5} e_{i_1 \cdots i_I} \,.
 \eea
It remains to convert the above expressions from the oscillator basis (\ref{oscbasis}) to the canonical basis
(\ref{canonical-basis}). This can be done as in (\ref{genform}) and we shall not repeat the formula here.
The explicit expressions of the integrability conditions in the canonical basis can be found in the appendices.

\newsection{Generic $N=2$ backgrounds with $SU(4)\ltimes \bR^8$-invariant spinors}

\subsection{Preliminaries}

Applying the results of section \ref{invspin} to this case, one finds that the two Killing spinors of $N=2$ backgrounds
with $SU(4)\ltimes \bR^8$-invariant spinors can be written as
\bea
\epsilon=z~\eta
\eea
where $\eta_1=1+e_{1234}$ and $\eta_2=i(1-e_{1234})$ and  $z$ is a complex
$2\times 2$ matrix.
There are two classes of such  backgrounds. For {\it generic} half-maximal
$SU(4)\ltimes \bR^8$-backgrounds the matrix $z$ is non-degenerate, ${\rm det}z\not=0$,
whereas the {\it degenerate} half-maximal
$SU(4)\ltimes \bR^8$-backgrounds have ${\rm det}z=0$ but the two Killing spinors are linearly independent
over the real numbers. To solve the Killing spinor
equations in the generic case, we adapt the formalism developed  in
section \ref{invspin}.  In this section, we shall investigate
the generic class of half-maximal $SU(4)\ltimes \bR^8$-backgrounds.
The degenerate half-maximal $SU(4)\ltimes \bR^8$-backgrounds
will be examined in section \ref{degencase}.

\subsection{The solution of the linear system}

The linear system and its solution are described in appendix \ref{conconap}.
It turns out that to solve the linear system for generic half-maximal $SU(4)\ltimes\bR^8$-backgrounds one has to
consider two cases depending on whether or not $(A_{11}- A_{22})^2+ (A_{12}+A_{21})^2=0$,
where $A=z^{-1} z^*$. In both cases
after solving for the fluxes, the resulting equations can be separated into two classes. One class are the
algebraic equations which do not contain spacetime derivatives of the functions $z$, e.g.
(\ref{sfive}) and (\ref{seight}) .
The other case
are first order equations for the functions $z$ which however
are {\it non-linear} in $z$, e.g. (\ref{minuscomp}), (\ref{pluscomp}),
(\ref{genericcon1}) and (\ref{genericcon2}).
These first order equations can be viewed as the parallel transport equations of the restriction of the
supercovariant connection on the bundle of the Killing spinors ${\cal K}$. However, since the system
 is non-linear,
the analysis of the general case is rather involved. So instead of solving the system in general,
we shall investigate two examples. In the first example $(A_{11}- A_{22})^2+ (A_{12}+A_{21})^2\not=0$ while
in the second $(A_{11}- A_{22})^2+ (A_{12}+A_{21})^2=0$.

\subsubsection{Special cases}

The matrix $z$ in the example with $(A_{11}- A_{22})^2+ (A_{12}+A_{21})^2\not=0$ is chosen as
\bea
z=\begin{pmatrix}
z_{11}& 0\cr
            0&z_{22}
\end{pmatrix}
~,~~~z_{11}\not=z_{22}\not=0~,
\eea
where $z_{11}$ and $z_{22}$ are complex.
Without loss of generality, we shall also work in the gauge for which
 $\det z=1$ and so $z_{11} z_{22}=1$.  This gauge can always be locally attained
 by an appropriate $U(1)$ gauge transformation
 and an appropriate $Spin(9,1)$ transformation along $\Gamma_{05}$. We therefore can  set
\be
z_{11} = \rho_1 e^{i \varphi}, \qquad z_{22}= \rho_2 e^{-i \varphi}, \qquad \rho_1\rho_2=1
\ee
for $\rho_1=\rho, \rho_2=\rho^{-1}, \phi \in \bR$, $\rho> 0$. Consequently,
$A={\rm diag}(e^{-2i\varphi}, e^{2i\varphi})$.

We shall not go through a detailed analysis of the solution. This has been done in \ref{coconapsp}.
 Instead, we shall summarize the conditions
on the geometry and fluxes. In particular we find that the conditions on the geometry are
\bea
&&\Omega_{\a,\b+}=0~,~~~\Omega_{+,\a\b}=0~,~~~\Omega_{+,+\a}=0~,~~~Q_+=Q_\a=0~,~~~\Omega_{+,-+}=0
\cr
&&\partial_+\varphi=0~,~~~\partial_A\rho=0~,~~~\Omega_{+,\a}{}^\a=0~,~~~\Omega_{-,-+}=0~,~~~
\Omega_{\a,\bar\b+}=0~,
\cr
&&\Omega_{\b,}{}^\b{}_{\bar\a}=-{3\over \cos(4\varphi)+2} \Omega_{-,+\bar\a}~,~~~\Omega_{[\b_1, \b_2\b_3]}=0~,
\cr
&&\Omega_{\a,-+}+\Omega_{-,\a+}=0~,~~~\partial_{\bar\a}\varphi=-{\sin(4\varphi)\over \cos(4\varphi)+2} \Omega_{\bar\a,-+}~,
\cr
&&\Omega_{\bar\a,\b}{}^\b+{\cos(4\varphi)\over \cos(4\varphi)+2} \Omega_{-,+\bar\a}=0~,~~~
\Omega_{\a,\bar\b\bar\g}={2\over \cos(4\varphi)+2} \Omega_{-,+[\bar\b} g_{\bar\g]\a}~,
\la{complgeom}
\eea
the conditions on the $G$ and $P$ fluxes are
\bea
&&G_{+\a}{}^\a=0~,~~~P_+=0~,~~~G_{+\a\b}=G_{+\bar\a\bar\b}=0~,~~~G_{-\a}{}^\a=-{2\over\cos(2\varphi)} \Omega_{-,\a}{}^\a
\cr
&&G_{-\bar\a\bar\b}=-2 \cos(2\varphi) (\Omega_{-,\bar\a\bar\b}+i F_{-\bar\a\bar\b\g}{}^\g)
+i\sin(2\varphi) \epsilon_{\bar\a\bar\b}{}^{\g\d} (\Omega_{-,\g\d}-i F_{-\g\d\zeta}{}^\zeta)
\cr
&&G_{-\a\b}=-2 \cos(2\varphi) (\Omega_{-,\a\b}-i F_{-\a\b\g}{}^\g)
+i\sin(2\varphi) \epsilon_{\a\b}{}^{\bar\g\bar\d} (\Omega_{-,\bar\g\bar\d}+i F_{-\bar\g\bar\d\zeta}{}^\zeta)
\cr
&&P_{\bar\a}=(P_\a)^*={2\over \cos(4\varphi)+2} \Omega_{-,+\bar\a}~,~~~
G_{-+\bar\a}=(G_{-+\a})^*=-8 {\cos(2\varphi)\over \cos(4\varphi)+2} \Omega_{-,+\bar\a}
\cr
&&G_{\bar\a\b}{}^\b=G_{a\b}{}^\b=0~,~~~\epsilon_{\bar\a}{}^{\b_1\b_2\b_3} G_{\b_1\b_2\b_3}=
-(\epsilon_\a{}^{\bar\b_1\bar\b_2\bar\b_3} G_{\bar\b_1\bar\b_2\bar\b_3})^*=24 i
{\sin(2\varphi)\over \cos(4\varphi)+2}
\Omega_{-,+\bar\a}
\cr
&&G_{+\a\bar\b}=0~,~~~G_{\bar\a\b\g}=-(G_{\a\bar\b\bar\g})^*=i{1\over \sin(2\varphi)} \Omega_{\bar\a, \bar\d_1\bar\d_2} \epsilon_{\b\g}{}^{\bar\d_1\bar\d_2}
\la{complgp}
\eea
and the conditions on the $F$ fluxes, in addition to the self-duality,  are
\bea
&&F_{+\a\bar\b_1\bar\b_2\bar\b_3}=0~,~~~F_{-\a}{}^\a{}_\b{}^\b=2 Q_-~,~~~F_{-\b_1\b_2\b_3\b_4} \epsilon^{\b_1\b_2\b_3\b_4}
=-12\partial_-\varphi-6 \tan(2\varphi) \Omega_{-,\a}{}^\a~,
\cr
&&\epsilon_{\bar\a}{}^{\b_1\b_2\b_3} F_{-+\b_1\b_2\b_3}
=-3 {\sin(4\varphi) \over \cos (4\varphi)+2} \Omega_{-,+\bar\a}~,~~~F_{-+\bar\a\b}{}^\b=0~,~~~
F_{+\a_1\a_2\bar\b_1\bar\b_2}=0
\cr
&&
F_{-+\a\bar\b\bar\g}=-{1\over4}\mathrm{cotan} (2\varphi) \Omega_{\a,\g_1\g_2}
 \epsilon^{\g_1\g_2}{}_{\bar\b\bar\g}~,
 \cr
 &&i {\rm Im}(F_{-\a_1\a_2\a_3\a_4}
 \epsilon^{\a_1\a_2\a_3\a_4})=-6 \tan(2\varphi) \Omega_{-,\a}{}^\a~.
\la{complf}
\eea
The conditions above have an explicit dependence on the angle $\varphi$. This is due to the
non-linearity of the Killing spinor equation on the functions $z$.

The other special case that we shall consider is to take $z$ to be a real matrix. In this case,
$A=1_{2\times2}$ is the identity  matrix and the non-linear system becomes linear.
This case closely resembles  the maximally supersymmetric $SU(4)\ltimes \bR^8$ case that we
have investigated in \cite{ugjggpb}. As in the previous case, we shall simply summarize the
solution of the linear system. The conditions on the geometry
are
\bea
&&(z^{-1}\partial_+ z)_{11}=(z^{-1}\partial_+ z)_{22}=-{1\over2} \Omega_{+,-+}~,~~~
\cr
&&(z^{-1}\partial_+ z)_{12}=-(z^{-1}\partial_+ z)_{21}={i\over2} \Omega_{+, \a}{}^\a~,
\cr
&&(z^{-1}\partial_-z)_{11}=(z^{-1}\partial_-z)_{22}=-{1\over2} \Omega_{-,-+}~,
\cr
&&(z^{-1}\partial_-z)_{12}=-(z^{-1}\partial_-z)_{21}={i\over2} \Omega_{-,\a}{}^\a+{i\over4} G_{-\a}{}^\a~,
\cr
&&(z^{-1}\partial_{\bar\a}z)_{11}=(z^{-1}\partial_{\bar\a}z)_{22}=-{1\over2} \Omega_{\bar\a,-+}
-{1\over2} \Omega_{\b,}{}^\b{}_{\bar\a}~,
\cr
&&(z^{-1}\partial_{\bar\a}z)_{12}=-(z^{-1}\partial_{\bar\a}z)_{21}=
{i\over2} \Omega_{\bar\a,\b}{}^\b-{i\over6} \Omega_{\b,}{}^\b{}_{\bar\a}
+{i\over12} (G_{\bar\a\b}{}^\b-(G_{\a\b}{}^\b)^*)~,
\cr
&&Q_+=0~,~~~\Omega_{\a,\b+}=0~,~~~\Omega_{+,\a\b}=0~,~~~\Omega_{+,+\a}=0~,
\cr
&&\Omega_{\a,+\bar\b}+\Omega_{\bar\b,+\a}=0~,~~~\Omega_{\a_1,\a_2\a_3}=0~,~~~\Omega_{-+\bar\a}+\Omega_{\b,}{}^\b{}_{\bar\a}=0~,
\la{realgeom}
\eea
the conditions on the $P$ and $G$ fluxes are
\bea
&&P_+=0~,~~~G_{+\a}{}^\a=0~,~~~G_{+\a\b}=G_{+\bar\a\bar\b}=0~,~~~G_{-\a}{}^\a+(G_{-\a}{}^\a)^*=0~,
\cr
&&G_{-\a\b}=-2 (\Omega_{-,\a\b}-iF_{-\a\b\g}{}^\g)~,~~~G_{-\bar\a\bar\b}=-2 (\Omega_{-\bar\a\bar\b}+iF_{-\bar\a\bar\b\g}{}^\g)~,
\cr
&&G_{+\a\bar\b}=\Omega_{\a,+\bar\b}-\Omega_{\bar\b,+\a}~,~~~P_{\bar\a}=-{1\over3} G_{\bar\a\b}{}^\b-{2\over3}
(\Omega_{\b,}{}^\b{}_{\bar\a}-i F_{-+\bar\a\b}{}^\b)~,
\cr
&&G_{-+\bar\a}={1\over3} G_{\bar\a\b}{}^\b+{8\over3} (\Omega_{\b,}{}^\b{}_{\bar\a}-i F_{-+\bar\a\b}{}^\b)~,~~~G_{\a\b\g}=
G_{\bar\a\bar\b\bar\g}=0~,
\cr
&&P_\a={1\over3} G_{\a\b}{}^\b-{2\over3} (\Omega_{\bar\b,}{}^{\bar\b}{}_\a+iF_{-+\a\b}{}^\b)~,
~~~
G_{-+\a}=-{1\over3} G_{\a\b}{}^\b+{8\over3} (\Omega_{\bar\b,}{}^{\bar\b}{}_\a+iF_{-+\a\b}{}^\b)~,
\cr
&&G_{\a\bar\b_1\bar\b_2}=-2 (\Omega_{\a, \bar\b_1\bar\b_2}+2i F_{-+\a\bar\b_1\bar\b_2})-
\delta_{\a[\bar\b_1}\big(-{2\over3} G_{\bar\b_2]\g}{}^\g-{4\over3} \Omega_{|\gamma|,}{}^\g{}_{\bar\b_2]}-
{8i\over3} F_{\bar\b_2]-+\g}{}^\g\big)
\cr
&&G_{\bar\a\b_1\b_2}=-2 (\Omega_{\bar\a,\b_1\b_2}+2i F_{-+\bar\a\b_1\b_2})- \delta_{\bar\a[\b_1}
\big({2\over3} G_{\b_2]\g}{}^\g-{4\over3} \Omega_{|\bar\g|,}{}^{\bar\g}{}_{\bar\b_2]}+{8i\over3}
F_{\bar\b_2]-+\g}{}^\g\big)~
\la{realpg}
\eea
and the conditions on the $F$ fluxes, in addition to the self-duality, are
\bea
&&F_{+\a\bar\b_1\bar\b_2\bar\b_3}=0~,~~F_{-\a}{}^\a{}_\b{}^\b=2Q_-~,~~~F_{-\b_1\b_2\b_3\b_4}=0~,
\cr
&&F_{+\a\bar\b\g}{}^\g=0~,~~~F_{-+\a\b}{}^\b={1\over2} Q_\a~,~~~
F_{-+\bar\a\b}{}^\b={i\over8} (G_{\bar\a\b}{}^\b+(G_{\a\b}{}^\b)^*)~,
\cr
&&F_{-+\a_1\a_2\a_3}=0~.~~~
\la{realf}
\eea
Observe that the conditions have been expressed in representations of $SU(4)\ltimes\bR^8$ as
 may have been expected.

\subsection{The geometry}

To investigate the geometry of generic half-maximally supersymmetric $SU(4)\ltimes\bR^8$ backgrounds, one
has to solve the parallel transport equation
\bea
z^{-1} dz+C=0~.
\eea
However as we have seen that the connection depends on $z$, $C= C(z)$,
the first order system is non-linear.
Because of this, we shall focus on the geometry of two examples we have described in the previous section.

First, let us consider the case for which $z$ is diagonal. The conditions on
the geometry (\ref{complgeom}) imply that
the Killing spinors can be written as
\bea
\epsilon_1= e^{i\varphi} \eta_1~,~~~
\epsilon_2=e^{-i\varphi} \eta_2~,
\la{kscompl}
\eea
where $\varphi$ depends on the spacetime coordinates.
In addition $\varphi$ satisfies a parallel
transport equation
\bea
d\varphi+C=0
\eea
where
\bea
C_+&=&0~,~~~C_-={1\over2} \tan(2\varphi) \Omega_{-,\a}{}^\a+{1\over12} F_{-\b_1\b_2\b_3\b_4}
\epsilon^{\b_1\b_2\b_3\b_4}~,
\cr
C_{\bar\a}&=&{\sin(4\varphi)\over \cos(4\varphi)+2} \Omega_{\bar\a,-+}~.
\eea
Observe that $C$ depends on the angle $\varphi$ as  expected.
The dependence of $\epsilon_1, \epsilon_2$
on the angle $\varphi$ cannot be eliminated with a $Spin(9,1)\times U(1)$ gauge transformation
because we have already used such  transformations to simplify the Killing spinors\footnote{We do not expect
additional $Spin(9,1)\times U(1)$ gauge transformations that preserve the space spanned by $\epsilon_1, \epsilon_2$
to exist, apart from the subgroup $Spin(1,1)\times U(1)$ that we have already used.}.  The angle $\varphi$ is determined
by the field equations.

To investigate further the geometry of these backgrounds, one can compute the spacetime form bi-linears
associated with the Killing spinors (\ref{kscompl}). This can be easily done using the results in appendix A.
For this we introduce a light-cone frame and write the spacetime metric as
\bea
ds^2= 2 e^- e^+ +\delta_{IJ} e^I e^J~,~~~I,J=1,\dots,8~.
\la{frametr}
\eea
Then after an appropriate normalization, one finds that the ring of Killing spinor bi-linears is
generated by
\bea
\kappa=e^-~,~~~ \xi=e^-\wedge \omega~,~~~\tau=e^-\wedge\chi~,~~~\tau^*=e^-\wedge \chi^*~,~~~\lambda=e^-\wedge \omega\wedge\omega
\la{genring}
\eea
as  can be seen from the bi-linears $\kappa(\epsilon_1, \tilde\epsilon_1)$, $\xi(\epsilon_1, \epsilon_2)$,
$\tau(\epsilon_1, \epsilon_2)$, $\tau(\epsilon_1, \tilde \epsilon_1)$ and $\tau(\epsilon_2, \tilde \epsilon_2)$.
Observe that the remaining bi-linears in appendix A depend on the angle $\varphi$.
Clearly the ring of bi-linears is two step nilpotent\footnote{Compare this with the ring  of invariant forms on 2n-dimensional
manifolds with an  $SU(n)$-structure which is not nilpotent.}.
The one-form $\kappa$ is associated with a {\it null Killing} vector field $X=e_+$. However unlike the case
of maximally supersymmetric $SU(4)\ltimes \bR^8$-backgrounds, $X$ is not
rotation free. It is straightforward to express the various components of the Levi-Civita connection $\Omega$
that appear in the geometry relations (\ref{complgeom}) in terms of the covariant derivatives of the generators (\ref{genring}).
Then one could rewrite (\ref{complgeom}) in terms of the covariant derivatives. However, we shall not do this here.
Observe  that the various geometry conditions in (\ref{complgeom}) depend on the angle $\varphi$ and that the
covariant derivatives of (\ref{genring}) do not depend on $\varphi$. As a result, the geometry conditions (\ref{complgeom})
are not simply
linear relations between the various components of the covariant derivatives of (\ref{genring})
as in the case of Hermitian manifolds in \cite{grayhervella}.

Next, let us take $z$ to be a real invertible matrix. The parallel transport equation for $z$
is given in (\ref{realgeom}). It is easy to see that the connection $C$ can be written as
\bea
C=\hat \Omega^0 t_0+ \hat \Omega^1 t_1
\eea
where $t_0=1_{2\times2}$, $t_1$ is skew-symmetric with $(t_1)_{12}=1$, and $\hat\Omega^0$ and $\hat\Omega^1$ are easily
computed from (\ref{realgeom}). Since $t_0$ and $t_1$
commute, $C$ is an abelian connection. In addition $C$ does not depend on $z$ because $A=1_{2\times2}$
in this example. A necessary condition for the existence of a solution to the parallel transport
problem is that the curvature of $C$, $F=dC$, must vanish. In addition, it turns out that
$C$ can be trivialized with the $e^{a\Gamma_{05}+b \Gamma_{16}}$ gauge $Spin(9,1)$ transformation
for suitable choices of the parameters $a,b$. Therefore, we have shown that up to a $Spin(9,1)$
gauge transformation, we can set $z=1_{2\times 2}$ and so the two Killing spinors\footnote{It is not always
possible to find a gauge such that the Killing spinors of a supersymmetric
 background are all constant. This has to be shown in each case. An explanation of this   has
 been given
in \cite{ugjggpb}. As an example, it can be shown that the only maximally supersymmetric background of IIB
supergravity  with
constant Killing spinors is locally isometric to Minkowski spacetime but it is known that there are two more
maximally supersymmetric backgrounds the
$AdS_5\times S^5$ and the plane wave.}
\bea
\epsilon_1&=&\eta_1
\cr
\epsilon_2&=&\eta_2~.
\eea
Setting $z=1_{2\times 2}$ in (\ref{realgeom}), the resulting equations are interpreted either
as restrictions on the geometry or as conditions that relate components of the fluxes to the
geometry.

To further investigate the geometry, we write the spacetime metric in a light-cone frame as in (\ref{frametr}).
Then it can be easily seen from the results of appendix A that the ring of $SU(4)\ltimes \bR^8$ forms
is generated by the forms in (\ref{genring}). However unlike the previous case, all spinor bi-linears
are constant in the frame $e^-, e^+, e^I$.
The ring of spinor bi-linears  is again two step nilpotent.
It is easy to see from the conditions on the geometry that $\kappa$ is a {\it null Killing} vector field.
Unlike the maximally supersymmetric $SU(4)\ltimes\bR^8$-backgrounds, $\kappa$ is not rotation free.
In particular observe that $(d\kappa)_{\a\bar\b}$ is proportional to  $G_{+\a\bar\b}$.
One can proceed further to re-express the conditions on the geometry in terms of the Levi-Civita
covariant derivatives of the spacetime forms $\kappa$, $\xi$, $\tau$, $\tau^*$ and $\lambda$ as suggested
in \cite {grayhervella}. However, we shall not do this here.

\newsection{ Degenerate $N=2$ backgrounds with $SU(4)\ltimes \bR^8$-invariant spinors}\la{degencase}

\subsection{Preliminaries}

The Killing spinors of degenerate half-maximal $SU(4)\ltimes \bR^8$-backgrounds can be written as
\bea
\epsilon_1&=&(f-g_2+ig_1) 1+ (f+g_2+ig_1) e_{1234}
\cr
\epsilon_2&=& w \epsilon_1~,~~~~~~~~~~~~~~~~~~~~~~~~~~~~~~~~~ {\rm Im} w\not=0~,
\eea
where $f, g_1, g_2$ are real functions, $f, g_2\not=0$.
To see this, it can always be arranged such that $z_{11}\not=0$. If this is
the case, then solving the condition ${\rm det}z=0$ for $z_{22}$ and substituting
it into the second spinor, one finds that $\epsilon_2= w \epsilon_1$, where $w=z_{21}/z_{11}$. If $w$
is real, then $\eta_1$ and $\eta_2$ are linearly dependent and so this case is excluded. Thus, we take
$w$ to be complex with ${\rm Im}~w\not=0$.

The algebraic Killing spinor equation for $\epsilon_2$ can be written
as
\bea
w^{-1} w^* P_A\Gamma^AC\epsilon_1^*+{1\over24} G_{ABC} \Gamma^{ABC}\epsilon_1=0~.
\eea
Subtracting this from the algebraic Killing spinor equation of $\epsilon_1$ and using ${\rm Im}~w\not=0$, one finds
that
\bea
P_A\Gamma^A C\epsilon_1^*=0~,~~~~G_{ABC} \Gamma^{ABC} \epsilon_1=0~.
\la{spone}
\eea
Similarly using ${\cal D}\epsilon_1=0$, the Killing spinor equations of  $\epsilon_2$
associated with the supercovariant derivative ${\cal D}$ become
\bea
\partial_A w \epsilon_1-{1\over 96} ( w^*-w)[ \Gamma_{A}{}^{B_1B_2B_3} G_{B_1B_2B_3}-
9 \Gamma^{B_1B_2} G_{AB_1B_2}] C\epsilon_1^*=0~.
\la{sptwo}
\eea
In turn the Killing spinor equation associated with the supercovariant derivative of $\epsilon_1$ can be rewritten as
\bea
\tilde\nabla_A \epsilon_1-(w^*-w)^{-1} \partial_A w \epsilon_1+{i\over 48} \Gamma^{N_1\dots N_4 } \epsilon_1
 F_{N_1\dots N_4 M}=0~.
 \la{spthree}
\eea
Therefore the independent equations that have to be solved in this case are (\ref{spone}), (\ref{sptwo}) and
(\ref{spthree}). It is clear that in this special case, the Killing spinor equations
factorize in a way similar to that we have encountered for maximally supersymmetric $H$-backgrounds.
The linear system associated with the above Killing spinor equations is given in appendix \ref{degencaseap}.

  \subsection{The solution of the linear system}

 The linear system associated with the
 Killing spinor equations has been presented  in appendix \ref{degencaseap}.
 We shall not explain in detail the solution of this system. It turns out that  it is simpler than that
 of
 $N=1$ backgrounds with a $SU(4)\ltimes \bR^8$ invariant Killing spinor \cite{ugjggpa}. First, we
 shall summarize the conditions that are implied from the equations involving the
 $G$ and $P$ fluxes and then we shall give the conditions that are implied by the rest of the equations.
 In particular, we have
 \bea
 \partial_+ w=\partial_- w=\partial_\a w=\partial_{\bar\a}w&=&0
 \cr
P_+= G_{-\b}{}^\b=G_{+\b}{}^\b=G_{-+\a}=G_{\bar\a_1\bar\a_2\bar\a_3}=G_{\a\b}{}^\b&=&0
 \cr
G_{+\a\b}=G_{+\bar\a\bar\b}=G_{-+\bar\a}=G_{\a_1\a_2\a_3}=G_{\bar\a\b}{}^\b=G_{+\a\bar\b}&=&0
 \cr
(f+g_2-ig_1)P_{\bar\a}=0~,~~~(f-g_2-ig_1) P_{\a}=0~,~~~
 \cr
 (f-g_2-ig_1) G_{\bar\a\b\g}=0~,~~~(f+g_2-ig_1) G_{\a\bar\b\bar\g}&=&0
 \cr
 G_{-\bar\b_1\bar\b_2} (f+g_2-i g_1) -{1\over2} (f-g_2-i g_1) G_{-\g_1\g_2} \epsilon^{\g_1\g_2}{}_{\bar\b_1\bar\b_2}&=&0~.
 \eea
 The first equation implies that the complex function $w$ is constant. Therefore $\epsilon_2$ is linearly dependent
 to $\epsilon_1$ over the complex numbers as expected. The last three equations require some explanation. If the Killing spinor
 $\epsilon_1$ is not pure, then $P_\a=P_{\bar\a}=0$ and similarly  $G_{\bar\a\b\g}=G_{\a\bar\b\bar\g}=0$.
 However if the Killing spinor $\epsilon_1$ is pure, then either $P_{\bar\a}=0$ or $P_\a=0$ and similarly
 either the (1,2) or the (2,1) component of
 $G$ vanishes depending on whether $\epsilon_1$ is proportional to either $1$ or $e_{1234}$, respectively. In addition,
 if $\epsilon_1$ is a  pure spinor, the last equation implies that either $G_{-\a\b}$ or  $G_{-\bar\a\bar\b}$ will vanish.
 The Killing spinor equations
do  not determine the traceless $G_{-\a\bar\b}$ component of $G$.

Next, we summarize the conditions on the flux $F$. In addition to the self-duality condition on $F$, we find that
\bea
i F_{-+\bar\b\g}{}^\g&=&{f^2+g_2^2+g_1^2\over 2f g_2} \Omega_{-,+\bar\b}
\cr
i F_{-+\a\b\g}&=&-{f^2-(g_2-i g_1)^2\over 4 f g_2} \Omega_{-,+\bar\d} \epsilon^{\bar\d}{}_{\a\b\g}
\cr
i F_{-\bar\a_1\bar\a_2\g}{}^\g&=&-{1\over 2fg_2} [-(f^2+g_2^2+g_1^2)\Omega_{-,\bar\a_1\bar\a_2}
\cr
&& + {1\over2} (f^2-(g_2+i g_1)^2)
\Omega_{-, \b_1\b_2} \epsilon^{\b_1\b_2}{}_{\bar\a_1\bar\a_2}]
\cr
i F_{-\b_1\b_2\b_3\b_4} \epsilon^{\b_1\b_2\b_3\b_4}&=&-{3\over f g_2} [ (\partial_- +\Omega_{-,\g}{}^\g+\Omega_{-,-+})
(f^2-(g_2-ig_1)^2)]
\cr
i F_{-\b}{}^\b{}_\g{}^\g&=&{1\over f g_2} [2i Q_- f g_2-2i g_2\partial_-g_1+2i g_1 \partial_-g_2
+\Omega_{-,\g}{}^\g (f^2+g_2^2+g_1^2)]
\cr
F_{+\a\bar\b_1\bar\b_2\bar\b_3}&=&0~,
\cr
iF_{-+\a\bar\b_1\bar\b_2}&=&{f g_2\over f^2+g_2^2+g_1^2} [\Omega_{\a, \bar\b_1\bar\b_2}+2 \Omega_{\g,}{}^\g{}_{[\bar\b_1}
g_{\bar\b_2]\a}]
\la{sinFeqn}
\eea
and if $\epsilon_1$ is not pure
\bea
F_{+\a\bar\b\g}{}^\g=0~.
\eea
Alternatively, if $\epsilon_1$ is pure the last equation becomes
\bea
i F_{+\a\bar\b\g}{}^\g=\pm \Omega_{\a,+\bar\b}
\eea
where the sign depends on whether the pure spinor is proportional to $1$ or to $e_{1234}$, respectively.
Taking the complex conjugate of the above relation, we find
\bea
\Omega_{\a,+\bar\b}+\Omega_{\bar\b,+\a}=0~.
\eea
Observe that in the pure spinor case some of the components of $F$ in (\ref{sinFeqn}) vanish.

Finally, the conditions on the geometry are
\bea
\Omega_{+,\a\b}=0~,~~~i Q_+ g_2+\Omega_{+,\g}{}^\g f&=&0~,
\cr
2\partial_+f+ Q_+ g_1+\Omega_{+,-+} f&=&0~,
\cr
2\partial_+ g_2-i g_1 \Omega_{+,\g}{}^\g+\Omega_{+,-+} g_2&=&0~,
\cr
2 \partial_+ g_1- Q_+ f+i \Omega_{+,\g}{}^\g g_2+ g_1 \Omega_{+,-+}&=&0~,
\cr
\partial_- (f^2+g_2^2+g_1^2)+\Omega_{-,-+} (f^2+g_2^2+g_1^2)&=&0~,
\cr
\Omega_{\a,+}{}^\a=0~,~~~\Omega_{\a,+\b}&=&0~,
\cr
4 f^2 g_2^2 \Omega_{\g,}{}^\g{}_{\bar\b}+(f^2+g_2^2+g_1^2)^2 \Omega_{-,+\bar\b}&=&0~,
\cr
{1\over2} (f^2+g_2^2+g_1^2) \Omega_{\a, \g_1\g_2} \epsilon^{\g_1\g_2}{}_{\bar\b_1\bar\b_2}- \Omega_{\a,\bar\b_1\bar\b_2}
[ f^2-(g_2-i g_1)^2]&=&0~,
\eea
and
\bea
Q_\alpha &=& {if \over g_2}\Omega_{\alpha, \beta}{}^\beta -i{(g_2 -ig_1) \over f g_2{}^2}(f^2+g_1{}^2-ig_1 g_2)
\Omega_{-,+\alpha}
\cr
\partial_\alpha f &=& -i{g_1 f \over 2 g_2}\Omega_{\alpha, \beta}{}^\beta
-{1 \over 2}f \Omega_{\alpha,-+}
\cr
&+&{i(g_2-ig_1) \over 2f g_2{}^2}(g_1(f^2+g_1{}^2+g_2{}^2)
-ig_2(g_1{}^2+g_2{}^2)) \Omega_{-,+\alpha}
\cr
\partial_\alpha g_1 &=& i{(f^2-g_2{}^2) \over 2 g_2}\Omega_{\alpha, \beta}{}^\beta
-{1 \over 2}g_1 \Omega_{\alpha,-+}
\cr
&-&{i \over 2 g_2{}^2}(g_2(f^2-g_2{}^2)-ig_1(f^2+g_1{}^2+2g_2{}^2)) \Omega_{-,+\alpha}
\cr
\partial_\alpha g_2 &=&{i \over 2}g_1 \Omega_{\alpha, \beta}{}^\beta -{1 \over 2}g_2
\Omega_{\alpha,-+}+{1 \over 2g_2}(f^2+g_1{}^2-ig_1 g_2) \Omega_{-,+\alpha}
\eea

and if $\epsilon_1$ is not pure
\bea
\Omega_{\a,+\bar\b}=0~.
\eea

Observe that some of the conditions on the fluxes $F$ can be interpreted as conditions on the geometry because they restrict the
$\partial_-$ derivative of functions $f, g_1, g_2$ which determine the spinor. In turn, the integrability conditions
restrict both the fluxes and the geometry.

\subsection{The geometry}

The linearly independent forms on the spacetime associated with the Killing spinor bi-linears are those
that we have computed in \cite{ugjggpa} for the case of $N=1$ $SU(4)\ltimes \bR^8$-backgrounds.
Because of this, we shall not present them here. It turns out that the vector field $X$
associated to the form
\bea
\kappa=(f^2+g_1^2+g_2^2) e^-
\eea
is a {\it null Killing} vector field. Moreover one can choose the gauge $f^2+g_1^2+g_2^2=1$. This is achieved
by an  $Spin(9,1)$ transformation $e^{a\Gamma_{05}}$ for an appropriate choice of the parameter $a$.
The metric then can be put into the form (\ref{frametr}). Observe  that the first order system
for the functions $f,g_1, g_2$ is again non-linear.

\newsection{Examples of integrability conditions}

In this section we will solve the linear systems associated with the
integrability conditions for maximal\footnote{The Killing spinor equations for these cases were
solved in \cite{ugjggpb}. The same holds for the case with maximal
$G_2$ supersymmetry, but since this yielded a purely gravitational
solution we will not discuss it.} $Spin(7)\ltimes\bR^8$- and $SU(4)\ltimes\bR^8$-backgrounds.
This will determine which
field equations are implied by the Killing spinor equations. The
remaining field equations, which still need to be imposed on the
supersymmetric background, will be given explicitly.

There are linear systems associated to the Killing spinor equations
and to  integrability conditions of any supersymmetric background.
However, as it was explained in section~\ref{invspin}, these systems
factorize for maximal $H$-backgrounds. In particular, the linear
system of integrability conditions splits up in three separate parts
involving only two types of field equations: $\E$ and $\LF$, $\LP$
and $\BP$ and $\LG$ and $\BG$. This considerably simplifies the
analysis of the linear systems. In what follows, we shall apply the
formalism to the linear systems of the maximal $Spin(7) \ltimes
\bR^8$- and $SU(4) \ltimes \bR^8$-backgrounds.

\subsection{Maximal $Spin(7) \ltimes \bR^8$-backgrounds}

The field equations that are {\it not} implied by the Killing spinor equations
of maximally supersymmetric $Spin(7) \ltimes \bR^8$-backgrounds
are (where the tilde
indicates traceless components)
 \bea
  & & \E_{--}, \LP, \BP_{\a_1 \a_2},~~ \tilde{\BP}_{\a \bar \b},~~
\BP_{\bar
\b_1 \bar \b_2}, \BP_{\a -},~~ \BP_{\bar
  \b -}, \BP_{-+}, \LG_{\a_1 \a_2}, ~~\tilde{\LG}_{\a \bar \b}, \LG_{\bar
\b_1
\bar \b_2}, ~~  \nonumber \\
  &&
\LG_{\a -},~~  \LG_{\bar
  \b -}, \BG_{\a_1 \cdots \a_4},~~ \BG_{\a_1 \cdots \a_3 \bar \b}, ~~\BG_{\a_1
\a_2 \bar \b_1 \bar \b_2},~~
  \BG_{\a \bar \b_1 \cdots \bar \b_3}, ~~\BG_{\bar \b_1 \cdots \bar \b_4},~~
\nonumber \\
  &&
 \BG_{\a_1 \cdots \a_3 -},~~ \BG_{\a_1 \a_2 \bar \b -},~~ \BG_{\a \bar \b_1 \bar \b_2 -},~~
   \BG_{\bar \b_1 \cdots \bar \b_3 -},~~ \BG_{\a_1 \a_2 -+}, \tilde{\BG}_{\a
\bar \b
  -+},~~\BG_{\bar \b_1 \bar \b_2 -+},~~
  \nonumber \\
  &&
\LF_{\a_1 \cdots \a_4}, \LF_{\a_1 \cdots \a_3 \bar \b},~~
  \LF_{\a_1 \a_2 \bar \b_1 \bar \b_2},~~  \LF_{\a_1 \cdots \a_3 -},~~ \tilde{\LF}_{\a_1 \a_2 \bar \b -},~~
\tilde{\LF}_{\a \bar \b -+} \,,
 \eea
subject to the relations
 \begin{align}
  & \LP + 2 \BP_{-+} 
  = \BP_{\a_1 \a_2} - \tfrac{1}{2} \e_{\a_1 \a_2}{}^{\bar \b_1
  \bar \b_2} \BP_{\bar \b_1 \bar \b_2}  = 0 \,, \nonumber \\
  & \LG_{\a_1 \a_2} + 24 \BG_{\a_1 \a_2 -+} 
   = \tilde{\LG}_{\a \bar \b} + 24 \tilde{\BG}_{\a \bar \b -+} = 0 \,, \nonumber \\
  & \LG_{\bar \b_1 \bar \b_2} + 24 \BG_{\bar \b_1 \bar \b_2 -+}
   = \LG_{\a -} - 24 \BG_{\a} \cont{\g} {}_- + 8 \e_{\a}{}^{\bar \b_1
\cdots \bar \b_3}
  \BG_{\bar \b_1 \cdots \bar \b_3 -} = 0 \,, \displaybreak[2] \nonumber \\
  & \LG_{\bar \b -} + 24 \BG_{\bar \b} \cont{\g} {}_- + 8 \e_{\bar
\b}{}^{\g_1 \cdots \g_3}
  \BG_{\g_1 \cdots \g_3 -} 
   = 8 \BG_{\a_1 \cdots \a_4} + \e_{\a_1 \cdots \a_4} \BG
  \cont{\g_1} \cont{\g_2}  = 0\,, \nonumber \\
  & 8 \BG_{\bar \b_1 \cdots \bar \b_4} + \e_{\bar \b_1 \cdots \bar
\b_4}
\BG
  \cont{\g_1} \cont{\g_2} 
  = \BG_{\a \bar \b_1 \cdots \bar \b_3} - \tfrac{3}{2} g_{\a [
  \bar \b_1} \BG_{\bar \b_2 \bar \b_3 ]} \cont{\g} = 0 \,, \displaybreak[2] \nonumber \\
  & \BG_{\bar \b \g_1 \cdots \g_3} + \tfrac{3}{2} g_{\bar \b [
  \g_1} \BG_{\g_2 \g_3 ]} \cont{\g_4} 
  = \BG_{\a_1 \a_2} \cont{\g} + \tfrac{1}{2} \e_{\a_1
  \a_2}{}^{\bar \b_1 \bar \b_2} \BG_{\bar \b_1 \bar \b_2} \cont{\g} = 0
  \,, \nonumber \\
  &  4 \BG_{\a \bar \b} \cont{\g} - g_{\a \bar \b} \BG
  \cont{\g_1} \cont{\g_2} 
  = \BG_{\a_1 \a_2 -+} - \tfrac{1}{2} \e_{\a_1
  \a_2}{}^{\bar \b_1 \bar \b_2} \BG_{\bar \b_1 \bar \b_2 -+} = 0  \,,
  \displaybreak[2] \nonumber \\
  & \LF_{\a \bar \b_1 \cdots \bar \b_3} - \tfrac{3}{2} g_{\a [ \bar
  \b_1} \LF_{\bar \b_2 \bar \b_3 ]} \cont{\g}
  = \LF_{\a_1 \a_2} \cont{\g} + \tfrac{1}{2} \e_{\a_1 \a_2}{}^{\bar
\b_1
\bar \b_2}
  \LF_{\bar \b_1 \bar \b_2} \cont{\g} = 0
  \,, \nonumber \\
  & 3 {\LF}_{\a \bar \b} \cont{\g} + 3 \tilde{\LF}_{\a \bar \b
  -+} + \e_{\bar \b}{}^{\g_1 \cdots \g_3} \LF_{\a \g_1 \cdots \g_3} 
  = 3 \LF_{\bar \b} \cont{\g} {}_{-} + \e_{\bar \b}{}^{\g_1 \cdots
  \g_3} \LF_{\g_1 \cdots \g_3 -} = 0 \,.
 \end{align}
Using these relations, one can show that all the field equations and Bianchi identities are satisfied
provided one imposes the vanishing of the equations
 \bea
  && E_{--},~~ \BP_{\a_1 \a_2},~~ \tilde{\BP}_{\a \bar \b}~~
  \BP_{\a -}, ~~\BP_{\bar \b -},~~ \BP_{-+}, \tilde{\LF}_{\a_1 \a_2 \bar \b_1 \bar
\b_2},~~ \LF \cont{\g_1} \cont{\g_2},~~
  \LF_{\a_1 \a_2} \cont{\g}, \nonumber \\
  && \LF_{\a_1 \a_2 \bar \b -}, ~~ \tilde{\BG}_{\a_1
  \a_2 \bar \b_1 \bar \b_2},~~ \BG \cont{\g_1} \cont{\g_2},~~ \BG_{\a_1 \a_2} \cont{\g},~~\BG_{\a_1 \a_2 \a_3 -},
  \BG_{\a_1 \a_2 \bar \b -},~~ \BG_{\a \bar \b_1 \bar \b_2 -},~~
  \nonumber \\
  && \BG_{\bar \b_1 \bar \b_2 \bar \b_3 -},~~  \BG_{\a_1 \a_2 -+}, ~~\tilde{\BG}_{\a \bar \b -+} \,.
 \eea
Since the Bianchi identity and field equation for $F$ are interchangeable because of the self-duality
of $F$, the only field equation that needs to be imposed is the $E_{--}$ component of the field
equations.

We now turn to the corresponding supersymmetric background. As
it has been shown in \cite{ugjggpb}, the metric of
 $Spin(7) \ltimes \mathbb{R}^8$-backgrounds can be written as
 \bea
  ds^2 = 2 dv (du + \a dv + \b_I dy^I) + \g_{IJ} dy^I dy^J \,,
  \label{Spin(7)-metric}
 \eea
with $\a$, $\b$ and $\g_{IJ}$ functions of $v$ and $y^I$ only, and
$I = (1, \ldots, 8)$.  This is a pp-wave metric with rotation, see also \cite{hull}. A natural frame is given by
 \bea
  e^- = dv \,, \quad e^+ = du + \a dv + \b_I dy^I \,, \quad e^\a =
  e^\a{}_I dy^I \,, \quad e^{\bar \a} = e^{\bar \a}{}_I dy^I \,.
  \label{Spin(7)-frame}
 \eea
The components of the spin connection
are
 \bea
  && \Omega_{P,Q-} = e^I{}_{(P} \partial_v e_{Q) I} + \partial_{[P}
  \b_{Q]} \,, \qquad \Omega_{-,-P} = \partial_P \a - \partial_v \b_P \,, \nonumber \\
  && \Omega_{-,PQ} = e^I{}_{[P} \partial_v e_{Q] I} - \partial_{[P} \beta_{Q]} \,, \label{Spin(7)-spinconnection}
 \eea
and $\Omega_{P,QR}$, where $P = (\a , \bar \a)$. In addition, the components of
$\Omega_{P,QR}$ and $\Omega_{-,PQ}$  take values in $spin(7)$,
i.e.
  \bea
   \Omega_{P,\a_1 \a_2} = \tfrac{1}{2} \e_{\a_1 \a_2}{}^{\bar \b_1
   \bar \b_2} \Omega_{P, \bar \b_1 \bar \b_2} \,, \qquad
   \Omega_{P,} \cont{\a} = 0 \,,
  \eea
and similarly for $\Omega_{-,QR}$.

The Killing spinor equations restrict the fluxes as follows. The  non-vanishing components of the fluxes are
$P_-$, $G_{PQ-}$ and $F_{P_1 \cdots P_4 -}$ which in addition satisfy  the following
conditions. $G_{PQ-}$ takes values in $spin(7)$, i.e. in the
decomposition $\Lambda^2 (\mathbb{R}^8) = \Lambda_{\bf 7}^2
(\mathbb{R}^8) \oplus \Lambda_{\bf 21}^2 (\mathbb{R}^8)$ into
$Spin(7)$ representations, only the $\Lambda_{\bf 21}^2 (\mathbb{R}^8)$ is allowed by the
Killing spinor equations. Similarly, the components
of $F_{P_1 \cdots P_4 -}$ lie in  $\Lambda_{\bf 1}^4
(\mathbb{R}^8)$ and $\Lambda_{\bf 27}^4 (\mathbb{R}^8)$  in the decomposition
$\Lambda^4 (\mathbb{R}^8) = \Lambda_{\bf 1}^4 (\mathbb{R}^8) \oplus
\Lambda_{\bf 7}^4 (\mathbb{R}^8) \oplus \Lambda_{\bf 27}^4
(\mathbb{R}^8) \oplus \Lambda_{\bf 35}^4 (\mathbb{R}^8)$. The
singlet is given by
 \bea
  F_{P_1 \cdots P_4 -} \psi^{P_1 \cdots P_4} = 24 Q_- \,,
 \eea
where $\psi$ is the $Spin(7)$-invariant four-form, whose definition
can be found in \cite{ugjggpb}.

Among the field equations that still need to be imposed on the
solution to the $N=2$ $Spin(7) \ltimes \mathbb{R}^8$ Killing spinor
equations is the Einstein equation $\E_{--}$, which is given by
 \bea
  && - (\partial^P + \Omega_{Q,}{}^{QP})(\partial_P \alpha -
  \partial_v \b_P) + \partial_{[P} \beta_{Q]} \partial^P \beta^Q -
  \tfrac{1}{2} \gamma^{IJ} \partial_v{}^2 \gamma_{IJ}
  - \tfrac{1}{4} \partial_v \gamma^{IJ} \partial_v \gamma_{IJ} \nonumber \\
  &&
 - \tfrac{1}{6} F_{- P_1 \cdots P_4} F_-{}^{P_1 \cdots P_4}
  - \tfrac{1}{4} G_{-}{}^{P_1 P_2} G^*_{- P_1 P_2} -2 P_- P_-^* = 0 \,.
 \label{Einstein-eq}
 \eea
where $\gamma^{IJ}$ is the inverse of the metric $\gamma_{IJ}$
defined in ({\ref{Spin(7)-metric}}). For the special
case of $\a$, $\b_I$ and $\g_{IJ}$ independent of $v$ this equation
becomes
 \bea
  - \Box_8 \a + \partial_{[P} \beta_{Q]} \partial^P
  \beta^Q - \tfrac{1}{6} F_{- P_1 \cdots P_4} F_-{}^{P_1 \cdots P_4}
  - \tfrac{1}{4} G_{-}{}^{P_1 P_2} G^*_{- P_1 P_2} -2 P_- P_-^* = 0 \,,
 \label{Einstein-eq-trunc}
 \eea
where $\Box_8$ is the Laplacian on the eight-dimensional space and
$\partial_{[P} \b_{Q]}$ only takes values in $Spin(7)$.

In addition one needs to impose several components of the Bianchi
identities on the
fluxes $P_-$, $G_{PQ-}$ and $F_{P_1 \cdots P_4 -}$. For example, the
remaining $BP$ components imply that $P_-$ is a function of $v$
only. We will not analyze the rest of these restrictions in detail.

Observe that the contribution of the rotation in the
Einstein equations has a different sign from that of the contribution
of the fluxes. Because of this and assuming that the transverse space of the
pp-wave is a compact $Spin(7)$ manifold, the Einstein equation can be solved provided that
the total rotation cancels the contributions from the fluxes. This means that
the integral of the expression in the right-hand-side of (\ref{Einstein-eq-trunc})
must vanish. For a detailed similar argument see \cite{ugjggpa}.
The above solution resembles flux-tube type of configurations \cite{mateos}
but without the backreaction of the branes. There are also similarities
with the solutions of \cite{pope, chen}.

\subsection{Maximal $SU(4) \ltimes \bR^8$backgrounds}

The field equations that are {\it not} implied by the Killing spinor equations
of maximally supersymmetric $SU(4) \ltimes \bR^8$-backgrounds
are
 \bea
  & & \E_{--},~~ \LP,~~ \tilde{\BP}_{\a \bar \b},~~ \BP_{\a -},~~ \BP_{\bar
  \b -},~~ \BP_{-+}, ~~\tilde{\LG}_{\a \bar \b},~~ \LG_{\a -},~~ \LG_{\bar
  \b -},
  \nonumber \\
  &&
  \tilde{\BG}_{\a_1 \a_2 \bar \b_1 \bar \b_2},~~  \BG_{\a_1 \a_2 \bar \b -},~~ \BG_{\a
  \bar \b_1 \bar \b_2 -}, ~~\tilde{\BG}_{\a \bar \b -+}, ~~\tilde{\LF}_{\a_1 \a_2 \bar
\b_1 \bar \b_2},~~
  \tilde{\LF}_{\a_1 \a_2 \bar \b -},~~
 \eea
subject to the relations
 \begin{align}
  & \LP + 2 \BP_{-+} = \tilde{\LG}_{\a \bar \b} + 24 \tilde{\BG}_{\a \bar \b -+} = 0 \,, \nonumber \\
  & \LG_{\a -} - 24 \BG_{\a} \cont{\g} {}_- = \LG_{\bar \b -} + 24 \BG_{\bar \b} \cont{\g} {}_- = 0
   \,.
 \end{align}
Using these relations, one can show that all field equations are satisfied provided
that the field equations
 \bea
  & & \E_{--},~~ \tilde{\BP}_{\a \bar \b},~~ \BP_{\a -},~~ \BP_{\bar
  \b -},~~ \BP_{-+},~~ \tilde{\BG}_{\a_1 \a_2 \bar \b_1 \bar \b_2},~~
  \BG_{\a_1 \a_2 \bar \b -},~~
  \nonumber \\
  &&
   \BG_{\a
  \bar \b_1 \bar \b_2 -}, ~~\tilde{\BG}_{\a \bar \b -+},~~
   \tilde{\LF}_{\a_1 \a_2 \bar \b_1 \bar \b_2},~~
  \tilde{\LF}_{\a_1 \a_2 \bar \b -}
  \,,
 \eea
are satisfied. It is easy to see that apart from the Bianchi identities, the only field
equation that one has to impose is the $E_{--}$ component of the Einstein equation.

The investigation of the field equations of the maximal $SU(4) \ltimes \mathbb{R}^8$-backgrounds is related
 to that of maximal $Spin(7) \ltimes \mathbb{R}^8$-backgrounds.
 In particular, there is a gauge  for the Killing spinors
such that $\Omega_{A,-+} = 0$ and $\Omega_{A,} \cont{\b} = 0$
\cite{ugjggpb}. In addition the metric can be written  in Penrose coordinates as in  \eqref{Spin(7)-metric}
and so one can introduce the frame
\eqref{Spin(7)-frame}. One can compute the spin connection which has components
 $\Omega_{-,-P}$, $\Omega_{P,Q-}$,
$\Omega_{-,PQ}$ and $\Omega_{P,QR}$. The first three are given in
\eqref{Spin(7)-spinconnection}, and in addition the latter two take
values in $SU(4)$, i.e.
 \bea
  \Omega_{P, \a_1 \a_2} = 0 \,, \qquad \Omega_{P,} \cont{\a} =0 \,,
 \eea
and similarly for $\Omega_{-,PQ}$. The non-vanishing components of
the fluxes are
 \bea
 P_-~, ~~\tilde{G}_{\a \bar \b -}~,~~ G \cont{\a} {}_- (v)~,~~ F_{\a_1 \cdots \a_4 -} (v)~,~~
  \tilde{F}_{\a_1 \a_2 \bar \b_1 \bar \b_2 -}~,~~
  F \cont{\a} \cont{\b} {}_- = 2 Q_- \,.
 \eea

Using these, one can easily compute the Einstein equation
 $E_{--}$. It is easy to see that it takes the same form as
\eqref{Einstein-eq}. In addition the
remaining Bianchi identities will impose closure of the remaining
fluxes $P_-$, $G_{PQ-}$ and $F_{P_1 \cdots P_4 -}$.

In the case that the components of the metric are independent of the
$v$ coordinate, the transverse space of the pp-wave is a Calabi-Yau
manifold. To find a solution, one can use the Donaldson theorem for
$U(1)$ connections and require cancelation of the rotation and flux
charges. Using a similar argument to that in \cite{ugjggpa}, one can
show that there is a smooth solution for pp-waves with transverse
space a compact Calabi-Yau manifold.

\newsection{Concluding remarks}

We have shown that the  Killing spinor equations of any IIB
supergravity background can be written as a linear system for the
fluxes, geometry and spacetime derivatives of the functions that
determine the Killing spinors. This has been achieved by using the
spinorial geometry techniques of \cite{uggp}. We have also shown
that another linear system, constructed in a similar way, can be
used to determine the field equations  and Bianchi identities of IIB
supergravity that are determined by the Killing spinors for any
supersymmetric background. These two linear systems can be used to
systematically investigate all supersymmetric backgrounds of IIB
supergravity.

For general  supersymmetric backgrounds these two linear systems are
rather complicated.
 However, we have shown that these linear systems
simplify for backgrounds that admit $H$-invariant spinors, $H\subset Spin(9,1)$.
We have mostly focused on two cases, those for
which the background admits a maximal number of $H$-invariant Killing spinors, maximally
supersymmetric
$H$-backgrounds, and those that admit half the number of maximal
$H$-invariant spinors, half-maximally supersymmetric $H$-backgrounds.
In the former case, the Killing spinor equations factorize and the resulting linear
systems are easy to solve. We have demonstrated that the system associated
with the Killing spinor equations gives rise to a flatness condition for the connection which
is identified as the restriction of the supercovariant connection on the bundle
of Killing spinors ${\cal K}$.

There are several cases of half-maximal $H$-backgrounds which should be considered. The generic
case consists of those backgrounds for which the Killing spinors are linearly independent
over the complex numbers. There are also several degenerate cases for which the Killing spinors
are linearly dependent over the complex numbers but linearly independent over the real numbers.
The degenerate cases are of co-dimension two or more relative to the generic case.
We have demonstrated that the Killing spinor equations of half-maximal $H$-backgrounds
do not factorize. The linear system of the Killing spinor equations gives rise to a flatness condition
for the restriction of the supercovariant connection on the bundle of Killing spinors ${\cal K}$.
However, the restricted connection depends non-linearly on the functions that determine
the Killing spinors.

To give an overview of the current status of the problem in IIB supergravity, we summarize
some of the results
in the table 1 below.  In this table, we indicate the cases that have been investigated as well
as the maximal and half-maximal $H$-backgrounds that remain to be tackled.

\begin{table}[ht]
\hspace{-.5cm}
\begin{tabular}{|c|c|c|c|c|c|c|c|c|c}\hline
$\mathrm{H}$ & $\mathrm{N=1}$ & $\mathrm{N=2}$ & $\mathrm{N=3}$  & $\mathrm{N=4}$ & $\mathrm{N=6}$
  & $\mathrm{N=8}$  & $\mathrm{N=16}$ & $\mathrm{N=32}$
 \\\hline
$Spin(7)\ltimes \bR^8$ & $\surd$ & $\surd$ &-  &- &- &- & - & - \\
 $SU(4)\ltimes \bR^8$ & $\surd$   & $\surd$  & & $\surd$  & -&- &-& - \\
$G_2$ & $\surd$  & $\odot$ &  & $\surd$  & - &- & - & -\\
$Sp(2)\ltimes \bR^8$&- & &$\odot$ &   & $\odot$ & - & -& - \\
$(SU(2)\times SU(2))\ltimes \bR^8$&-  & & & $\odot$ & & $\odot$ & - & - \\
$SU(3)$&- &  & & $\odot$ & & $\odot$ & - & -\\
$\bR^8$ &-  & &  & & & $\odot$ &$\odot$& -\\
$SU(2)$&- & & &   &  & $\odot$ &$\odot$& -
\\
1&-&&&&&&$\odot$& $\surd$
\\ \hline
\end{tabular}
\caption{$\surd$ denotes the cases for which the Killing spinor equations have already been solved.
$\odot$ denotes the remaining cases that correspond to backgrounds with $H$-invariant spinors
and can be tackled with the techniques described in this paper.
$-$ denotes the cases that do not occur, e.g.  there are no backgrounds with $N>2$ and $Spin(7)\ltimes \bR^8$-
invariant Killing spinors.
The remaining  entries may occur but it is expected that the associated linear
systems are more involved.}
\end{table}

In table 1 below, the list  of  cases that   remain to
 be tackled contains  the $N=4$ and $N=8$ $SU(3)$-backgrounds.
It is expected that the former includes all backgrounds which are dual to  ${\cal N}=1$ ($N=4$) four-dimensional gauge theories.
The list also includes all supersymmetric backgrounds that preserve 1/2  of the supersymmetry ($N=16$). There are three
classes of 1/2  supersymmetric backgrounds. The maximal $\bR^8$-backgrounds, the maximal $SU(2)$-backgrounds and
the half-maximal $1$-backgrounds. It would be of interest to investigate all these cases.

\section*{Acknowledgements}

The work of U.G. is funded by the The Swedish Research Council
and in addition the research of both U.G. and D.R. is funded
by the PPARC grant PPA/G/O/2002/00475.

\setcounter{section}{0}

\appendix{Spinors}

The description of IIB supergravity spinors that we used in this paper
 can be found in \cite{ugjggpa}. For the general
theory see \cite{wang, lawson, harvey}.  Here for completeness,
we shall briefly summarize the main formulae without explanation.

Consider the vector space $U=\bR<e_1,\dots,e_5>$,  where $e_1,\dots,e_5$ is an orthonormal basis.
The space of Dirac spinors of $Spin(9,1)$  is
$\Delta_c=\Lambda^*(U\otimes \bC)$. This decomposes into
two complex chiral representations
according to the degree of the form $\Delta_c^+=\Lambda^{{\rm even}}(U\otimes \bC)$
and  $\Delta_c^-=\Lambda^{{\rm odd}}(U\otimes \bC)$.
These are the complex Weyl representations
of $Spin(9,1)$.
The gamma matrices are represented on $\Delta_c$ as
\bea
\Gamma_0\eta&=& -e_5\wedge\eta +e_5\lc\eta~,~~~~
\Gamma_5\eta= e_5\wedge\eta+e_5\lc \eta~,
\cr
\Gamma_i\eta&=& e_i\wedge \eta+ e_i\lc \eta~,~~~~~~i=1,\dots,4
\cr
\Gamma_{5+i}\eta&=& i e_i\wedge\eta-ie_i\lc\eta~,
\eea
where $\Gamma_0$ is the gamma matrix along the time direction.
The above gamma matrices
satisfy the Clifford algebra relations
$\Gamma_A\Gamma_B+\Gamma_B \Gamma_A=2 \eta_{AB}$ with respect to the
Lorentzian inner product as expected.

The Dirac inner product on the space of spinors $\Delta_c$ is
\bea
D(\eta,\theta)=<\Gamma_0\eta, \theta>~,
\eea
where
\be
<z^a e_a, w^b e_b>=\sum_{a=1}^{5}  (z^a)^* w^a~,~~~~
\ee
on $U\otimes \bC$  and then extended to $\Delta_c$. Note that
 $(z^a)^*$ is the standard
complex conjugate of $z^a$.

 The Majorana $Spin (9,1)$-invariant inner product that we use is
\be
B(\eta,\theta)= <B(\eta^*), \theta>~,~~~~~~~~
\ee
where  $B=\Gamma_{06789}$.
The Majorana-Weyl representations $\Delta^\pm$ of $Spin(9,1)$ are constructed by imposing the
reality condition
\be
\eta=-\Gamma_0 B(\eta^*)~,
\ee
or equivalently
\be
\eta^*=\Gamma_{6789}\eta~,
\la{rcon}
\ee
on $\Delta_c^\pm$.
The map $C=\Gamma_{6789}$ is the charge conjugation matrix.

The oscillator basis in $\Delta_c$ that we use in this paper  is
\be
\Gamma_{\bar\a}= {1\over \sqrt {2}}(\Gamma_\a+i \Gamma_{\a+5})~,~~~~~~~~~
\Gamma_\pm={1\over \sqrt{2}} (\Gamma_5\pm\Gamma_0)
~,~~~~~~~~~\Gamma_{\a}= {1\over \sqrt {2}}(\Gamma_\a-i \Gamma_{\a+5})~.
\la{hbasis}
\ee
Observe that the Clifford algebra relations in the above basis are
$\Gamma_A\Gamma_B+\Gamma_B\Gamma_A=2 \eta_{AB}$,   where the non-vanishing
components of the metric are
$\eta_{\a\bar\b}=\delta_{\a\bar\b}, \eta_{+-}=1$. In addition, we define
$\Gamma^B=g^{BA} \Gamma_A$.
The $1$ spinor is a Clifford  vacuum, $\Gamma^{\a}1=\Gamma^- 1=0$
and  the representation $\Delta_c$
can be constructed by acting on $1$ with the creation operators
$\Gamma^{\bar\a}, \Gamma^+$
or equivalently any spinor can be written as
\be
\eta= \sum_{k=0}^5 {1\over k!}~ \phi_{\bar a_1\dots \bar a_k}~
 \Gamma^{\bar a_1\dots\bar a_k} 1~,~~~~\bar a=\bar\a, +~,
 \la{hbasisa}
\ee
i.e. $\Gamma^{\bar a_1\dots\bar a_k} 1$, for $k=0,\dots,5$, is a basis in the
space of (Dirac) spinors.

The spacetime form bi-linears associated with the pair of spinors $(\eta,\epsilon)$.
are
\be
\alpha(\eta, \epsilon)={1\over k!} B(\eta,\Gamma_{A_1\dots A_k} \epsilon)
e^{A_1}\wedge\dots\wedge e^{A_k}~,~~~~~~~k=0,\dots, 9~.
\la{forms}
\ee
For the application to backgrounds with $SU(4)\ltimes \bR^8$ spinors, we use
the form bi-linear of
$1$ and $e_{1234}$ spinors. These are the following (see also \cite{ugjggpa}): a one-form
\bea
\kappa(e_{1234},1)=\kappa(1, e_{1234})= e^0-e^5~,
\eea
a three-form
\bea
\xi(e_{1234}, 1)=- \xi(1,e_{1234})=i (e^0-e^5)\wedge \omega~,
\eea
and  five-forms
\bea
\tau(1,1)=(e^0-e^5)\wedge \chi
\cr
\tau(e_{1234}, e_{1234})= (e^0-e^5)\wedge \chi^*
\cr
\tau(e_{1234}, 1)=\tau(1,e_{1234})=- {1\over2} (e^0-e^5)\omega\wedge \omega~,
\eea
where
\bea
\omega= e^1\wedge e^6+ e^2\wedge e^7+e^3\wedge e^8+e^4\wedge e^9
\cr
\chi=(e^1+i e^6)\wedge
(e^2+i e^7)\wedge (e^3+i e^8)\wedge (e^4+i e^9)~.
\eea
Note that $\chi$ and $\omega$ are the familiar $SU(4)$ invariant forms.

Using the above results, we can easily compute
the spacetime form bilinears of the half-maximally supersymmetric $SU(4)\ltimes\bR^8$ backgrounds. The
Killing spinors are $\epsilon= z~\eta$, where $\eta_1=1+e_{1234}$ and $\eta_2=i(1-e_{1234})$.
 Since the Killing spinors $\epsilon_1$ and $\epsilon_2$ are generically complex, $\tilde\epsilon_1=C (\epsilon_1)^*$
and $\tilde\epsilon_2=C (\epsilon_2)^*$ are also defined on the
spacetime although they are not necessarily Killing. Because of
this, the forms that are defined on the spacetime are associated
with the bi-linears $(\epsilon_i, \epsilon_j)$, $(\epsilon_i,
\tilde\epsilon_j)$ and $(\tilde\epsilon_i, \tilde\epsilon_j)$,
$i,j=1,2$. It suffices to compute the forms associated with the
first two bilinears because the forms of the last bilinear can be
easily computed from those of the first. To see this observe that
since $\epsilon= z\eta$ and $\eta_i$ are Majorana-Weyl spinors, then
$\tilde\epsilon=z^*\eta$. Thus the effect of the charge conjugation
operation is to replace the matrix $z$ with its complex conjugate
$z^*$.
 In particular, we find
the one-forms
\bea
\kappa(\epsilon_1, \epsilon_1)&=&2 (z_{11}^2+ z_{12}^2) (e^0-e^5)~,
\cr
\kappa(\epsilon_2, \epsilon_2)&=&2 (z_{21}^2+ z_{22}^2) (e^0-e^5)~,
\cr
\kappa(\epsilon_1, \epsilon_2)&=&2 (z_{11} z_{21}+ z_{12} z_{22}) (e^0- e^5)~,
\cr
\kappa(\epsilon_1, \tilde\epsilon_1)&=&2 (|z_{11}|^2+ |z_{12}|^2) (e^0-e^5)~,
\cr
\kappa(\epsilon_2, \tilde\epsilon_2)&=&2 (|z_{21}|^2+ |z_{22}|^2) (e^0-e^5)~,
\cr
\kappa(\epsilon_1, \tilde \epsilon_2)&=&2 (z_{11} z_{21}^*+z_{12} z_{22}^*) (e^0-e^5)~,
\cr
\kappa(\tilde\epsilon_1,  \epsilon_2)&=& 2 (z_{11}^* z_{21}+z_{12}^* z_{22}) (e^0-e^5)~,
\la{oneformbi}
\eea
the three-forms
\bea
\xi(\epsilon_1, \epsilon_2)&=&-2 {\rm det} z (e^0-e^5)\wedge \omega~,
\cr
\xi(\epsilon_1, \tilde\epsilon_1)&=& 4i {\rm Im} (z_{11}^* z_{12}) (e^0-e^5)\wedge \omega~,
\cr
\xi(\epsilon_1, \tilde\epsilon_2)&=& -2 (z_{11} z_{22}^*-z_{12} z_{21}^*) (e^0-e^5)\wedge \omega~,
\cr
\xi(\tilde\epsilon_1, \epsilon_2)&=&-2 (z_{11}^* z_{22}-z_{12}^* z_{21}) (e^0-e^5)\wedge \omega~,
\la{threeformbi}
\eea
and the five-forms
\bea
\tau(\epsilon_1, \epsilon_1)&=&(e^0-e^5)\wedge [ (z_{11}+i z_{12})^2  \chi+ (z_{11}-iz_{12})^2  \chi^*-
 (z_{11}^2+ z_{12}^2)  \omega\wedge \omega]~,
 \cr
 \tau(\epsilon_2, \epsilon_2)&=&(e^0-e^5)\wedge [(z_{21}+i z_{22})^2  \chi+ (z_{21}-iz_{22})^2  \chi^*-
 (z_{21}^2+ z_{22}^2)  \omega\wedge \omega]~,
 \cr
\tau(\epsilon_1, \epsilon_2)&=&(e^0-e^5)\wedge [ (z_{11}+i z_{12})  (z_{21}+i z_{22})\chi
+(z_{11}-i z_{12})  (z_{21}-i z_{22}) \chi^*
\cr
~~~~~~~&&- (z_{11} z_{21}+ z_{12} z_{22}) \omega\wedge \omega]~,
\cr
\tau(\epsilon_1, \tilde\epsilon_1)&=&(e^0-e^5)\wedge [(z_{11}+i z_{12}) (z^*_{11}+i z^*_{12})\chi+
(z_{11}-iz_{12}) (z^*_{11}-iz^*_{12}) \chi^*
\cr
~~~~~~~&&- (|z_{11}|^2+ |z_{12}|^2 )\omega\wedge \omega]~,
\cr
\tau(\epsilon_2, \tilde\epsilon_2)&=&(e^0-e^5)\wedge [(z_{21}+i z_{22}) (z^*_{21}+i z^*_{22})\chi+
(z_{21}-iz_{22}) (z^*_{21}-iz^*_{22}) \chi^*
\cr
~~~~~~~&&- (|z_{21}|^2+ |z_{22}|^2 )\omega\wedge \omega]~,
\cr
\tau(\epsilon_1, \tilde \epsilon_2)&=&(e^0-e^5)\wedge [(z_{11}+i z_{12}) (z_{21}^*+i z_{22}^*) \chi+
(z_{11}-iz_{12}) (z^*_{21}-iz^*_{22}) \chi^*
\cr
~~~~~~~&&-(z_{11} z_{21}^*+z_{12} z_{22}^*) \omega\wedge \omega]~,
\cr
\tau(\tilde\epsilon_1,  \epsilon_2)&=&(e^0-e^5)\wedge [(z_{11}^*+iz_{12}^*) (z_{21}+iz_{22}) \chi
+(z_{11}^*-i z_{12}^*) (z_{21}-i z_{22}) \chi^*
\cr
~~~~~~~&&- (z_{11}^* z_{21}+z_{12}^* z_{22}) \omega\wedge \omega]~.
\la{fiveformbi}
\eea
It is worth mentioning that all the one-forms are along the same null direction. The same applies
for the three-forms. However, the five-forms point to different directions spanned by
$(e^0-e^5)\wedge \chi$, $(e^0-e^5)\wedge \chi^*$ and  $(e^0-e^5)\wedge\omega\wedge\omega$.

\appendix{Killing spinor equations}

\subsection{Killing spinor equations on  $1$}

The first spinor basis element we consider is $1$. Substituting this spinor into the (algebraic) Killing spinor equation
({\ref{kseqnb}}) and expanding the resulting expression in the basis ({\ref{hbasisa}}), we find
 \bea
  1 &:& 0 \,, \cr
  \Gamma^{\bar \b} &:& \tfrac{1}{4} G_{\bar{\b}} \cont{\g} + \tfrac{1}{4} G_{\bar{\b}-+} \,, \cr
  \Gamma^{+} &:& \tfrac{1}{4} G_{+} \cont{\g} \,, \cr
  \Gamma^{(2)} &:& 0 \,, \cr
  \Gamma^{\bar \b_1 \cdots \bar \b_3} &:& \tfrac{1}{24} G_{\bar \b_1 \cdots \bar \b_3} - \tfrac{1}{12} \e_{\bar \b_1 \cdots \bar \b_3}{}^{\g} P_{\g} \,, \cr
  \Gamma^{\bar \b_1 \bar \b_2 +} &:& \tfrac{1}{8} G_{\bar \b_1 \bar \b_2 +} \,, \cr
  \Gamma^{(4)} &:& 0 \,, \cr
  \Gamma^{\bar \b_1 \cdots \bar \b_4 +} &:& \tfrac{1}{96} \e_{\bar \b_1 \cdots \bar \b_4} P_+ .
 \eea
Next we turn to the Killing spinor equation associated
with the supercovariant derivative (\ref{kseqna}).
The components along the $\a$-frame derivative of the supercovariant connection are
\bea
  1 &:& D_{\a} + \tfrac{1}{2} \Omega_{\a,} \cont{\g} + \tfrac{1}{2} \Omega_{\a,-+} + \tfrac{i}{4} F_{\a} \cont{\g_1} \cont{\g_2} + \tfrac{i}{2} F_{\a} \cont{\g_1}{}_{-+} \,, \cr
  \Gamma^{(1)} &:& 0 \,, \cr
  \Gamma^{\bar \b_1 \bar \b_2} &:& \tfrac{1}{4} \Omega_{\a, \bar \b_1 \bar \b_2} + \tfrac{i}{4} F_{\a \bar \b_1 \bar \b_2} \cont{\g} + \tfrac{i}{4} F_{\a \bar \b_1 \bar \b_2 -+} - \tfrac{1}{32} \e_{\bar \b_1 \bar \b_2}{}^{\gamma_1 \gamma_2} G_{\a \gamma_1 \gamma_2} \,, \cr
  \Gamma^{\bar \b +} &:& \tfrac{1}{2} \Omega_{\a, \bar \b +} + \tfrac{i}{2} F_{\a \bar \b +} \cont{\g} \,, \cr
  \Gamma^{(3)} &:& 0 \,, \cr
  \Gamma^{\bar \b_1 \cdots \bar \b_4} &:& \tfrac{i}{48} F_{\a \bar \b_1 \cdots \bar \b_4} - \tfrac{1}{768} \e_{\bar \b_1 \cdots \bar \b_4}( G_{\a} \cont{\g} - G_{\a -+}) \,, \cr
  \Gamma^{\bar \b_1 \cdots \bar \b_3 +} &:& \tfrac{i}{12} F_{\a \bar \b_1 \cdots \bar \b_3 +} - \tfrac{1}{96} \e_{\bar \b_1 \cdots \bar \b_3}{}^{\g} G_{\a \g +} \,, \cr
   \Gamma^{(5)} &:& 0 .
\eea
Along the $\bar \a$-frame derivative of the supercovariant connection we find
\bea
  1 &:& D_{\bar \a} + \tfrac{1}{2} \Omega_{\bar \a,} \cont{\g} + \tfrac{1}{2} \Omega_{\bar \a,-+} + \tfrac{i}{4} F_{\bar \a} \cont{\g_1} \cont{\g_2} + \tfrac{i}{2} F_{\bar \a} \cont{\g_1}{}_{-+} - \tfrac{1}{24} \e_{\bar \a}{}^{\g_1 \cdots \g_3} G_{\g_1 \cdots \g_3} \,, \cr
  \Gamma^{(1)} &:& 0 \,, \cr
  \Gamma^{\bar \b_1 \bar \b_2} &:& \tfrac{1}{4} \Omega_{\bar \a, \bar \b_1 \bar \b_2} + \tfrac{i}{4} F_{\bar \a \bar \b_1 \bar \b_2} \cont{\g} + \tfrac{i}{4} F_{\bar \a \bar \b_1 \bar \b_2 -+} - \tfrac{1}{32} \e_{\bar \b_1 \bar \b_2}{}^{\gamma_1 \gamma_2} (2 G_{\bar \a \gamma_1 \gamma_2}+ g_{\bar \a \g_1} G_{\g_2} \cont{\d} - g_{\bar \a \g_1} G_{\g_2 -+}) \,, \cr
  \Gamma^{\bar \b +} &:& \tfrac{1}{2} \Omega_{\bar \a, \bar \b +} + \tfrac{i}{2} F_{\bar \a \bar \b +} \cont{\g} + \tfrac{1}{16} \e_{\bar \a \bar \b}{}^{\g_1 \g_2} G_{\g_1 \g_2 +} \,, \cr
  \Gamma^{(3)} &:& 0 \,, \cr
  \Gamma^{\bar \b_1 \cdots \bar \b_4} &:& - \tfrac{1}{384} \e_{\bar \b_1 \cdots \bar \b_4} ( G_{\bar \a} \cont{\g} - G_{\bar \a -+}) \,, \cr
  \Gamma^{\bar \b_1 \cdots \bar \b_3 +} &:& \tfrac{i}{12} F_{\bar \a \bar \b_1 \cdots \bar \b_3 +} - \tfrac{1}{192} \e_{\bar \b_1 \cdots \bar \b_3}{}^{\g} (4 G_{\bar \a \g +} + g_{\bar \a \g} G_{+} \cont{\d}) \,, \cr
   \Gamma^{(5)} &:& 0 .
\eea
The supercovariant derivative with $M=-$ gives
\bea
  1 &:& D_{-} + \tfrac{1}{2} \Omega_{-,} \cont{\g} + \tfrac{1}{2} \Omega_{-,-+} + \tfrac{i}{4} F_{-} \cont{\g_1} \cont{\g_2}\,, \cr
  \Gamma^{(1)} &:& 0 \,, \cr
  \Gamma^{\bar \b_1 \bar \b_2} &:& \tfrac{1}{4} \Omega_{-, \bar \b_1 \bar \b_2} + \tfrac{i}{4} F_{- \bar \b_1 \bar \b_2} \cont{\g} - \tfrac{1}{16} \e_{\bar \b_1 \bar \b_2}{}^{\gamma_1 \gamma_2} G_{- \gamma_1 \gamma_2} \,, \cr
  \Gamma^{\bar \b +} &:& \tfrac{1}{2} \Omega_{-, \bar \b +} - \tfrac{i}{2} F_{\bar \b - +} \cont{\g} + \tfrac{1}{48} \e_{\bar \b}{}^{\g_1 \cdots \g_3} G_{\g_1 \cdots \g_3} \,, \cr
  \Gamma^{(3)} &:& 0 \,, \cr
  \Gamma^{\bar \b_1 \cdots \bar \b_4} &:& \tfrac{i}{48} F_{- \bar \b_1 \cdots \bar \b_4} - \tfrac{1}{384} \e_{\bar \b_1 \cdots \bar \b_4} G_{-} \cont{\g} \,, \cr
  \Gamma^{\bar \b_1 \cdots \bar \b_3 +} &:& - \tfrac{i}{12} F_{\bar \b_1 \cdots \bar \b_3 -+} + \tfrac{1}{192} \e_{\bar \b_1 \cdots \bar \b_3}{}^{\g} (G_{\g} \cont{\d} + 3 G_{\g - +}) \,, \cr
   \Gamma^{(5)} &:& 0 .
\eea
Finally, for $M=+$ we find
\bea
  1 &:& D_{+} + \tfrac{1}{2} \Omega_{+,} \cont{\g} + \tfrac{1}{2} \Omega_{+,-+} + \tfrac{i}{4} F_{+} \cont{\g_1} \cont{\g_2}\,, \cr
  \Gamma^{(1)} &:& 0 \,, \cr
  \Gamma^{\bar \b_1 \bar \b_2} &:& \tfrac{1}{4} \Omega_{+, \bar \b_1 \bar \b_2} + \tfrac{i}{4} F_{+ \bar \b_1 \bar \b_2} \cont{\g} - \tfrac{1}{32} \e_{\bar \b_1 \bar \b_2}{}^{\gamma_1 \gamma_2} G_{+ \gamma_1 \gamma_2} \,, \cr
  \Gamma^{\bar \b +} &:& \tfrac{1}{2} \Omega_{+, \bar \b +} \,, \cr
  \Gamma^{(3)} &:& 0 \,, \cr
  \Gamma^{\bar \b_1 \cdots \bar \b_4} &:& \tfrac{i}{48} F_{+ \bar \b_1 \cdots \bar \b_4} - \tfrac{1}{768} \e_{\bar \b_1 \cdots \bar \b_4} G_{+} \cont{\g} \,, \cr
  \Gamma^{\bar \b_1 \cdots \bar \b_3 +} &:& 0 \,, \cr
   \Gamma^{(5)} &:& 0 .
\eea

\subsection{Killing spinor equations on   $e_{ij}$}

Next we consider the basis elements $e_{ij}$ with $i,j \leq 4$, i.e.~without holomorphic indices. We split up $\a$ into $a=(i,j)$ and $p$, which contains the remaining two holomorphic indices. Furthermore, two different two-dimensional Levi-Civita tensors will appear, which are defined by $\e_{ij} = +1$ and $\e_{p_1 p_2} = \e_{ij p_1 p_2}$. Substituting this spinor into the (algebraic) Killing spinor equation
({\ref{kseqnb}}) and expanding the resulting expression in the basis ({\ref{hbasisa}}), we find
 \bea
 1 &:& 0 \,, \cr
 \Gamma^{\bar b} &:& \tfrac{1}{4} \e_{\bar b}{}^{c_1} (G_{c_1} \cont{c_2} - G_{c_1} \cont{r} - G_{c_1 -+} ) \,, \cr
 \Gamma^{\bar q} &:& \e_{\bar q}{}^r P_r - \tfrac{1}{4} \e^{c_1 c_2} G_{c_1 c_2 \bar q} \,, \cr
 \Gamma^+ &:& - \tfrac{1}{4} \e^{c_1 c_2} G_{c_1 c_2 +} \,, \cr
 \Gamma^{(2)} &:& 0 \,, \cr
 \Gamma^{\bar b_1 \bar b_2 \bar q} &:& - \tfrac{1}{16} \e_{\bar b_1 \bar b_2} (G_{\bar q} \cont{c} - G_{\bar q} \cont{r} - G_{\bar q -+}) \,, \cr
 \Gamma^{\bar b_1 \bar b_2 +} &:& - \tfrac{1}{16} (G_{+} \cont{c} - G_{+} \cont{r}) \,, \cr
 \Gamma^{\bar b \bar q_1 \bar q_2} &:& - \tfrac{1}{4} \e_{\bar q_1 \bar q_2} P_{\bar b} - \tfrac{1}{8} \e_{\bar b}{}^c G_{c \bar q_1 \bar q_2} \,, \cr
 \Gamma^{\bar b \bar q +} &:& - \tfrac{1}{4} \e_{\bar b}{}^c G_{c \bar q +} \,, \cr
 \Gamma^{\bar q_1 \bar q_2 +} &:& - \tfrac{1}{4} \e_{\bar q_1 \bar q_2} P_+ \,, \cr
 \Gamma^{(4)} &:& 0 \,, \cr
 \Gamma^{\bar b_1 \bar b_2 \bar q_1 \bar q_2 +} &:& \tfrac{1}{32} \e_{\bar b_1 \bar b_2} G_{\bar q_1 \bar q_2 +} .
 \eea
Next we turn to the Killing spinor equation associated
with the supercovariant derivative (\ref{kseqna}).
The components along the $a$-frame derivative of the supercovariant connection are
\bea
 1 &:& -\tfrac{1}{2} \e^{c_1 c_2} \Omega_{a, c_1 c_2} + \tfrac{1}{4} \e^{r_1 r_2} G_{a r_1 r_2} \,, \cr
 \Gamma^{(1)} &:& 0 \,, \cr
 \Gamma^{\bar b_1 \bar b_2} &:& \tfrac{1}{4} \e_{\bar b_1 \bar b_2} (D_a - \tfrac{1}{2} \Omega_{a,} \cont{c} + \tfrac{1}{2} \Omega_{a,} \cont{r} + \tfrac{1}{2} \Omega_{a, -+} - \tfrac{i}{2} F_{a} \cont{c} \cont{r} - \tfrac{i}{2} F_{a -+} \cont{c} + \tfrac{i}{4} F_{a} \cont{r_1} \cont{r_2}
\cr
&&
+ \tfrac{i}{2} F_{a -+} \cont{r})
 - \tfrac{1}{16} \e^{r_1 r_2} g_{a [ \bar b_1} G_{\bar b_2 ] r_1 r_2} \,, \cr
 \Gamma^{\bar b \bar q} &:& - \tfrac{1}{2} \e_{\bar b}{}^c ( \Omega_{a,c \bar q} + i F_{ac \bar  q} \cont{r} + i F_{ac \bar q -+} ) + \tfrac{1}{16} g_{a \bar b} \e_{\bar q}{}^{r_1} ( 3 G_{r_1} \cont{c} + 3 G_{r_1} \cont{r_2} - G_{r_1 -+}) + \tfrac{1}{2} \e_{\bar q}{}^r G_{a \bar b r} \,, \cr
 \Gamma^{\bar b +} &:& - \tfrac{1}{2} \e_{\bar b}{}^c ( \Omega_{a,c+} + i F_{ac+} \cont{r}) - \tfrac{1}{16} g_{a \bar b} \e^{r_1 r_2} G_{r_1 r_2 +} \,, \cr
 \Gamma^{\bar q_1 \bar q_2} &:& - \tfrac{1}{16} \e_{\bar q_1 \bar q_2} ( G_{a} \cont{c} - G_{a} \cont{r} + G_{a -+}) \,, \cr
 \Gamma^{\bar q +} &:& \tfrac{1}{4} \e_{\bar q}{}^r G_{ar+} \,, \cr
 \Gamma^{(3)} &:& 0 \,, \cr
 \Gamma^{\bar b_1 \bar b_2 \bar q_1 \bar q_2} &:& \tfrac{1}{16} \e_{\bar b_1 \bar b_2} (\Omega_{a, \bar q_1 \bar q_2} - i F_{a \bar q_1 \bar q_2} \cont{r} + i F_{a \bar q_1 \bar q_2 -+}) - \tfrac{1}{64} \e_{\bar q_1 \bar q_2} g_{a [ \bar b_1}  (3 G_{\bar b_2]} \cont{c} + G_{\bar b_2]} \cont{r} - G_{\bar b_2] -+}) \,, \cr
 \Gamma^{\bar b_1 \bar b_2 \bar q +} &:& \tfrac{1}{8} \e_{\bar b_1 \bar b_2} ( \Omega_{a, \bar q +}- i F_{a \bar q +} \cont{c} + i F_{a \bar q +}\cont{r}) - \tfrac{1}{16} \e_{\bar q}{}^r g_{a [ \bar b_1} G_{\bar b_2 ] r +} \,, \cr
 \Gamma^{\bar b \bar q_1 \bar q_2 +} &:& - \tfrac{i}{4} \e_{\bar b}{}^c F_{ac \bar q_1 \bar q_2 +} - \tfrac{1}{64} \e_{\bar q_1 \bar q_2 } ( 4 G_{a \bar b +} - g_{a \bar b} (G_{+} \cont{c} - G_{+} \cont{r})) \,, \cr
 \Gamma^{(5)} &:& 0 .
 \eea
Along the $\bar a$-frame derivative of the supercovariant connection we find
\bea
 1 &:& -\tfrac{1}{2} \e^{c_1 c_2} (\Omega_{\bar a, c_1 c_2} + i F_{\bar a c_1 c_2} \cont{r} + i F_{\bar a c_1 c_2 -+}) + \tfrac{1}{8} \e^{r_1 r_2} G_{\bar a r_1 r_2} \,, \cr
 \Gamma^{(1)} &:& 0 \,, \cr
 \Gamma^{\bar b_1 \bar b_2} &:& \tfrac{1}{4} \e_{\bar b_1 \bar b_2} (D_{\bar a} - \tfrac{1}{2} \Omega_{\bar a,} \cont{c} + \tfrac{1}{2} \Omega_{\bar a,} \cont{r} + \tfrac{1}{2} \Omega_{\bar a, -+} - \tfrac{i}{2} F_{\bar a} \cont{c} \cont{r} - \tfrac{i}{2} F_{\bar a -+} \cont{c} + \tfrac{i}{4} F_{\bar a} \cont{r_1} \cont{r_2} + \tfrac{i}{2} F_{\bar a -+} \cont{r}) \,, \cr
 \Gamma^{\bar b \bar q} &:& - \tfrac{1}{2} \e_{\bar b}{}^c ( \Omega_{\bar a,c \bar q} - i F_{\bar a c \bar  q} \cont{c} + i F_{\bar a c \bar q} \cont{r} + i F_{\bar a c \bar q -+} ) + \tfrac{1}{8} \e_{\bar q}{}^{r} G_{\bar a \bar b r} \,, \cr
 \Gamma^{\bar b +} &:& - \tfrac{1}{2} \e_{\bar b}{}^{c_1} ( \Omega_{\bar a, c_1 +} - i F_{\bar a c_1 +} \cont{c_2} + i F_{\bar a c_2 +} \cont{r}) \,, \cr
 \Gamma^{\bar q_1 \bar q_2} &:& - \tfrac{i}{4} \e^{c_1 c_2} F_{\bar a c_1 c_2 \bar q_1 \bar q_2} + \tfrac{1}{32} \e_{\bar q_1 \bar q_2} ( G_{\bar a} \cont{c} + G_{\bar a} \cont{r} - G_{\bar a -+}) \,, \cr
 \Gamma^{\bar q +} &:& - \tfrac{i}{2} \e^{c_1 c_2} F_{\bar a c_1 c_2 \bar q +} + \tfrac{1}{16} \e_{\bar q}{}^r G_{\bar a r +} \,, \cr
 \Gamma^{(3)} &:& 0 \,, \cr
 \Gamma^{\bar b_1 \bar b_2 \bar q_1 \bar q_2} &:& \tfrac{1}{16} \e_{\bar b_1 \bar b_2} (\Omega_{\bar a, \bar q_1 \bar q_2} - i F_{\bar a \bar q_1 \bar q_2} \cont{c} + i F_{\bar a \bar q_1 \bar q_2 -+}) - \tfrac{1}{32} \e_{\bar q_1 \bar q_2} G_{\bar a \bar b_1 \bar b_2} \,, \cr
 \Gamma^{\bar b_1 \bar b_2 \bar q +} &:& \tfrac{1}{8} \e_{\bar b_1 \bar b_2} ( \Omega_{\bar a, \bar q +}- i F_{\bar a \bar q +} \cont{c} + i F_{\bar a \bar q +} \cont{r}) \,, \cr
 \Gamma^{\bar b \bar q_1 \bar q_2 +} &:& - \tfrac{i}{4} \e_{\bar b}{}^c F_{\bar a c \bar q_1 \bar q_2 +} - \tfrac{1}{32} \e_{\bar q_1 \bar q_2 } G_{\bar a \bar b +} \,, \cr
 \Gamma^{(5)} &:& 0 .
 \eea
The components along the $p$-frame derivative of the supercovariant connection are
\bea
 1 &:& -\tfrac{1}{2} \e^{c_1 c_2} (\Omega_{p, c_1 c_2} + i F_{p c_1 c_2} \cont{r} + i F_{p c_1 c_2 -+}) \,, \cr
 \Gamma^{(1)} &:& 0 \,, \cr
 \Gamma^{\bar b_1 \bar b_2} &:& \tfrac{1}{4} \e_{\bar b_1 \bar b_2} (D_{p} - \tfrac{1}{2} \Omega_{p,} \cont{c} + \tfrac{1}{2} \Omega_{p,} \cont{r} + \tfrac{1}{2} \Omega_{p, -+} + \tfrac{i}{4} F_{p} \cont{c_1} \cont{c_2} - \tfrac{i}{2} F_{p} \cont{c} \cont{r} - \tfrac{i}{2} F_{p -+} \cont{c} + \tfrac{i}{2} F_{p -+} \cont{r}) \,, \cr
 \Gamma^{\bar b \bar q} &:& - \tfrac{1}{2} \e_{\bar b}{}^{c_1} ( \Omega_{p, c_1 \bar q} + i F_{c_1 p \bar  q} \cont{c_2} - i F_{c_1 p \bar q} \cont{r} - i F_{c_1 p \bar q -+} ) - \tfrac{1}{8} \e_{\bar q}{}^{r} G_{\bar b p r} \,, \cr
 \Gamma^{\bar b +} &:& - \tfrac{1}{2} \e_{\bar b}{}^c ( \Omega_{p,c+} + i F_{a p +} \cont{c} - F_{a p +} \cont{r}) \,, \cr
 \Gamma^{\bar q_1 \bar q_2} &:& - \tfrac{i}{4} \e^{c_1 c_2} F_{c_1 c_2 p \bar q_1 \bar q_2} - \tfrac{1}{32} \e_{\bar q_1 \bar q_2} ( G_{p} \cont{c} - G_{p} \cont{r} + G_{p -+}) \,, \cr
 \Gamma^{\bar q +} &:& - \tfrac{i}{2} \e^{c_1 c_2} F_{c_1 c_2 p \bar q +} + \tfrac{1}{8} \e_{\bar q}{}^{r} G_{p r +} \,, \cr
 \Gamma^{(3)} &:& 0 \,, \cr
 \Gamma^{\bar b_1 \bar b_2 \bar q_1 \bar q_2} &:& \tfrac{1}{16} \e_{\bar b_1 \bar b_2} ( \Omega_{p, \bar q_1 \bar q_2} - i F_{p \bar q_1 \bar q_2} \cont{c} + i F_{p \bar q_1 \bar q_2 -+} ) - \tfrac{1}{64} \e_{\bar q_1 \bar q_2} G_{\bar b_1 \bar b_2 p} \,, \cr
 \Gamma^{\bar b_1 \bar b_2 \bar q +} &:& \tfrac{1}{8} \e_{\bar b_1 \bar b_2} ( \Omega_{p, \bar q +} - i F_{p \bar q +} \cont{c} + i F_{p \bar q +} \cont{r}) \,, \cr
 \Gamma^{\bar b \bar q_1 \bar q_2 +} &:& \tfrac{i}{4} \e_{\bar b}{}^c F_{c p \bar q_1 \bar q_2 +} + \tfrac{1}{64} \e_{\bar q_1 \bar q_2 } G_{\bar b p +} \,, \cr
 \Gamma^{(5)} &:& 0 \,,
 \eea
Similarly, the components along the $\bar p$-frame derivative are
\bea
 1 &:& -\tfrac{1}{2} \e^{c_1 c_2} (\Omega_{\bar p, c_1 c_2} + i F_{\bar p c_1 c_2} \cont{r} + i F_{\bar p c_1 c_2 -+}) + \tfrac{1}{8} \e^{r_1 r_2} (G_{\bar p r_1 r_2} - g_{\bar p q_1} ( G_{q_2} \cont{c} + G_{q_2} \cont{r} + G_{q_2 -+})) \,, \cr
 \Gamma^{(1)} &:& 0 \,, \cr
 \Gamma^{\bar b_1 \bar b_2} &:& \tfrac{1}{4} \e_{\bar b_1 \bar b_2} (D_{\bar p} - \tfrac{1}{2} \Omega_{\bar p,} \cont{c} + \tfrac{1}{2} \Omega_{\bar p,} \cont{r} + \tfrac{1}{2} \Omega_{\bar p, -+} + \tfrac{i}{4} F_{\bar p} \cont{c_1} \cont{c_2} - \tfrac{i}{2} F_{\bar p} \cont{c} \cont{r} - \tfrac{i}{2} F_{\bar p -+} \cont{c}
\cr
&& + \tfrac{i}{2} F_{\bar p -+} \cont{r})
- \tfrac{1}{16} \e_{\bar p}{}^{r} G_{\bar b_1 \bar b_2 r} \,, \cr
 \Gamma^{\bar b \bar q} &:& - \tfrac{1}{2} \e_{\bar b}{}^{c_1} ( \Omega_{\bar p, c_1 \bar q} + i F_{c_1 \bar p \bar  q} \cont{c_2} - i F_{c_1 \bar p \bar q -+} ) + \tfrac{1}{16} \e_{\bar p}{}^{r_1} ( 4 G_{\bar  b \bar q r_1} - g_{\bar q r_1} ( G_{\bar b} \cont{c} + 3 G_{\bar b} \cont{r_2} + G_{\bar b -+})) \,, \cr
 \Gamma^{\bar b +} &:& - \tfrac{1}{2} \e_{\bar b}{}^c ( \Omega_{\bar p,c+} + i F_{a \bar p +} \cont{c} - F_{a \bar p +} \cont{r}) + \tfrac{1}{8} \e_{\bar p}{}^{r} G_{\bar b r +} \,, \cr
 \Gamma^{\bar q_1 \bar q_2} &:& - \tfrac{1}{16} \e_{\bar q_1 \bar q_2} ( G_{\bar p} \cont{c} - G_{\bar p} \cont{r} + G_{\bar p -+}) \,, \cr
 \Gamma^{\bar q +} &:& - \tfrac{i}{2} \e^{c_1 c_2} F_{c_1 c_2 \bar p \bar q +} + \tfrac{1}{16} \e_{\bar q}{}^{r_1} ( 4 G_{\bar p r_1 +} - g_{\bar p r_1} (G_{+} \cont{c} - G_{+} \cont{r_2} ))) \,, \cr
 \Gamma^{(3)} &:& 0 \,, \cr
 \Gamma^{\bar b_1 \bar b_2 \bar q_1 \bar q_2} &:& \tfrac{1}{16} \e_{\bar b_1 \bar b_2} \Omega_{\bar p, \bar q_1 \bar q_2} - \tfrac{1}{32} \e_{\bar q_1 \bar q_2} G_{\bar b_1 \bar b_2 \bar p} \,, \cr
 \Gamma^{\bar b_1 \bar b_2 \bar q +} &:& \tfrac{1}{8} \e_{\bar b_1 \bar b_2} ( \Omega_{\bar p, \bar q +}- i F_{\bar p \bar q +} \cont{c} ) + \tfrac{1}{32} \e_{\bar p \bar q} G_{\bar b_1 \bar b_2 +} \,, \cr
 \Gamma^{\bar b \bar q_1 \bar q_2 +} &:& \tfrac{1}{16} \e_{\bar q_1 \bar q_2 } G_{\bar b \bar p +} \,, \cr
 \Gamma^{(5)} &:& 0 .
 \eea
The supercovariant derivative with $M=-$ gives
\bea
 1 &:& -\tfrac{1}{2} \e^{c_1 c_2} (\Omega_{-, c_1 c_2} + i F_{c_1 c_2 -} \cont{r}) + \tfrac{1}{4} \e^{r_1 r_2} G_{r_1 r_2 -} \,, \cr
 \Gamma^{(1)} &:& 0 \,, \cr
 \Gamma^{\bar b_1 \bar b_2} &:& \tfrac{1}{4} \e_{\bar b_1 \bar b_2} (D_{-} - \tfrac{1}{2} \Omega_{-,} \cont{c} + \tfrac{1}{2} \Omega_{-,} \cont{r} + \tfrac{1}{2} \Omega_{-, -+} + \tfrac{i}{4} F_{-} \cont{c_1} \cont{c_2} - \tfrac{i}{2} F_{-} \cont{c} \cont{r} + \tfrac{i}{4} F_{-} \cont{r_1} \cont{r_2}) \,, \cr
 \Gamma^{\bar b \bar q} &:& - \tfrac{1}{2} \e_{\bar b}{}^c ( \Omega_{-,c \bar q} - i F_{c \bar  q -} \cont{c} + i F_{c \bar q -} \cont{r} ) + \tfrac{1}{4} \e_{\bar q}{}^{r} G_{\bar b r -} \,, \cr
 \Gamma^{\bar b +} &:& - \tfrac{1}{2} \e_{\bar b}{}^{c_1} (\Omega_{-, c_1 +} + i F_{c_1 -+} \cont{c_2} - i F_{c_1 -+} \cont{r}) + \tfrac{1}{16} \e^{r_1 r_2} G_{\bar b r_1 r_2} \,, \cr
 \Gamma^{\bar q_1 \bar q_2} &:& - \tfrac{i}{4} \e^{c_1 c_2} F_{c_1 c_2 \bar q_1 \bar q_2 -} - \tfrac{1}{16} \e_{\bar q_1 \bar q_2} ( G_{-} \cont{c} - G_{-} \cont{r} ) \,, \cr
 \Gamma^{\bar q +} &:& \tfrac{i}{2} \e^{c_1 c_2} F_{c_1 c_2 \bar q -+} + \tfrac{1}{16} \e_{\bar q}{}^{r_1} ( G_{r_1} \cont{c} - G_{r_1} \cont{r_2} - 3 G_{r_1 -+}) \,, \cr
 \Gamma^{(3)} &:& 0 \,, \cr
 \Gamma^{\bar b_1 \bar b_2 \bar q_1 \bar q_2} &:& \tfrac{1}{16} \e_{\bar b_1 \bar b_2} (\Omega_{-, \bar q_1 \bar q_2} - i F_{\bar q_1 \bar q_2 -} \cont{c} ) - \tfrac{1}{32} \e_{\bar q_1 \bar q_2} G_{\bar b_1 \bar b_2 -} \,, \cr
 \Gamma^{\bar b_1 \bar b_2 \bar q +} &:& \tfrac{1}{8} \e_{\bar b_1 \bar b_2} (\Omega_{-, \bar q +} + i F_{\bar q -+} \cont{c} - i F_{\bar q -+} \cont{r}) + \tfrac{1}{32} \e_{\bar q}{}^r G_{\bar b_1 \bar b_2 r} \,, \cr
 \Gamma^{\bar b \bar q_1 \bar q_2 +} &:& \tfrac{i}{4} \e_{\bar b}{}^c F_{c \bar q_1 \bar q_2 -+} - \tfrac{1}{64} \e_{\bar q_1 \bar q_2} (G_{\bar b} \cont{c} - G_{\bar b} \cont{r} - 3 G_{\bar b -+}) \,, \cr
 \Gamma^{(5)} &:& 0 .
 \eea
Finally, for $M=+$ we find
\bea
 1 &:& -\tfrac{1}{2} \e^{c_1 c_2} (\Omega_{+, c_1 c_2} + i F_{+ c_1 c_2} \cont{r}) + \tfrac{1}{8} \e^{r_1 r_2} G_{r_1 r_2 +} \,, \cr
 \Gamma^{(1)} &:& 0 \,, \cr
 \Gamma^{\bar b_1 \bar b_2} &:& \tfrac{1}{4} \e_{\bar b_1 \bar b_2} (D_{+} - \tfrac{1}{2} \Omega_{+,} \cont{c} + \tfrac{1}{2} \Omega_{+,} \cont{r} + \tfrac{1}{2} \Omega_{+, -+} + \tfrac{i}{4} F_{+} \cont{c_1} \cont{c_2} - \tfrac{i}{2} F_{+} \cont{c} \cont{r} + \tfrac{i}{4} F_{+} \cont{r_1} \cont{r_2}) \,, \cr
 \Gamma^{\bar b \bar q} &:& - \tfrac{1}{2} \e_{\bar b}{}^c ( \Omega_{+,c \bar q} - i F_{c \bar  q +} \cont{c} + i F_{c \bar q +} \cont{r} ) + \tfrac{1}{8} \e_{\bar q}{}^{r} G_{\bar b r +} \,, \cr
 \Gamma^{\bar b +} &:& - \tfrac{1}{2} \e_{\bar b}{}^c \Omega_{+,c+} \,, \cr
 \Gamma^{\bar q_1 \bar q_2} &:& - \tfrac{i}{4} \e^{c_1 c_2} F_{c_1 c_2 \bar q_1 \bar q_2 +} - \tfrac{1}{32} \e_{\bar q_1 \bar q_2} ( G_{+} \cont{c} - G_{+} \cont{r} ) \,, \cr
 \Gamma^{\bar q +} &:& 0 \,, \cr
 \Gamma^{(3)} &:& 0 \,, \cr
 \Gamma^{\bar b_1 \bar b_2 \bar q_1 \bar q_2} &:& \tfrac{1}{16} \e_{\bar b_1 \bar b_2} (\Omega_{+, \bar q_1 \bar q_2} - i F_{\bar q_1 \bar q_2 +} \cont{c} ) - \tfrac{1}{64} \e_{\bar q_1 \bar q_2} G_{\bar b_1 \bar b_2 +} \,, \cr
 \Gamma^{\bar b_1 \bar b_2 \bar q +} &:& \tfrac{1}{8} \e_{\bar b_1 \bar b_2} \Omega_{+, \bar q +} \,, \cr
 \Gamma^{\bar b \bar q_1 \bar q_2 +} &:& 0 \,, \cr
 \Gamma^{(5)} &:& 0 .
 \eea

\subsection{Killing spinor equations on   $e_{k5}$}

We now consider the basis elements $e_{k5} = \tfrac{1}{2} \Gamma^{\bar k} \Gamma^{+} 1$ with $k \leq 4$. For this purpose we split up $\a$ into $\rho$ and $k$, where $\rho$ are the remaining three holomorphic indices: $\rho = (1, \ldots, \hat k, \ldots, 4)$. Thus, $k$ is a single element of $\{ 1,2,3,4 \}$ and not an index that should be summed over. Furthermore we will use the three-dimensional Levi-Civita symbol defined by $\e_{\rho_1 \cdots \rho_3} = \e_{k \rho_1 \cdots \rho_3}$. Substituting this spinor into the (algebraic) Killing spinor equation
({\ref{kseqnb}}) and expanding the resulting expression in the basis ({\ref{hbasisa}}), we find
 \bea
  1 &:& 0 \,, \cr
  \Gamma^{\bar k} &:& \tfrac{1}{4} ( G_{- k \bar k} - G_{-} \cont{\g} ) \,, \cr
  \Gamma^{\bar \tau} &:& \tfrac{1}{2} G_{k \bar \tau -} \,, \cr
  \Gamma^{+} &:& \tfrac{1}{4} ( G_{k} \cont{\g} - G_{k -+} ) \,, \cr
  \Gamma^{(2)} &:& 0 \,, \cr
  \Gamma^{\bar k \bar \tau_1 \bar \tau_2} &:& - \tfrac{1}{8} G_{\bar \tau_1 \bar \tau_2 -} \,, \cr
  \Gamma^{\bar k \bar \tau +} &:& \tfrac{1}{8} (G_{\bar \tau k \bar k} - G_{\bar \tau} \cont{\s} + G_{\bar \tau -+} ) \,, \cr
  \Gamma^{\bar \tau_1 \cdots \bar \tau_3} &:& - \tfrac{1}{12} \e_{\bar \tau_1 \cdots \bar \tau_3} P_- \,, \cr
  \Gamma^{\bar \tau_1 \bar \tau_2 +} &:& \tfrac{1}{4} \e_{\bar \tau_1 \bar \tau_2}{}^{\s} P_{\s} + \tfrac{1}{8} G_{k \bar \tau_1 \bar \tau_2} \,, \cr
  \Gamma^{(4)} &:& 0 \,, \cr
  \Gamma^{\bar k \bar \tau_1 \cdots \bar \tau_3 +} &:& \tfrac{1}{24} \e_{\bar \tau_1 \cdots \bar \tau_3} P_{\bar k} - \tfrac{1}{48} G_{\bar \tau_1 \cdots \bar \tau_3} .
 \eea
Next we turn to the Killing spinor equation associated
with the supercovariant derivative (\ref{kseqna}).
The components along the $k$-frame derivative of the supercovariant connection are
\bea
 1 &:& - \Omega_{k, k -} \,, \cr
 \Gamma^{(1)} &:& 0 \,, \cr
 \Gamma^{\bar k \bar \tau} &:& \tfrac{1}{2} \Omega_{k, \bar \tau -} + \tfrac{i}{2} F_{k \bar \tau -} \cont{\s}
 - \tfrac{1}{16} \e_{\bar \tau}{}^{\s_1 \s_2} G_{\s_1 \s_2 -} \,, \cr
 \Gamma^{\bar k +} &:& \tfrac{1}{2} ( D_{k} - \tfrac{1}{2} \Omega_{k, k \bar k}
 + \tfrac{1}{2} \Omega_{k,} \cont{\s} - \tfrac{1}{2} \Omega_{k,-+}
 + \tfrac{i}{4} F_{k} \cont{\s_1} \cont{\s_2} - \tfrac{i}{2} F_{k} \cont{\s} {}_{-+}) + \tfrac{1}{48} \e^{\s_1 \cdots \s_3} G_{\s_1 \cdots \s_3} \,, \cr
 \Gamma^{\bar \tau_1 \bar \tau_2} &:& -\tfrac{1}{8} \e_{\bar \tau_1 \bar \tau_2}{}^{\s} G_{k \s -} \,, \cr
 \Gamma^{\bar \tau +} &:&  - \tfrac{1}{2} \Omega_{k, k \bar \tau}
 - \tfrac{1}{8} \e_{\bar \rho}{}^{\s_1 \s_2} G_{k \s_1 \s_2} \,, \cr
 \Gamma^{(3)} &:& 0 \,, \cr
 \Gamma^{\bar k \bar \tau_1 \cdots \bar \tau_3} &:& \tfrac{i}{12} F_{k \bar \tau_1 \cdots \bar \tau_3 -} - \tfrac{1}{192} \e_{\bar \tau_1 \cdots \bar \tau_3} (3 G_{- k \bar k} + G_{-} \cont{\s}) \,, \cr
 \Gamma^{\bar k \bar \tau_1 \bar \tau_2 +} &:& \tfrac{1}{8} ( \Omega_{k, \bar \tau_1 \bar \tau_2} + i F_{k \bar \tau_1 \bar \tau_2} \cont{\s} - i F_{k \bar \tau_1 \bar \tau_2 -+}) + \tfrac{1}{64} \e_{\bar \tau_1 \bar \tau_2}{}^{\s_1} (3 G_{\s_1 k \bar k} + G_{\s_1} \cont{\s_2} + G_{\s_1 -+}) \,, \cr
 \Gamma^{\bar \tau_1 \cdots \bar \tau_3 +} &:& - \tfrac{1}{96} \e_{\bar \tau_1 \cdots \bar \tau_3} (G_{k} \cont{\s} + G_{k -+}) \,, \cr
   \Gamma^{(5)} &:& 0 .
\eea
Along the $\bar k$-frame derivative of the supercovariant connection we find
\bea
 1 &:& - \Omega_{\bar k, k -} + i F_{k \bar k-} \cont{\s} \,, \cr
 \Gamma^{(1)} &:& 0 \,, \cr
 \Gamma^{\bar k \bar \tau} &:& \tfrac{1}{2} \Omega_{\bar k, \bar \tau -}
 + \tfrac{i}{2} F_{ \bar k \bar \tau -} \cont{\s} \,, \cr
 \Gamma^{\bar k +} &:& \tfrac{1}{2} ( D_{\bar k} - \tfrac{1}{2} \Omega_{\bar k, k \bar k}
 + \tfrac{1}{2} \Omega_{\bar k} \cont{\s} - \tfrac{1}{2} \Omega_{\bar k,-+}
 + \tfrac{i}{4} F_{\bar k} \cont{\s_1} \cont{\s_2} - \tfrac{i}{2} F_{\bar k} \cont{\s} {}_{-+}) \,, \cr
 \Gamma^{\bar \tau_1 \bar \tau_2} &:& \tfrac{i}{2} F_{k \bar k \bar \tau_1 \bar \tau_2 -} - \tfrac{1}{16} \e_{\bar \tau_1 \bar \tau_2}{}^{\s}  G_{\bar k \s -} \,, \cr
 \Gamma^{\bar \tau +} &:&  - \tfrac{1}{2} \Omega_{\bar k, k \bar \tau}
 + \tfrac{i}{2} F_{k \bar k \bar \tau} \cont{\s} - \tfrac{i}{2} F_{k \bar k \bar \tau -+}
 - \tfrac{1}{16} \e_{\bar \tau}{}^{\s_1 \s_2} G_{\bar k \s_1 \s_2} \,, \cr
 \Gamma^{(3)} &:& 0 \,, \cr
 \Gamma^{\bar k \bar \tau_1 \cdots \bar \tau_3} &:& \tfrac{i}{12} F_{\bar k \bar \tau_1 \cdots \bar \tau_3 -} \,, \cr
 \Gamma^{\bar k \bar \tau_1 \bar \tau_2 +} &:& \tfrac{1}{8} ( \Omega_{\bar k, \bar \tau_1 \bar \tau_2} + i F_{\bar k \bar \tau_1 \bar \tau_2} \cont{\s}- i F_{\bar k \bar \tau_1 \bar \tau_2 -+}) \,, \cr
 \Gamma^{\bar \tau_1 \cdots \bar \tau_3 +} &:& \tfrac{i}{12} F_{k \bar k \bar \tau_1 \cdots \bar \tau_3} - \tfrac{1}{192} \e_{\bar \tau_1 \cdots \bar \tau_3} (G_{\bar k} \cont{\s} + G_{\bar k -+})  \,, \cr
   \Gamma^{(5)} &:& 0 .
\eea
The components along the $\rho$-frame derivative of the supercovariant connection are
\bea
 1 &:& - \Omega_{\rho, k -} + i F_{k \rho -} \cont{\s} \,, \cr
 \Gamma^{(1)} &:& 0 \,, \cr
 \Gamma^{\bar k \bar \tau} &:& \tfrac{1}{2} \Omega_{ \rho, \bar \tau -}
 - \tfrac{i}{2} F_{ \rho \bar \tau - k \bar k} + \tfrac{i}{2} F_{\rho \bar \tau -} \cont{\s} \,, \cr
 \Gamma^{\bar k +} &:& \tfrac{1}{2} ( D_{ \rho} - \tfrac{1}{2} \Omega_{ \rho, k \bar k}
 + \tfrac{1}{2} \Omega_{ \rho,} \cont{\s} - \tfrac{1}{2} \Omega_{ \rho,-+}
 - \tfrac{i}{2} F_{ \rho k \bar k} \cont{\s} + \tfrac{i}{2} F_{ \rho k \bar k -+}
 + \tfrac{i}{4} F_{ \rho} \cont{\s_1} \cont{\s_2} - \tfrac{i}{2} F_{ \rho} \cont{\s} {}_{-+}) \,, \cr
 \Gamma^{\bar \tau_1 \bar \tau_2} &:& \tfrac{i}{2} F_{k  \rho \bar \tau_1 \bar \tau_2 -} - \tfrac{1}{16} \e_{\bar \tau_1 \bar \tau_2}{}^{\s}  G_{\rho \s -} \,, \cr
 \Gamma^{\bar \tau +} &:&  - \tfrac{1}{2} \Omega_{ \rho, k \bar \tau}
 + \tfrac{i}{2} F_{k  \rho \bar \tau} \cont{\s} - \tfrac{i}{2} F_{k  \rho \bar \tau -+}
 - \tfrac{1}{16} \e_{\bar \tau}{}^{\s_1 \s_2} G_{\rho \s_1 \s_2} \,, \cr
 \Gamma^{(3)} &:& 0 \,, \cr
 \Gamma^{\bar k \bar \tau_1 \cdots \bar \tau_3} &:& \tfrac{i}{12} F_{\rho \bar \tau_1 \cdots \bar \tau_3 -} + \tfrac{1}{96} \e_{\bar \tau_1 \cdots \bar \tau_3} G_{\bar k \rho -} \,, \cr
 \Gamma^{\bar k \bar \tau_1 \bar \tau_2 +} &:& \tfrac{1}{8} ( \Omega_{ \rho, \bar \tau_1 \bar \tau_2} - i F_{k \bar k  \rho \bar \tau_1 \bar \tau_2} + i F_{ \rho \bar \tau_1 \bar \tau_2} \cont{\s}- i F_{ \rho \bar \tau_1 \bar \tau_2 -+}) - \tfrac{1}{32} \e_{\bar \tau_1 \bar \tau_2}{}^{\s} G_{\bar k \rho \s} \,, \cr
 \Gamma^{\bar \tau_1 \cdots \bar \tau_3 +} &:& \tfrac{i}{12} F_{k \rho \bar \tau_1 \cdots \bar \tau_3} + \tfrac{1}{192} \e_{\bar \tau_1 \cdots \bar \tau_3} (G_{\rho k \bar k} - G_{\rho} \cont{\s} - G_{\rho -+})  \,, \cr
   \Gamma^{(5)} &:& 0 .
\eea
Similarly, the components along the $\bar \rho$-frame derivative are
\bea
 1 &:& - \Omega_{\bar \rho, k -}+ i F_{k \bar \rho -} \cont{\s} - \tfrac{1}{8} \e_{\bar \rho}{}^{\s_1 \s_2} G_{\s_1 \s_2 -} \,, \cr
 \Gamma^{(1)} &:& 0 \,, \cr
 \Gamma^{\bar k \bar \tau} &:& \tfrac{1}{2} \Omega_{\bar \rho, \bar \tau -}
 - \tfrac{i}{2} F_{\bar \rho \bar \tau - k \bar k} + \tfrac{i}{2} F_{\bar \rho \bar \tau -} \cont{\s}
 - \tfrac{1}{8} \e_{\bar \rho \bar \tau}{}^{\s} G_{\bar k \s -} \,, \cr
 \Gamma^{\bar k +} &:& \tfrac{1}{2} ( D_{\bar \rho} - \tfrac{1}{2} \Omega_{\bar \rho, k \bar k}
 + \tfrac{1}{2} \Omega_{\bar \rho,} \cont{\s} - \tfrac{1}{2} \Omega_{\bar \rho,-+}
 - \tfrac{i}{2} F_{\bar \rho k \bar k} \cont{\s} + \tfrac{i}{2} F_{\bar \rho k \bar k -+}
 + \tfrac{i}{4} F_{\bar \rho} \cont{\s_1} \cont{\s_2}
\cr
&&- \tfrac{i}{2} F_{\bar \rho} \cont{\s} {}_{-+})  - \tfrac{1}{16} \e_{\bar \rho}{}^{\s_1 \s_2} G_{\bar k \s_1 \s_2} \,, \cr
 \Gamma^{\bar \tau_1 \bar \tau_2} &:& \tfrac{i}{2} F_{k \bar \rho \bar \tau_1 \bar \tau_2 -} + \tfrac{1}{32} \e_{\bar \tau_1 \bar \tau_2}{}^{\s_1} (g_{\bar \rho \s_1} G_{- k \bar k} - g_{\bar \rho \s_1} G_{-} \cont{\s_2} - 4 G_{\bar \rho \s_1 -}) \,, \cr
 \Gamma^{\bar \tau +} &:&  - \tfrac{1}{2} \Omega_{\bar \rho, k \bar \tau}
 + \tfrac{i}{2} F_{k \bar \rho \bar \tau} \cont{\s} - \tfrac{i}{2} F_{k \bar \rho \bar \tau -+}
 - \tfrac{1}{16} \e_{\bar \rho}{}^{\s_1 \s_2} ( 2 G_{\bar \tau \s_1 \s_2} + g_{\bar \tau \s_1} (G_{\s_2 k \bar k}
 + 3 G_{\s_2} \cont{\s_3} - G_{\s_2 - +} )) \,, \cr
 \Gamma^{(3)} &:& 0 \,, \cr
 \Gamma^{\bar k \bar \tau_1 \cdots \bar \tau_3} &:& + \tfrac{1}{48} \e_{\bar \tau_1 \cdots \bar \tau_3} G_{\bar k \bar \rho -} \,, \cr
 \Gamma^{\bar k \bar \tau_1 \bar \tau_2 +} &:& \tfrac{1}{8} ( \Omega_{\bar \rho, \bar \tau_1 \bar \tau_2} - i F_{k \bar k \bar \rho \bar \tau_1 \bar \tau_2} - i F_{\bar \rho \bar \tau_1 \bar \tau_2 -+}) - \tfrac{1}{64} \e_{\bar \tau_1 \bar \tau_2}{}^{\s_1} (4 G_{\bar k \bar \rho \s_1} + g_{\bar \rho \s_1} ( G_{\bar k} \cont{\s_2} + G_{\bar k -+} )) \,, \cr
 \Gamma^{\bar \tau_1 \cdots \bar \tau_3 +} &:& \tfrac{1}{96} \e_{\bar \tau_1 \cdots \bar \tau_3} (G_{\bar \rho k \bar k} - G_{\bar \rho} \cont{\s} - G_{\bar \rho -+} ) \,, \cr
   \Gamma^{(5)} &:& 0 .
\eea
The supercovariant derivative with $M=-$ gives
\bea
 1 &:& - \Omega_{-, k -} \,, \cr
 \Gamma^{(1)} &:& 0 \,, \cr
 \Gamma^{\bar k \bar \tau} &:& \tfrac{1}{2} \Omega_{-, \bar \tau -} \,, \cr
 \Gamma^{\bar k +} &:& \tfrac{1}{2} ( D_{-} - \tfrac{1}{2} \Omega_{-, k \bar k}
 + \tfrac{1}{2} \Omega_{-,} \cont{\s} - \tfrac{1}{2} \Omega_{-,-+}
 - \tfrac{i}{2} F_{-} \cont{\s} {}_{k \bar k}+ \tfrac{i}{4} F_{-} \cont{\s_1} \cont{\s_2} ) \,, \cr
 \Gamma^{\bar \tau_1 \bar \tau_2} &:& 0 \,, \cr
 \Gamma^{\bar \tau +} &:&  - \tfrac{1}{2} \Omega_{-, k \bar \tau} - \tfrac{i}{2} F_{k \bar \tau} \cont{\s} {}_{-}
 - \tfrac{1}{16} \e_{\bar \rho}{}^{\s_1 \s_2} G_{- \s_1 \s_2} \,, \cr
 \Gamma^{(3)} &:& 0 \,, \cr
 \Gamma^{\bar k \bar \tau_1 \cdots \bar \tau_3} &:& 0 \,, \cr
 \Gamma^{\bar k \bar \tau_1 \bar \tau_2 +} &:& \tfrac{1}{8} ( \Omega_{-, \bar \tau_1 \bar \tau_2} + i F_{- \bar \tau_1 \bar \tau_2} \cont{\s} - i F_{- \bar \tau_1 \bar \tau_2 k \bar k}) + \tfrac{1}{32} \e_{\bar \tau_1 \bar \tau_2}{}^{\s} G_{\bar k \s -} \,, \cr
 \Gamma^{\bar \tau_1 \cdots \bar \tau_3 +} &:& - \tfrac{i}{12} F_{k \bar \tau_1 \cdots \bar \tau_3 - } + \tfrac{1}{192} \e_{\bar \tau_1 \cdots \bar \tau_3} (G_{- k \bar k} - G_{-} \cont{\s})  \,, \cr
   \Gamma^{(5)} &:& 0 .
 \eea
Finally, for $M=+$ we find
\bea
  1 &:& - \Omega_{+, k -} - i F_{k -+} \cont{\s} + \tfrac{1}{24} \e^{\s_1 \cdots \s_3} G_{\s_1 \cdots \s_3} \,, \cr
 \Gamma^{(1)} &:& 0 \,, \cr
 \Gamma^{\bar k \bar \tau} &:& \tfrac{1}{2} \Omega_{+, \bar \tau -}
 - \tfrac{i}{2} F_{\bar \tau - + k \bar k} + \tfrac{i}{2} F_{\bar \tau -+} \cont{\s} + \tfrac{1}{16} \e_{\bar \tau}{}^{\s_2 \s_2} G_{\bar k \s_1 \s_2} \,, \cr
 \Gamma^{\bar k +} &:& \tfrac{1}{2} ( D_{+} - \tfrac{1}{2} \Omega_{+, k \bar k}
 + \tfrac{1}{2} \Omega_{+,} \cont{\s} - \tfrac{1}{2} \Omega_{+,-+}
 - \tfrac{i}{2} F_{+ k \bar k} \cont{\s}
 + \tfrac{i}{4} F_{+} \cont{\s_1} \cont{\s_2}) \,, \cr
 \Gamma^{\bar \tau_1 \bar \tau_2} &:& - \tfrac{i}{2} F_{k  \bar \tau_1 \bar \tau_2 -+} - \tfrac{1}{32} \e_{\bar \tau_1 \bar \tau_2}{}^{\s_1}  (G_{\s_1 k \bar k} - G_{\s_1} \cont{\s_2} + 3 G_{\s_1 -+} )\,, \cr
 \Gamma^{\bar \tau +} &:&  - \tfrac{1}{2} \Omega_{+, k \bar \tau}
 - \tfrac{i}{2} F_{k \bar \tau +} \cont{\s}
 - \tfrac{1}{8} \e_{\bar \tau}{}^{\s_1 \s_2} G_{+ \s_1 \s_2} \,, \cr
 \Gamma^{(3)} &:& 0 \,, \cr
 \Gamma^{\bar k \bar \tau_1 \cdots \bar \tau_3} &:& \tfrac{i}{12} F_{\bar \tau_1 \cdots \bar \tau_3 -+} + \tfrac{1}{192} \e_{\bar \tau_1 \cdots \bar \tau_3} (G_{\bar k} \cont{\s} - 3 G_{\bar k -+} ) \,, \cr
 \Gamma^{\bar k \bar \tau_1 \bar \tau_2 +} &:& \tfrac{1}{8} ( \Omega_{+, \bar \tau_1 \bar \tau_2} - i F_{k \bar k  \bar \tau_1 \bar \tau_2 +} + i F_{\bar \tau_1 \bar \tau_2 +} \cont{\s}) + \tfrac{1}{16} \e_{\bar \tau_1 \bar \tau_2}{}^{\s} G_{\bar k \s + } \,, \cr
 \Gamma^{\bar \tau_1 \cdots \bar \tau_3 +} &:& - \tfrac{i}{12} F_{k \bar \tau_1 \cdots \bar \tau_3 +} + \tfrac{1}{96} \e_{\bar \tau_1 \cdots \bar \tau_3} (G_{+ k \bar k} - G_{+} \cont{\s})  \,, \cr
   \Gamma^{(5)} &:& 0 .
\eea

\subsection{Killing spinor equations on   $e_{1234}$}

The next spinor basis element we consider is $e_{1234}$. Substituting this spinor into the (algebraic) Killing spinor equation
({\ref{kseqnb}}) and expanding the resulting expression in the basis ({\ref{hbasisa}}), we find
 \bea
  1 &:& 0 \,, \cr
  \Gamma^{\bar \b} &:& P_{\bar \b} + \tfrac{1}{12} \e_{\bar \b}{}^{\g_1 \cdots \g_3} G_{\g_1 \cdots \g_3} \,, \cr
  \Gamma^{+} &:& P_+ \,, \cr
  \Gamma^{(2)} &:& 0 \,, \cr
  \Gamma^{\bar \b_1 \cdots \bar \b_3} &:& \tfrac{1}{48} \e_{\bar \b_1 \cdots \bar \b_3}{}^{\g} (G_{\g} \cont{\d} - G_{\g -+}) \,, \cr
  \Gamma^{\bar \b_1 \bar \b_2 +} &:& -\tfrac{1}{16} \e_{\bar \b_1 \bar \b_2}{}^{\g_1 \g_2} G_{\g_1 \g_2 +} \,, \cr
  \Gamma^{(4)} &:& 0 \,, \cr
  \Gamma^{\bar \b_1 \cdots \bar \b_4 +} &:& - \tfrac{1}{384} \e_{\bar \b_1 \cdots \bar \b_4} G_{+} \cont{\g} .
 \eea
Next we turn to the Killing spinor equation associated
with the supercovariant derivative (\ref{kseqna}).
The components along the $\a$-frame derivative of the supercovariant connection are
\bea
  1 &:& \tfrac{1}{4} (G_{\a} \cont{\g} + G_{\a -+}) \,, \cr
  \Gamma^{(1)} &:& 0 \,, \cr
  \Gamma^{\bar \b_1 \bar \b_2} &:& - \tfrac{1}{8} \epsilon_{\bar \b_1 \bar \b_2}{}^{\g_1 \g_2} (\Omega_{\a, \g_1 \g_2} - i F_{\a \g_1 \g_2} \cont{\d} + i F_{\a \g_1 \g_2 -+}) + \tfrac{1}{8} G_{\a \bar \b_1 \bar \b_2} - \tfrac{1}{16} g_{\a [\bar \b_1} (G_{\bar \b_2]} \cont{\g} + G_{\bar \b_2] -+}) \,, \cr
  \Gamma^{\bar \b +} &:& \tfrac{i}{6} \e_{\bar \b}{}^{\g_1 \cdots \g_3} F_{\a \g_1 \cdots \g_3 +} + \tfrac{1}{4} G_{\a \bar \b +} - \tfrac{1}{16} g_{\a \bar \b} G_{+} \cont{\g} \,, \cr
  \Gamma^{(3)} &:& 0 \,, \cr
  \Gamma^{\bar \b_1 \cdots \bar \b_4} &:& \tfrac{1}{96} \e_{\bar \b_1 \cdots \bar \b_4} (D_{\a} - \tfrac{1}{2} \Omega_{\a,} \cont{\g} + \tfrac{1}{2} \Omega_{\a,-+} + \tfrac{i}{4} F_{\a} \cont{\g_1} \cont{\g_2} - \tfrac{i}{2} F_{\a} \cont{\g_1}{}_{-+}) - \tfrac{1}{96} g_{\a [ \bar \b_1} G_{\bar \b_2 \cdots \bar \b_4]} \,, \cr
  \Gamma^{\bar \b_1 \cdots \bar \b_3 +} &:& - \tfrac{1}{24} \e_{\bar \b_1 \cdots \bar \b_3}{}^{\g} (\Omega_{\a, \g +} - i F_{\a \g +} \cont{\d}) - \tfrac{1}{32} g_{\a [ \bar \b_1} G_{\bar \b_2 \bar \b_3] +} \,, \cr
   \Gamma^{(5)} &:& 0 .
\eea
Along the $\bar \a$-frame derivative of the supercovariant connection we find
\bea
  1 &:& \tfrac{i}{12} \e^{\b_1 \cdots \b_4} F_{\bar \a \b_1 \cdots \b_4} + \tfrac{1}{8} (G_{\bar \a} \cont{\g} + G_{\bar \a -+}) \,, \cr
  \Gamma^{(1)} &:& 0 \,, \cr
  \Gamma^{\bar \b_1 \bar \b_2} &:& - \tfrac{1}{8} \epsilon_{\bar \b_1 \bar \b_2}{}^{\g_1 \g_2} (\Omega_{\bar \a, \g_1 \g_2} - i F_{\bar \a \g_1 \g_2} \cont{\d} + i F_{\bar \a \g_1 \g_2 -+}) + \tfrac{1}{16} G_{\bar \a \bar \b_1 \bar \b_2} \,, \cr
  \Gamma^{\bar \b +} &:& \tfrac{i}{6} \e_{\bar \b}{}^{\g_1 \cdots \g_3} F_{\bar \a \g_1 \cdots \g_3 +} + \tfrac{1}{8} G_{\bar \a \bar \b +} \,, \cr
  \Gamma^{(3)} &:& 0 \,, \cr
  \Gamma^{\bar \b_1 \cdots \bar \b_4} &:& \tfrac{1}{96} \e_{\bar \b_1 \cdots \bar \b_4} (D_{\bar \a} - \tfrac{1}{2} \Omega_{\bar \a,} \cont{\g} + \tfrac{1}{2} \Omega_{\bar \a,-+} + \tfrac{i}{4} F_{\bar \a} \cont{\g_1} \cont{\g_2} - \tfrac{i}{2} F_{\bar \a} \cont{\g_1}{}_{-+}) \,, \cr
  \Gamma^{\bar \b_1 \cdots \bar \b_3 +} &:& - \tfrac{1}{24} \e_{\bar \b_1 \cdots \bar \b_3}{}^{\g} (\Omega_{\bar \a, \g +} - i F_{\bar \a \g +} \cont{\d}) \,, \cr
   \Gamma^{(5)} &:& 0 .
\eea
The supercovariant derivative with $M=-$ gives
\bea
  1 &:& \tfrac{i}{12} \e^{\b_1 \cdots \b_4} F_{- \b_1 \cdots \b_4} + \tfrac{1}{4} G_{-} \cont{\g} \,, \cr
  \Gamma^{(1)} &:& 0 \,, \cr
  \Gamma^{\bar \b_1 \bar \b_2} &:& - \tfrac{1}{8} \epsilon_{\bar \b_1 \bar \b_2}{}^{\g_1 \g_2} (\Omega_{-, \g_1 \g_2} - i F_{- \g_1 \g_2} \cont{\d}) + \tfrac{1}{8} G_{- \bar \b_1 \bar \b_2} \,, \cr
  \Gamma^{\bar \b +} &:& - \tfrac{i}{6} \e_{\bar \b}{}^{\g_1 \cdots \g_3} F_{\g_1 \cdots \g_3 -+} + \tfrac{1}{16} (G_{\bar \b} \cont{\g} - 3 G_{\bar \b -+} ) \,, \cr
  \Gamma^{(3)} &:& 0 \,, \cr
  \Gamma^{\bar \b_1 \cdots \bar \b_4} &:& \tfrac{1}{96} \e_{\bar \b_1 \cdots \bar \b_4} (D_{-} - \tfrac{1}{2} \Omega_{-,} \cont{\g} + \tfrac{1}{2} \Omega_{-,-+} + \tfrac{i}{4} F_{-} \cont{\g_1} \cont{\g_2} ) \,, \cr
  \Gamma^{\bar \b_1 \cdots \bar \b_3 +} &:& - \tfrac{1}{24} \e_{\bar \b_1 \cdots \bar \b_3}{}^{\g} (\Omega_{-, \g +} + i F_{\g - +} \cont{\d}) + \tfrac{1}{96} G_{\bar \b_1 \cdots \bar \b_3} \,, \cr
   \Gamma^{(5)} &:& 0 .
\eea
Finally, for $M=+$ we find
\bea
  1 &:& \tfrac{i}{12} \e^{\b_1 \cdots \b_4} F_{\b_1 \cdots \b_4 +} + \tfrac{1}{8} G_{+} \cont{\g} \,, \cr
  \Gamma^{(1)} &:& 0 \,, \cr
  \Gamma^{\bar \b_1 \bar \b_2} &:& - \tfrac{1}{8} \epsilon_{\bar \b_1 \bar \b_2}{}^{\g_1 \g_2} (\Omega_{+, \g_1 \g_2} - i F_{\g_1 \g_2 +} \cont{\d}) + \tfrac{1}{16} G_{\bar \b_1 \bar \b_2 +} \,, \cr
  \Gamma^{\bar \b +} &:& 0 \,, \cr
  \Gamma^{(3)} &:& 0 \,, \cr
  \Gamma^{\bar \b_1 \cdots \bar \b_4} &:& \tfrac{1}{96} \e_{\bar \b_1 \cdots \bar \b_4} (D_{+} - \tfrac{1}{2} \Omega_{+,} \cont{\g} + \tfrac{1}{2} \Omega_{+,-+} + \tfrac{i}{4} F_{+} \cont{\g_1} \cont{\g_2}) \,, \cr
  \Gamma^{\bar \b_1 \cdots \bar \b_3 +} &:& - \tfrac{1}{24} \e_{\bar \b_1 \cdots \bar \b_3}{}^{\g} \Omega_{+, \g +} \,, \cr
   \Gamma^{(5)} &:& 0 .
\eea

\subsection{Killing spinor equations on   $e_{i_1 \cdots i_3 5}$}

We now consider the basis elements $e_{i_1 \cdots i_3 5} = \tfrac{1}{4} \Gamma^{i_1 \cdots i_3} \Gamma^{+} 1$ with $i_1, i_2, i_3 \leq 4$. For this purpose we split up $\a$ into $\rho$ and $k$, where $\rho = (i_1, \ldots, i_3)$ and $k$ is the missing fourth holomorphic coordinate (and is not an index that should be summed over). Again we will use the three-dimensional Levi-Civita symbol defined by $\e_{\rho_1 \cdots \rho_3} = \e_{k \rho_1 \cdots \rho_3}$. Substituting this spinor into the (algebraic) Killing spinor equation
({\ref{kseqnb}}) and expanding the resulting expression in the basis ({\ref{hbasisa}}), we find
 \bea
  1 &:& 0 \,, \cr
  \Gamma^{\bar k} &:& - P_- \,, \cr
  \Gamma^{\bar \tau} &:& \tfrac{1}{4} \e_{\bar \tau}{}^{\s_1 \s_2} G_{\s_1 \s_2 -} \,, \cr
  \Gamma^{+} &:& P_k - \tfrac{1}{12} \e^{\s_1 \cdots \s_3} G_{\s_1 \cdots \s_3} \,, \cr
  \Gamma^{(2)} &:& 0 \,, \cr
  \Gamma^{\bar k \bar \tau_1 \bar \tau_2} &:& - \tfrac{1}{8} \e_{\bar \tau_1 \bar \tau_2}{}^{\s} G_{\bar k \s -} \,, \cr
  \Gamma^{\bar k \bar \tau +} &:& - \tfrac{1}{2} P_{\bar \tau} - \tfrac{1}{8} \e_{\bar \tau}{}^{\s_1 \s_2} G_{\bar k \s_1 \s_2} \,, \cr
  \Gamma^{\bar \tau_1 \cdots \bar \tau_3} &:& - \tfrac{1}{48} \e_{\bar \tau_1 \cdots \bar \tau_3} ( G_{- k \bar k} - G_{-} \cont{\s} ) \,, \cr
  \Gamma^{\bar \tau_1 \bar \tau_2 +} &:& \tfrac{1}{16} \e_{\bar \tau_1 \bar \tau_2}{}^{\s_1} (G_{\s_1 k \bar k} - G_{\s_1} \cont{\s_2} - G_{\s_2 -+} ) \,, \cr
  \Gamma^{(4)} &:& 0 \,, \cr
  \Gamma^{\bar k \bar \tau_1 \cdots \bar \tau_3 +} &:& - \tfrac{1}{96} \e_{\bar \tau_1 \cdots \bar \tau_3} (G_{\bar k} \cont{\s} + G_{\bar k -+} ) .
 \eea
Next we turn to the Killing spinor equation associated
with the supercovariant derivative (\ref{kseqna}).
The components along the $k$-frame derivative of the supercovariant connection are
\bea
 1 &:& \tfrac{i}{3} \e^{\s_1 \cdots \s_3} F_{k \s_1 \cdots \s_3 -} \,, \cr
 \Gamma^{(1)} &:& 0 \,, \cr
 \Gamma^{\bar k \bar \tau} &:& \tfrac{i}{2} \e_{\bar \tau}{}^{\s_1 \s_2} F_{k \bar k \s_1 \s_2 -} + \tfrac{1}{8} G_{k \bar \tau -} \,, \cr
 \Gamma^{\bar k +} &:& - \tfrac{i}{6} \e^{\s_1 \cdots \s_3} F_{k \bar k \s_1 \cdots \s_3} +\tfrac{1}{16} (G_{k} \cont{\s} - G_{k -+}) \,, \cr
 \Gamma^{\bar \tau_1 \bar \tau_2} &:& -\tfrac{1}{4} \e_{\bar \tau_1 \bar \tau_2}{}^{\s_1}  (\Omega_{k, \s_1 -} - i F_{k \s_1 -} \cont{\s_2}) \,, \cr
 \Gamma^{\bar \tau +} &:&  - \tfrac{1}{4} \e_{\bar \tau}{}^{\s_1 \s_2} (\Omega_{k , \s_1 \s_2}  - i F_{k \s_1 \s_2} \cont{\s_3} - i F_{k \s_1 \s_2 -+}) \,, \cr
 \Gamma^{(3)} &:& 0 \,, \cr
 \Gamma^{\bar k \bar \tau_1 \cdots \bar \tau_3} &:& - \tfrac{1}{24} \e_{\bar \tau_1 \cdots \bar \tau_3} (\Omega_{k, \bar k -} - i F_{k \bar k -} \cont{\s}) \,, \cr
 \Gamma^{\bar k \bar \tau_1 \bar \tau_2 +} &:& \tfrac{1}{8} \e_{\bar \tau_1 \bar \tau_2}{}^{\s_1} (\Omega_{k, \bar k \s_1} - i F_{k \bar k \s_1} \cont{\s_2} - i F_{k \bar k \s_1 -+}) +\tfrac{1}{32} G_{k \bar \tau_1 \bar \tau_2} \,, \cr
 \Gamma^{\bar \tau_1 \cdots \bar \tau_3 +} &:&  \tfrac{1}{24} \e_{\bar \tau_1 \cdots \bar \tau_3} (D_{k} + \tfrac{1}{2} \Omega_{k, k \bar k}
 - \tfrac{1}{2} \Omega_{k,} \cont{\s} - \tfrac{1}{2} \Omega_{k,-+}
 + \tfrac{i}{4} F_{k} \cont{\s_1} \cont{\s_2} + \tfrac{i}{2} F_{k} \cont{\s} {}_{-+}) \,, \cr
   \Gamma^{(5)} &:& 0 .
\eea
Along the $\bar k$-frame derivative of the supercovariant connection we find
\bea
 1 &:& \tfrac{i}{3} \e^{\s_1 \cdots \s_3} F_{\bar k \s_1 \cdots \s_3 -} + \tfrac{1}{8} (3 G_{- k \bar k} + G_{-} \cont{\s}) \,, \cr
 \Gamma^{(1)} &:& 0 \,, \cr
 \Gamma^{\bar k \bar \tau} &:& \tfrac{1}{4} G_{\bar k \bar \tau -} \,, \cr
 \Gamma^{\bar k +} &:& \tfrac{1}{8} (G_{\bar k} \cont{\s} - G_{\bar k -+}) \,, \cr
 \Gamma^{\bar \tau_1 \bar \tau_2} &:& -\tfrac{1}{4} \e_{\bar \tau_1 \bar \tau_2}{}^{\s_1}  (\Omega_{\bar k, \s_1 -} - i F_{\bar k \s_1 -}\cont{\s_2}) + \tfrac{1}{16} G_{\bar \tau_1 \bar \tau_2 -} \,, \cr
 \Gamma^{\bar \tau +} &:&  - \tfrac{1}{4} \e_{\bar \tau}{}^{\s_1 \s_2} (\Omega_{\bar k, \s_1 \s_2} - i F_{\bar k \s_1 \s_2} \cont{\s_3} - i F_{\bar k \s_1 \s_2 -+}) + \tfrac{1}{16} ( 3 G_{\bar \tau k \bar k} + G_{\bar \tau} \cont{\s} - G_{\bar \tau -+}) \,, \cr
 \Gamma^{(3)} &:& 0 \,, \cr
 \Gamma^{\bar k \bar \tau_1 \cdots \bar \tau_3} &:& - \tfrac{1}{24} \e_{\bar \tau_1 \cdots \bar \tau_3} \Omega_{\bar k, \bar k -} \,, \cr
 \Gamma^{\bar k \bar \tau_1 \bar \tau_2 +} &:& \tfrac{1}{8} \e_{\bar \tau_1 \bar \tau_2}{}^{\s} \Omega_{\bar k, \bar k \s} + \tfrac{1}{16} G_{\bar k \bar \tau_1 \bar \tau_2} \,, \cr
 \Gamma^{\bar \tau_1 \cdots \bar \tau_3 +} &:&  \tfrac{1}{24} \e_{\bar \tau_1 \cdots \bar \tau_3} (D_{ \bar k} + \tfrac{1}{2} \Omega_{ \bar k, k \bar k}
 - \tfrac{1}{2} \Omega_{ \bar k,} \cont{\s} - \tfrac{1}{2} \Omega_{ \bar k,-+}
 + \tfrac{i}{4} F_{ \bar k} \cont{\s_1} \cont{\s_2} + \tfrac{i}{2} F_{ \bar k} \cont{\s} {}_{-+}) + \tfrac{1}{96} G_{\bar \tau_1 \cdots \bar \tau_3}   \,, \cr
   \Gamma^{(5)} &:& 0 .
\eea
The components along the $\rho$-frame derivative of the supercovariant connection are
\bea
 1 &:& \tfrac{1}{2} G_{k \rho -} \,, \cr
 \Gamma^{(1)} &:& 0 \,, \cr
 \Gamma^{\bar k \bar \tau} &:& - \tfrac{i}{2} \e_{\bar \tau}{}^{\s_1 \s_2} F_{\bar k \rho \s_1 \s_2 -} + \tfrac{1}{4} G_{\rho \bar \tau -} + \tfrac{1}{16} g_{\rho \bar \tau} (G_{- k \bar k} - G_{-} \cont{\s}) \,, \cr
 \Gamma^{\bar k +} &:& - \tfrac{1}{8} (G_{\rho k \bar k} - G_{k} \cont{\s} + G_{\rho -+}) \,, \cr
 \Gamma^{\bar \tau_1 \bar \tau_2} &:& -\tfrac{1}{4} \e_{\bar \tau_1 \bar \tau_2}{}^{\s_1}  (\Omega_{\rho, \s_1 -} + i F_{\rho \s_1 - k \bar k} - i F_{\rho \s_1 -} \cont{\s_2}) + \tfrac{1}{8} g_{\rho [ \bar \tau_1} G_{\bar \tau_2 ] k -} \,, \cr
 \Gamma^{\bar \tau +} &:&  - \tfrac{1}{4} \e_{\bar \tau}{}^{\s_1 \s_2} (\Omega_{\rho, \s_1 \s_2} + i F_{\rho \s_1 \s_2 k \bar k} - i F_{\rho \s_1 \s_2 -+}) + \tfrac{1}{4} G_{k \rho \bar \tau} + \tfrac{1}{16} g_{\rho \bar \tau} ( - G_{k} \cont{\s} + G_{k -+}) \,, \cr
 \Gamma^{(3)} &:& 0 \,, \cr
 \Gamma^{\bar k \bar \tau_1 \cdots \bar \tau_3} &:& - \tfrac{1}{24} \e_{\bar \tau_1 \cdots \bar \tau_3} (\Omega_{\rho, \bar k -} + i F_{\bar k \rho -} \cont{\s}) - \tfrac{1}{32} g_{\rho [ \bar \tau_1} G_{\bar \tau_2 \bar \tau_3 ] -} \,, \cr
 \Gamma^{\bar k \bar \tau_1 \bar \tau_2 +} &:& \tfrac{1}{8} \e_{\bar \tau_1 \bar \tau_2}{}^{\s_1} (\Omega_{\rho, \bar k \s_1} + i F_{\bar k \rho \s_1} \cont{\s_2} + i F_{\bar k \rho \s_1 -+}) + \tfrac{1}{16} G_{\rho \bar \tau_1 \bar \tau_2}
\cr
&& + \tfrac{1}{32} g_{\rho [ \bar \tau_1} ( G_{\bar \tau_2 ] k \bar k} - G_{\bar \tau_2 ]} \cont{\s_2} + G_{\bar \tau_2 ] -+} ) \,, \cr
 \Gamma^{\bar \tau_1 \cdots \bar \tau_3 +} &:&  \tfrac{1}{24} \e_{\bar \tau_1 \cdots \bar \tau_3} (D_{ \rho} + \tfrac{1}{2} \Omega_{ \rho, k \bar k}
 - \tfrac{1}{2} \Omega_{ \rho,} \cont{\s} - \tfrac{1}{2} \Omega_{ \rho,-+}
 - \tfrac{i}{2} F_{ \rho k \bar k} \cont{\s} - \tfrac{i}{2} F_{ \rho k \bar k -+}
\cr
&& + \tfrac{i}{4} F_{ \rho} \cont{\s_1} \cont{\s_2} + \tfrac{i}{2} F_{ \rho} \cont{\s} {}_{-+})
 - \tfrac{1}{32} g_{\rho [\bar \tau_1} G_{\bar \tau_2 \bar \tau_3 ] k}   \,, \cr
   \Gamma^{(5)} &:& 0 .
\eea
Similarly, the components along the $\bar \rho$-frame derivative are
\bea
 1 &:& \tfrac{i}{3} \e^{\s_1 \cdots \s_3} F_{\bar \rho \s_1 \cdots \s_3 -} + \tfrac{1}{4} G_{k \bar \rho -} \,, \cr
 \Gamma^{(1)} &:& 0 \,, \cr
 \Gamma^{\bar k \bar \tau} &:& - \tfrac{i}{2} \e_{\bar \tau}{}^{\s_1 \s_2} F_{\bar k \bar \rho \s_1 \s_2 -} + \tfrac{1}{8} G_{\bar \rho \bar \tau -} \,, \cr
 \Gamma^{\bar k +} &:& \tfrac{i}{6} \e^{\s_1 \cdots \s_3} F_{\bar k \bar \rho \s_1 \cdots \s_3} - \tfrac{1}{16} (G_{\bar \rho k \bar k} - G_{\bar k} \cont{\s} + G_{\bar \rho -+}) \,, \cr
 \Gamma^{\bar \tau_1 \bar \tau_2} &:& -\tfrac{1}{4} \e_{\bar \tau_1 \bar \tau_2}{}^{\s_1}  (\Omega_{\bar \rho, \s_1 -} + i F_{\bar \rho \s_1 - k \bar k} - i F_{\bar \rho \s_1 -} \cont{\s_2}) \,, \cr
 \Gamma^{\bar \tau +} &:&  - \tfrac{1}{4} \e_{\bar \tau}{}^{\s_1 \s_2} (\Omega_{\bar \rho, \s_1 \s_2} + i F_{\bar \rho \s_1 \s_2 k \bar k} - i F_{\bar \rho \s_1 \s_2} \cont{\s_3} - i F_{\bar \rho \s_1 \s_2 -+}) + \tfrac{1}{8} G_{k \bar \rho \bar \tau} \,, \cr
 \Gamma^{(3)} &:& 0 \,, \cr
 \Gamma^{\bar k \bar \tau_1 \cdots \bar \tau_3} &:& - \tfrac{1}{24} \e_{\bar \tau_1 \cdots \bar \tau_3} (\Omega_{\bar \rho, \bar k -} + i F_{\bar k \bar \rho -} \cont{\s}) \,, \cr
 \Gamma^{\bar k \bar \tau_1 \bar \tau_2 +} &:& \tfrac{1}{8} \e_{\bar \tau_1 \bar \tau_2}{}^{\s_1} (\Omega_{\bar \rho, \bar k \s_1} + i F_{\bar k \bar \rho \s_1} \cont{\s_2} + i F_{\bar k \bar \rho \s_1 -+}) + \tfrac{1}{32} G_{\bar \rho \bar \tau_1 \bar \tau_2} \,, \cr
 \Gamma^{\bar \tau_1 \cdots \bar \tau_3 +} &:&  \tfrac{1}{24} \e_{\bar \tau_1 \cdots \bar \tau_3} (D_{\bar \rho} + \tfrac{1}{2} \Omega_{\bar \rho, k \bar k}
 - \tfrac{1}{2} \Omega_{\bar  \rho,} \cont{\s} - \tfrac{1}{2} \Omega_{\bar  \rho,-+}
\cr
&&- \tfrac{i}{2} F_{\bar  \rho k \bar k} \cont{\s} - \tfrac{i}{2} F_{\bar  \rho k \bar k -+}
 + \tfrac{i}{4} F_{\bar  \rho} \cont{\s_1} \cont{\s_2} + \tfrac{i}{2} F_{\bar  \rho} \cont{\s} {}_{-+}) \,, \cr
   \Gamma^{(5)} &:& 0 .
\eea
The supercovariant derivative with $M=-$ gives
\bea
 1 &:& 0 \,, \cr
 \Gamma^{(1)} &:& 0 \,, \cr
 \Gamma^{\bar k \bar \tau} &:& 0 \,, \cr
 \Gamma^{\bar k +} &:& - \tfrac{i}{6} \e^{\s_1 \cdots \s_3} F_{\bar k \s_1 \cdots \s_3 -} - \tfrac{1}{16} (G_{- k \bar k} - G_{-} \cont{\s}) \,, \cr
 \Gamma^{\bar \tau_1 \bar \tau_2} &:& - \tfrac{1}{4} \e_{\bar \tau_1 \bar \tau_2}{}^{\s}  \Omega_{-, \s -} \,, \cr
 \Gamma^{\bar \tau +} &:&  - \tfrac{1}{4} \e_{\bar \tau}{}^{\s_1 \s_2} (\Omega_{-, \s_1 \s_2} + i F_{- \s_1 \s_2 k \bar k} - i F_{-\s_1 \s_2} \cont{\s_3} ) - \tfrac{1}{8} G_{k \bar \tau -} \,, \cr
 \Gamma^{(3)} &:& 0 \,, \cr
 \Gamma^{\bar k \bar \tau_1 \cdots \bar \tau_3} &:& - \tfrac{1}{24} \e_{\bar \tau_1 \cdots \bar \tau_3} \Omega_{-, \bar k -} \,, \cr
 \Gamma^{\bar k \bar \tau_1 \bar \tau_2 +} &:& \tfrac{1}{8} \e_{\bar \tau_1 \bar \tau_2}{}^{\s_1} (\Omega_{-, \bar k \s_1}
 - i F_{\bar k \s_1 -} \cont{\s_2}) + \tfrac{1}{32} G_{- \bar \tau_1 \bar \tau_2} \,, \cr
 \Gamma^{\bar \tau_1 \cdots \bar \tau_3 +} &:&  \tfrac{1}{24} \e_{\bar \tau_1 \cdots \bar \tau_3} (D_{-}
  + \tfrac{1}{2} \Omega_{-, k \bar k}
 - \tfrac{1}{2} \Omega_{-,} \cont{\s} - \tfrac{1}{2} \Omega_{-,-+} - \tfrac{i}{2} F_{-} \cont{\s} {}_{k \bar k}
 + \tfrac{i}{4} F_{-} \cont{\s_1} \cont{\s_2})  \,, \cr
   \Gamma^{(5)} &:& 0 .
 \eea
Finally, for $M=+$ we find
\bea
 1 &:& \tfrac{i}{3} \e^{\s_1 \cdots \s_3} F_{\s_1 \cdots \s_3 -+} - \tfrac{1}{8} (G_{k} \cont{\s} + 3 G_{k -+}) \,, \cr
 \Gamma^{(1)} &:& 0 \,, \cr
 \Gamma^{\bar k \bar \tau} &:& \tfrac{i}{2} \e_{\bar \tau}{}^{\s_1 \s_2} F_{\bar k \s_1 \s_2 -+}
 - \tfrac{1}{16} (G_{\bar \tau k \bar k} - G_{\bar \tau} \cont{\s} - 3 G_{\bar \tau -+}) \,, \cr
 \Gamma^{\bar k +} &:& - \tfrac{i}{6} \e^{\s_1 \cdots \s_3} F_{\bar k \s_1 \cdots \s_3 +}
 - \tfrac{1}{8} (G_{+ k \bar k} - G_{+} \cont{\s} ) \,, \cr
 \Gamma^{\bar \tau_1 \bar \tau_2} &:& -\tfrac{1}{4} \e_{\bar \tau_1 \bar \tau_2}{}^{\s_1}  (\Omega_{+, \s_1 -}
 + i F_{ \s_1 k \bar k -+} - i F_{\s_1 -+} \cont{\s_2}) - \tfrac{1}{16} G_{k \bar \tau_1 \bar \tau_2} \,, \cr
 \Gamma^{\bar \tau +} &:&  - \tfrac{1}{4} \e_{\bar \tau}{}^{\s_1 \s_2} (\Omega_{+, \s_1 \s_2} + i F_{+ \s_1 \s_2 k \bar k}
 - i F_{+ \s_1 \s_2} \cont{\s_3}) - \tfrac{1}{4} G_{k \bar \tau +} \,, \cr
 \Gamma^{(3)} &:& 0 \,, \cr
 \Gamma^{\bar k \bar \tau_1 \cdots \bar \tau_3} &:& - \tfrac{1}{24} \e_{\bar \tau_1 \cdots \bar \tau_3} (\Omega_{+, \bar k -}
 - i F_{\bar k -+} \cont{\s}) + \tfrac{1}{96} G_{\bar \tau_1 \cdots \bar \tau_3} \,, \cr
 \Gamma^{\bar k \bar \tau_1 \bar \tau_2 +} &:& \tfrac{1}{8} \e_{\bar \tau_1 \bar \tau_2}{}^{\s_1} (\Omega_{+, \bar k \s_1}
 - i F_{\bar k \s_1 +} \cont{\s_2} ) + \tfrac{1}{16} G_{\bar \tau_1 \bar \tau_2 +} \,, \cr
 \Gamma^{\bar \tau_1 \cdots \bar \tau_3 +} &:&  \tfrac{1}{24} \e_{\bar \tau_1 \cdots \bar \tau_3} (D_{+} + \tfrac{1}{2} \Omega_{+, k \bar k}
 - \tfrac{1}{2} \Omega_{+,} \cont{\s} - \tfrac{1}{2} \Omega_{+,-+}
 - \tfrac{i}{2} F_{ k \bar k +} \cont{\s}
 + \tfrac{i}{4} F_{ +} \cont{\s_1} \cont{\s_2}) \,, \cr
   \Gamma^{(5)} &:& 0 .
\eea

\newsection{Integrability conditions}

\subsection{Integrability conditions on $1$}

The expressions for the integrability condition $\mathcal{I}$ on the
basis element $1$ read
 \bea
  1 &:& \LG \cont{\g} + \LG_{-+} + 12 \BG \cont{\g_1} \cont{\g_2} +
  24 \BG \cont{\g} {}_{-+} \,, \cr
  \Gamma^{\bar \b_1 \bar \b_2} &:& \tfrac{1}{2} \LG_{\bar \b_1 \bar
  \b_2} + 12 \BG_{\bar \b_1 \bar \b_2} \cont{\g} + 12 \BG_{\bar \b_1 \bar
  \b_2 -+} - \tfrac{1}{2} \e_{\bar \b_1 \bar
  \b_2}{}^{\g_1 \g_2} \BP_{\g_1 \g_2} \,, \cr
  \G^{\bar \b +} &:& \LG_{\bar \b+} + 24 \BG_{\bar \b} \cont{\g}
  {}_+
  \,, \cr
  \G^{\bar \b_1 \cdots \bar \b_4} &:& \BG_{\bar \b_1 \cdots \bar
  \b_4} + \tfrac{1}{96} \e_{\bar \b_1 \cdots \bar \b_4} (\LP - 2 \BP
  \cont{\g} + 2 \BP_{-+} ) \,, \cr
  \G^{\bar \b_1 \cdots \bar \b_3 +} &:& 4 \BG_{\bar \b_1 \cdots \bar
  \b_3 +} - \tfrac{1}{6} \e_{\bar \b_1 \cdots \bar \b_3}{}^{\g}
  \BP_{\g +} \,.
 \eea
The integrability conditions $\mathcal{I}_{\a}$ are
 \bea
 \Gamma^{\bar \b} &:& \tfrac{1}{2} \E_{\a \bar \b} - 6 i \LF_{\a
 \bar \b} \cont{\g} - 6 i \LF_{\a \bar \b -+} + 8 \e_{\bar \b}{}^{\g_1 \cdots \g_3} \BG_{\a \g_1 \cdots \g_3} \,, \cr
 \Gamma^+ &:& \tfrac{1}{2} \E_{\a +} - 6 i \LF_{\a +} \cont{\g} \,,
 \cr
 \Gamma^{\bar \b_1 \cdots \bar \b_3} &:& - i \LF_{\a \bar \b_1 \cdots \bar
 \b_3} + \tfrac{1}{12} \e_{\bar \b_1 \cdots \bar \b_3}{}^\g (\LG_{
 \a \g} + 24 \BG_{\a \g_1} \cont{\g_2} - 24 \BG_{\a \g_1 -+})  \,, \cr
 \G^{\bar \b_1 \bar \b_2 +} &:& - 3 i \LF_{\a \bar \b_1 \bar \b_2 +}
 - 6 \e_{\bar \b_1 \bar \b_2}{}^{\g_1 \g_2} \BG_{\a \g_1 \g_2 +} \,, \cr
 \G^{\bar \b_1 \cdots \bar \b_4 +} &:& - \tfrac{1}{96} \e_{\bar \b_1 \cdots \bar
 \b_4} (\LG_{\a +} + 24 \BG_{\a +} \cont{\g} ) \,.
 \eea
The integrability conditions $\mathcal{I}_{\bar \a}$ read
 \bea
  \Gamma^{\bar \b} &:& \tfrac{1}{2} \E_{\bar \a \bar \b} - 6 i
  (\LF_{\bar \a \bar \b} \cont{\g} + \LF_{\bar \a \bar \b -+} ) - 8
  \e_{\bar \b}{}^{\g_1 \cdots \g_3} \BG_{\bar \a \g_1 \cdots \g_3} + \cr
  && +
  24 \e_{\bar \a \bar \b}{}^{\g_1 \g_2} (\BG_{\g_1 \g_2} \cont{\g_3}
  - \BG_{\g_1 \g_2 -+}) \,, \cr
  \G^+ &:& \tfrac{1}{2} \E_{\bar \a +} - 6 i \LF_{\bar \a +} \cont{\g}
   - 16 \e_{\bar \a}{}^{\g_1 \cdots \g_3} \BG_{+ \g_1 \cdots \g_3}
  \,, \cr
  \G^{\bar \b_1 \cdots \bar \b_3} &:& - i \LF_{\bar \a \bar \b_1 \cdots \bar
  \b_3} + \tfrac{1}{12} \e_{\bar \b_1 \cdots \bar \b_3}{}^{\g_1}
  (\LG_{\bar \a \g_1} - 24 \BG_{\bar \a \g_1} \cont{\g_2} + 24
  \BG_{\bar \a \g_1 -+}) + \cr
  && +  \e_{\bar \a \bar \b_1 \cdots \bar
  \b_3} (\BG \cont{\g_1} \cont{\g_2} -2 \BG \cont{\g_1} {}_{-+}) \,,
  \cr
  \G^{\bar \b_1 \bar \b_2 +} &:& - 3 i \LF_{\bar \a \bar \b_1 \bar
  \b_2 +} + 12 \e_{\bar \a \bar \b_1 \bar \b_2}{}^{\g_1} \BG_{\g_1
  +} \cont{\g_2} + 6 \e_{\bar \b_1 \bar \b_2}{}^{\g_1 \g_2} \BG_{\bar
  \a \g_1 \g_2 +} \,, \cr
  \G^{\bar \b_1 \cdots \bar \b_4 +} &:& - \tfrac{1}{96} \e_{\bar \b_1 \cdots \bar
  \b_4} (\LG_{\bar \a +} - 24 \BG_{\bar \a +} \cont{\g}) \,.
 \eea
Similarly, $\mathcal{I}_A$ with $A=-$ is given by
 \bea
 \Gamma^{\bar \b} &:& \tfrac{1}{2} \E_{-  \bar \b} - 6 i \LF_{-
 \bar \b} \cont{\g} + 8 \e_{\bar \b}{}^{\g_1 \cdots \g_3} \BG_{\g_1 \cdots \g_3 -} \,, \cr
 \Gamma^+ &:& \tfrac{1}{2} \E_{-  +} - 6 i \LF_{-  +} \cont{\g} + 4 \e^{\g_1 \cdots \g_4} \BG_{\g_1 \cdots \g_4} \,,
 \cr
 \Gamma^{\bar \b_1 \cdots \bar \b_3} &:& - i \LF_{-  \bar \b_1 \cdots \bar
 \b_3} - \tfrac{1}{12} \e_{\bar \b_1 \cdots \bar \b_3}{}^\g (\LG_{
 \g -} - 24 \BG_{\g_1} \cont{\g_2} {}_{-} ) \,, \cr
 \G^{\bar \b_1 \bar \b_2 +} &:& - 3 i \LF_{-  \bar \b_1 \bar \b_2 +}
 + 6 \e_{\bar \b_1 \bar \b_2}{}^{\g_1 \g_2} \BG_{\g_1 \g_2} \cont{\g_3} \,, \cr
 \G^{\bar \b_1 \cdots \bar \b_4 +} &:& - \tfrac{1}{96} \e_{\bar \b_1 \cdots \bar
 \b_4} (\LG_{-  +} -12 \BG \cont{\g_1} \cont{\g_2}) \,.
 \eea
Finally, the expressions for $\mathcal{I}_+$ read
 \bea
  \Gamma^{\bar \b} &:& \tfrac{1}{2} \E_{+ \bar \b} - 6 i
  \LF_{+ \bar \b} \cont{\g}  + 8
  \e_{\bar \b}{}^{\g_1 \cdots \g_3} \BG_{+ \g_1 \cdots \g_3} \,, \cr
  \G^+ &:& \tfrac{1}{2} \E_{+ +} \,, \cr
  \G^{\bar \b_1 \cdots \bar \b_3} &:& - i \LF_{+ \bar \b_1 \cdots \bar
  \b_3} + \tfrac{1}{12} \e_{\bar \b_1 \cdots \bar \b_3}{}^{\g_1}
  (\LG_{+ \g_1} + 24 \BG_{+ \g_1} \cont{\g_2}) \,,
  \cr
  \G^{\bar \b_1 \bar \b_2 +} &:& 0 \,, \cr
  \G^{\bar \b_1 \cdots \bar \b_4 +} &:& 0 \,.
 \eea

\subsection{Integrability conditions on $e_{ij}$}

The expressions for the integrability condition $\mathcal{I}$ on the
basis element $e_{ij}$ read
 \bea
  1 &:& - \e^{b_1 b_2} ( \LG_{b_1 b_2} + 24 \BG_{b_1 b_2} \cont{r} + 24 \BG_{b_1 b_2 -+} ) + 2 \e^{r_1 r_2} \BP_{r_1 r_2} \,, \cr
  \G^{\bar b_1 \bar b_2} &:& \tfrac{1}{4} \e_{\bar b_1 \bar b_2} ( - \LG \cont{c} + \LG \cont{r} + \LG_{-+} + 12 \BG \cont{c_1} \cont{c_2} - 24 \BG \cont{c} \cont{r} - 24 \BG \cont{c} {}_{-+} + \cr
 && + 12 \BG \cont{r_1} \cont{r_2} + 24 \BG \cont{r} {}_{-+} ) \,, \cr
  \G^{\bar b \bar q} &:& - \e_{\bar b}{}^{c_1} ( \LG_{c_1 \bar q} - 24 \BG_{c_1} \cont{c_2} {}_{\bar q} + 24 \BG_{c_1 \bar q} \cont{r} + 24 \BG_{c_1 \bar q -+} ) + 2 \e_{\bar q}{}^r \BP_{\bar b r} \,, \cr
  \G^{\bar b +} &:& - \e_{\bar b}{}^{c_1} ( \LG_{c_1 +} - 24 \BG_{c_1} \cont{c_2} {}_{+} + 24 \BG_{c_1 +} \cont{r} ) \,, \cr
 \G^{\bar q_1 \bar q_2} &:& - 12 \e^{b_1 b_2} \BG_{b_1 b_2 \bar q_1 \bar q_2} - \tfrac{1}{4} \e_{\bar q_1 \bar q_2} ( \LP + 2 \BP \cont{c} - 2 \BP \cont{r} + 2 \BP_{-+} ) \,, \cr
 \G^{\bar q +} &:& - 24 \e^{b_1 b_2} \BG_{b_1 b_2 \bar q +} + 2 \e_{\bar q}{}^r \BP_{r +} \,, \cr
 \G^{\bar b_1 \bar b_2 \bar q_1 \bar q_1} &:& \tfrac{1}{8} \e_{\bar b_1 \bar b_2} ( \LG_{\bar q_1 \bar q_2} - 24 \BG_{\bar q_1 \bar q_2} \cont{c} + 24 \BG_{\bar q_1 \bar q_2 -+} ) - \tfrac{1}{4} \e_{\bar q_1 \bar q_2} \BP_{\bar b_1 \bar b_2} \,, \cr
 \G^{\bar b_1 \bar b_2 \bar q +} &:& \tfrac{1}{4} \e_{\bar b_1 \bar b_2} ( \LG_{\bar q +} - 24 \BG_{\bar q +} \cont{c} + 24 \BG_{\bar q +} \cont{r} ) \,, \cr
 \G^{\bar b \bar q_1 \bar q_2 +} &:& - 12 \e_{\bar b}{}^c \BG_{c \bar q_1 \bar q_2 +} - \tfrac{1}{2} \e_{\bar q_1 \bar q_2} \BP_{\bar b +} \,.
 \eea
The integrability conditions $\mathcal{I}_{a}$ are given by
 \bea
  \G^{\bar b} &:& - \tfrac{1}{2} \e_{\bar b}{}^c ( \E_{ac} - 12 i \LF_{ac} \cont{r} - 12 i \LF_{ac-+}) - 24 \e^{r_1 r_2} \BG_{a \bar b r_1 r_2} + \cr
 && + 24 g_{a \bar b} \e^{r_1 r_2} ( \BG_{r_1 r_2} \cont{c} + \BG_{r_1 r_2 -+})  \,, \cr
  \G^{\bar q} &:& - \e_{\bar q}{}^{r_1} (\LG_{a r_1} + 24 \BG_{a r_1} \cont{c} - 24 \BG_{a r_1} \cont{r_2} + 24 \BG_{a r_1 -+} ) \,, \cr
  \G^{+} &:& - 24 \e^{r_1 r_2} \BG_{a r_1 r_2 +} \,, \cr
  \G^{\bar b_1 \bar b_2 \bar q} &:& \tfrac{1}{8} \e_{\bar b_1 \bar b_2} ( \E_{a \bar q} + 12 i \LF_{a \bar q} \cont{c} - 12 i \LF_{a \bar q} \cont{r} - 12 i \LF_{a \bar q -+} ) + \cr
 &&  + 24 \e_{\bar q}{}^{r_1} g_{a [ \bar b_1} ( - 2 \BG_{\bar b_2 ] r_1} \cont{c} - \BG_{\bar b_2 ] r_1} \cont{r_2} + \BG_{\bar b_2 ] r_1 -+}) \,, \cr
  \G^{\bar b_1 \bar b_2 + } &:& \tfrac{1}{8} \e_{\bar b_1 \bar b_2} ( \E_{a +} + 12 i \LF_{a +} \cont{c} - 12 i \LF_{a +} \cont{r} ) + 24 g_{a [ \bar b_2} \e^{r_1 r_2} \BG_{\bar b_2 ] r_1 r_2 +} \,, \cr
  \G^{\bar b \bar q_1 \bar q_2} &:& 6 \e_{\bar q_1 \bar q_2} ( \BG_{a \bar b} \cont{c} - \BG_{a \bar b} \cont{r} + \BG_{a \bar b -+} ) - 3 g_{a \bar b} \e_{\bar q_1 \bar q_2} ( \BG \cont{c_1} \cont{c_2} - 2 \BG \cont{c} \cont{r} + \cr
  \G^{\bar b \bar q +} &:& 6 i \e_{\bar b}{}^c \LF_{a c \bar q +} + 24 g_{a \bar b} \e_{\bar q}{}^{r_1} (\BG \cont{c} {}_{r_1 +}  - \BG_{r_1} \cont{r_2} {}_+ ) - 24 \e_{\bar q}{}^r \BG_{a \bar b r +} \,, \cr
 && + 2 \BG \cont{c} {}_{-+} + \BG \cont{r_1} \cont{r_2} - 2 \BG \cont{r} {}_{-+} ) \,, \cr
  \G^{\bar q_1 \bar q_2 +} &:& \tfrac{1}{4} \e_{\bar q_1 \bar q_2} ( \LG_{a +} + 24 \BG_{a +} \cont{c} - 24 \BG_{a+} \cont{r} ) \,, \cr
  \G^{\bar b_1 \bar b_2 \bar q_1 \bar q_2 +} &:& - \tfrac{3}{4} i \e_{\bar b_1 \bar b_2} \LF_{a \bar q_1 \bar q_2 +} + 24 \e_{\bar q_1 \bar q_2} g_{a [ \bar b_1} \BG_{\bar b_2 ] +} \cont{r} \,.
 \eea
Similarly, $\mathcal{I}_{\bar a}$ reads
 \bea
  \G^{\bar b} &:& - \tfrac{1}{2} \e_{\bar b}{}^{c_1} ( \E_{\bar a c_1} + 12 i \LF_{\bar a c_1} \cont{c_2} - 12 i \LF_{\bar a c_1} \cont{r} - 12 i \LF_{\bar a c_1 -+} ) + 24 \e^{r_1 r_2} \BG_{\bar a \bar b r_1 r_2}  \,, \cr
  \G^{\bar q} &:& 6 i \e^{b_1 b_2} \LF_{\bar a b_1 b_2 \bar q} - \e_{\bar q}{}^{r_1} ( \LG_{\bar a r_1} - 24 \BG_{\bar a} \cont{c} {}_{r_1} + 24 \BG_{\bar a r_1} \cont{r_2} - 24 \BG_{\bar a r_1 -+} ) \,, \cr
  \G^{+} &:& 6 i \e^{b_1 b_2} \LF_{\bar a b_1 b_2 +} + 24 \e^{r_1 r_2} \BG_{\bar a r_1 r_2 +} \,, \cr
  \G^{\bar b_1 \bar b_2 \bar q} &:& \tfrac{1}{8} \e_{\bar b_1 \bar b_2} ( \E_{\bar a \bar q} + 12 i \LF_{\bar a \bar q} \cont{c} - 12 i \LF_{\bar a \bar q} \cont{r} - 12 i \LF_{\bar a \bar q -+} ) \,, \cr
  \G^{\bar b_1 \bar b_2 + } &:& \tfrac{1}{8} \e_{\bar b_1 \bar b_2} ( \E_{\bar a +} + 12 i \LF_{\bar a +} \cont{c} - 12 i \LF_{\bar a +} \cont{r} ) \,, \cr
  \G^{\bar b \bar q_1 \bar q_2} &:& 3 i \e_{\bar b}{}^c \LF_{\bar a c \bar q_1 \bar q_2} + \tfrac{1}{4} \e_{\bar q_1 \bar q_2} ( \LG_{\bar a \bar b} + 24 \BG_{\bar a \bar b} \cont{r} - 24 \BG_{\bar a \bar b -+} ) \,, \cr
  \G^{\bar b \bar q +} &:& 6 i \e_{\bar b}{}^c \LF_{\bar a c \bar q +} + 24 \e_{\bar q}{}^r \BG_{\bar a \bar b r +} \,, \cr
  \G^{\bar q_1 \bar q_2 +} &:& \tfrac{1}{4} \e_{\bar q_1 \bar q_2} (\LG_{\bar a +} - 24 \BG_{\bar a + } \cont{c} + 24 \BG_{\bar a +} \cont{r} ) \,, \cr
  \G^{\bar b_1 \bar b_2 \bar q_1 \bar q_2 +} &:& - \tfrac{3}{4} i \e_{\bar b_1 \bar b_2} \LF_{\bar a \bar q_1 \bar q_2 +} \,.
 \eea
The integrability condition $\mathcal{I}_M$ with $M = p$ is given by
 \bea
  \G^{\bar b} &:& - \tfrac{1}{2} \e_{\bar b}{}^{c_1} (\E_{c_1 p} - 12 i \LF_{c_1} \cont{c_2} {}_{p} + 12 i \LF_{c_1 p} \cont{r} + 12 i \LF_{c_1 p -+} ) \,, \cr
  \G^{\bar q} &:&  6 i \e^{b_1 b_2} \LF_{b_1 b_2 p \bar q} - \e_{\bar q}{}^r ( \LG_{pr} - 24 \BG_{pr} \cont{c} - 24 \BG_{pr -+} ) \,, \cr
  \G^{+} &:& 6 i \e^{b_1 b_2} \LF_{b_1 b_2 p +} \,, \cr
  \G^{\bar b_1 \bar b_2 \bar q} &:& \tfrac{1}{8} \e_{\bar b_1 \bar b_2} ( \E_{p \bar q} + 12 i \LF_{p \bar q} \cont{c} - 12 i \LF_{p \bar q} \cont{r} - 12 i \LF_{p \bar q -+} ) + 12 \e_{\bar q}{}^r \BG_{\bar b_1 \bar b_2 p r} \,, \cr
  \G^{\bar b_1 \bar b_2 + } &:& \tfrac{1}{8} \e_{\bar b_1 \bar b_2} (\E_{p+} + 12 i \LF_{p+} \cont{c} - 12 i \LF_{p+} \cont{r} ) \,, \cr
  \G^{\bar b \bar q_1 \bar q_2} &:& - 3 i \e_{\bar b}{}^c \LF_{c p \bar q_1 \bar q_2} - \tfrac{1}{4} \e_{\bar q_1 \bar q_2} ( \LG_{\bar b p} - 24 \BG_{\bar b p} \cont{c} + 24 \BG_{\bar b p} \cont{r} - 24 \BG_{\bar b p -+} ) \,, \cr
  \G^{\bar b \bar q +} &:& - 6 i \e_{\bar b}{}^c \LF_{c p \bar q +} - 24 \e_{\bar q}{}^r \BG_{\bar b p r +} \,, \cr
  \G^{\bar q_1 \bar q_2 +} &:& \tfrac{1}{4} \e_{\bar q_1 \bar q_2} ( \LG_{p+} - 24 \BG_{p+} \cont{c} + 24 \BG_{p+} \cont{r} )   \,, \cr
  \G^{\bar b_1 \bar b_2 \bar q_1 \bar q_2 +} &:& - \tfrac{3}{4} i \e_{\bar b_1 \bar b_2} \LF_{ p \bar q_1 \bar q_2 +}  - 3 \e_{\bar q_1 \bar q_2} \BG_{\bar b_1 \bar b_2 p +} \,.
 \eea
Furthermore, $\mathcal{I}_M$ with $M = \bar p$ is given by the expressions
 \bea
  \G^{\bar b} &:& - \tfrac{1}{2} \e_{\bar b}{}^{c_1} ( \E_{c_1 \bar p} - 12 i \LF_{c_1} \cont{c_2} {}_{\bar p} + 12 i \LF_{c_1 \bar p} \cont{r} + 12 i \LF_{c_1 \bar p -+} ) + \cr
 && - 48 \e_{\bar p}{}^{r_1} ( \BG_{\bar b r_1} \cont{c} + 2 \BG_{\bar b r_1} \cont{r_2} + \BG_{\bar c r_1 -+} )   \,, \cr
  \G^{\bar q} &:&  - 24 \e_{\bar q}{}^{r_1} ( \BG_{\bar p r_1} \cont{c} - \BG_{\bar p r_1} \cont{r_2} + \BG_{\bar p r_1 -+} ) + \cr
 && - 12 \e_{\bar p \bar q} ( \BG \cont{c_1} \cont{c_2} - 2 \BG \cont{c} \cont{r} + 2 \BG \cont{c} {}_{-+} + \BG \cont{r_1} \cont{r_2} -2 \BG \cont{r} {}_{-+} ) \,, \cr
  \G^{+} &:& 6 i \e^{b_1 b_2} \LF_{b_1 b_2 \bar p +} + 48 \e_{\bar p}{}^r \BG_{r +} \cont{c} \,, \cr
  \G^{\bar b_1 \bar b_2 \bar q} &:& \tfrac{1}{8} \e_{\bar b_1 \bar b_2} ( \E_{\bar p \bar q} + 12 i \LF_{\bar p \bar q} \cont{c} - 12 i \LF_{\bar p \bar q} {}_{-+} ) - 12 \e_{\bar q}{}^r \BG_{\bar b_1 \bar b_2 \bar p r} + \cr
 && + 12 \e_{\bar p \bar q} ( \BG_{\bar b_1 \bar b_2} \cont{r} - \BG_{\bar b_1 \bar b_2 -+} ) \,, \cr
  \G^{\bar b_1 \bar b_2 + } &:& \tfrac{1}{8} \e_{\bar b_1 \bar b_2} ( \E_{\bar p +} + 12 i \LF_{\bar p +} \cont{c} - 12 i \LF_{\bar p +} \cont{r} ) + 24 \e_{\bar p}{}^r \BG_{\bar b_1 \bar b_2 r +} \,, \cr
  \G^{\bar b \bar q_1 \bar q_2} &:& - \tfrac{1}{4} \e_{\bar q_1 \bar q_2} ( \LG_{\bar b \bar p} + 24 \BG_{\bar b \bar p} \cont{c} - 24 \BG_{\bar b \bar p} \cont{r} + 24 \BG_{\bar b \bar p -+} ) \,, \cr
  \G^{\bar b \bar q +} &:& - 6 i \e_{\bar b}{}^c \LF_{c \bar p \bar q +} + 24 \e_{\bar p \bar q} (\BG_{\bar b +} \cont{c} - \BG_{\bar b +} \cont{r} ) + 24 \e_{\bar q}{}^r \BG_{\bar b \bar p r +} \,, \cr
  \G^{\bar q_1 \bar q_2 +} &:&  \tfrac{1}{4} \e_{\bar q_1 \bar q_2} ( \LG_{\bar p +} + 24 \BG_{\bar p +} \cont{c} - 24 \BG_{\bar p +} \cont{r} ) \,, \cr
  \G^{\bar b_1 \bar b_2 \bar q_1 \bar q_2 +} &:& 3 \e_{\bar q_1 \bar q_2} \BG_{\bar b_1 \bar b_2 \bar p +} \,.
 \eea
The integrability conditions $\mathcal{I}_-$ read
 \bea
  \G^{\bar b} &:& - \tfrac{1}{2} \e_{\bar b}{}^{c_1} ( \E_{c_1 -} - 12 i \LF_{c_1} \cont{c_2} {}_- + 12 i \LF_{c_1 -} \cont{r} ) + 24 \e^{r_1 r_2} \BG_{\bar b r_1 r_2 -} \,, \cr
  \G^{\bar q} &:&  - 6i \e^{b_1 b_2} \LF_{b_1 b_2 \bar q -} + \e_{\bar q}{}^r LG_{r-} + 24 \e_{\bar q}{}^{r_1} (\BG \cont{c} {}_{r_1 -} - BG_{r_1} \cont{r_2} {}_- )  \,, \cr
  \G^{+} &:& 6 i \e^{b_1 b_2} \LF_{b_1 b_2 - + } + 24 \e^{r_1 r_2} \BG_{r_1 r_2} \cont{c} \,, \cr
  \G^{\bar b_1 \bar b_2 \bar q} &:& \tfrac{1}{8} \e_{\bar b_1 \bar b_2} ( \E_{\bar q -} - 12 i \LF_{\bar q -} \cont{c} + 12 i \LF_{\bar q -} \cont{r} ) + 12 \e_{\bar q}{}^r \BG_{\bar b_1 \bar b_2 r -} \,, \cr
  \G^{\bar b_1 \bar b_2 + } &:& \tfrac{1}{8} \e_{\bar b_1 \bar b_2} ( \E_{-+} + 12 i \LF \cont{c} {}_{-+} - 12 i \LF \cont{r} {}_{-+} ) + 12 \e^{r_1 r_2} \BG_{\bar b_1 \bar b_2 r_1 r_2} \,, \cr
  \G^{\bar b \bar q_1 \bar q_2} &:& - 3 i \e_{\bar b}{}^c \LG_{c \bar q_1 \bar q_2 -} - \tfrac{1}{4} \e_{\bar q_1 \bar q_2} (\LG_{\bar b -} + 24 \BG_{\bar b -} \cont{c} - 24 \BG_{\bar b -} \cont{r} ) \,, \cr
  \G^{\bar b \bar q +} &:& 6 i \e_{\bar b}{}^c \LF_{c \bar q -+} + 12 \e_{\bar q}{}^{r_1} (\BG_{\bar b r_1} \cont{c} - \BG_{\bar b r_1} \cont{r_2} ) \,, \cr
  \G^{\bar q_1 \bar q_2 +} &:&  \tfrac{1}{4} \e_{\bar q_1 \bar q_2} ( \LG_{-+} - 12 \BG \cont{c_1} \cont{c_2} + 24 \BG \cont{c} \cont{r}  - 12 \BG \cont{r_1} \cont{r_2} ) \,, \cr
  \G^{\bar b_1 \bar b_2 \bar q_1 \bar q_2 +} &:& - \tfrac{3}{4} i \e_{\bar b_1 \bar b_2} \LF_{\bar q_1 \bar q_2 -+} + 3 \e_{\bar q_1 \bar q_2} \BG_{\bar b_1 \bar b_2} \cont{r} \,.
 \eea
Finally, for $\mathcal{I}_+$ we find
 \bea
  \G^{\bar b} &:& - \tfrac{1}{2} \e_{\bar b}{}^{c_1} ( \E_{c_1 +} - 12 i \LF_{c_1} \cont{c_2} {}_{+} + 12 i \LF_{c_1 +} \cont{r} ) -24 \e^{r_1 r_2} \BG_{\bar b r_1 r_2 +}  \,, \cr
  \G^{\bar q} &:& - 6 i \e^{b_1 b_2} \LF_{b_1 b_2 \bar q +} + \e_{\bar q}{}^{r_1} ( \LG_{r_1 +} - 24 \BG_{r_1 +} \cont{c} + 24 \BG_{r_1} \cont{r_2} {}_{+} ) \,, \cr
  \G^{+} &:& 0 \,, \cr
  \G^{\bar b_1 \bar b_2 \bar q} &:& \tfrac{1}{8} \e_{\bar b_1 \bar b_2} ( \E_{\bar q+} - 12 i \LF_{\bar q +} \cont{c} + 12 i \LF_{\bar q +} \cont{r} ) -12 \e_{\bar q}{}^r \BG_{\bar b_1 \bar b_2 r +} \,, \cr
  \G^{\bar b_1 \bar b_2 + } &:& \tfrac{1}{8} \e_{\bar b_1 \bar b_2} \E_{++} \,, \cr
  \G^{\bar b \bar q +} &:& 0 \,, \cr
  \G^{\bar b \bar q_1 \bar q_2} &:& -3 i \e_{\bar b}{}^c \LF_{c \bar q_1 \bar q_2 +} - \tfrac{1}{4} \e_{\bar q_1 \bar q_2} ( \LG_{\bar b +} - 24 \BG_{\bar b +} \cont{c} + 24 \BG_{\bar b +} \cont{r} ) \,, \cr
  \G^{\bar q_1 \bar q_2 +} &:& 0 \,, \cr
  \G^{\bar b_1 \bar b_2 \bar q_1 \bar q_2 +} &:& 0 \,.
 \eea

\subsection{Integrability conditions on $e_{k5}$}

The expressions for the integrability condition $\mathcal{I}$ on the
basis element $e_{k5}$ read
 \bea
  1 &:& - 2 \LG_{k-} - 48 \LG_{k-} \cont{\s} \,, \cr
  \G^{\bar k \bar \t} &:& \LG_{\bar \t -} - 24 \BG_{k \bar k \bar \t -} + 24 \BG_{\bar \t -} \cont{\s} \,, \cr
  \G^{\bar k +} &:& - \tfrac{1}{2} (\LG_{k \bar k}- \LG \cont{\s} + \LG_{-+}) - 12 \BG_{k \bar k} \cont{\s} + 12 \BG_{k \bar k -+} + 6 \BG \cont{\s_1} \cont{\s_2} - 12 \BG \cont{\s} {}_{-+} \,, \cr
  \G^{\bar \t_1 \bar \t_2} &:& - 24 \BG_{k \bar \t_1 \bar \t_2 -} - \e_{\bar \t_1 \bar \t_2}{}^\s \BP_{\s -} \,, \cr
  \G^{\bar \t +} &:& - \LG_{k \bar \t} - 24 \BG_{k \bar \t} \cont{\s} + 24 \BG_{k \bar \t -+} - \e_{\bar \t}{}^{\s_1 \s_2} \BP_{\s_1 \s_2} \,, \cr
  \G^{\bar k \bar \t_1 \cdots \bar \t_3} &:& 4 \BG_{\bar \t_1 \cdots \bar \t_3 -} - \tfrac{1}{6} \e_{\bar \t_1 \cdots \bar \t_3} \BP_{\bar k -} \,, \cr
  \G^{\bar k \bar \t_1 \bar \t_2 +} &:& \tfrac{1}{4} \LG_{\bar \t_1 \bar \t_2} - 6 \BG_{k \bar k \bar \t_1 \bar \t_2} + 6 \BG_{\bar \t_1 \bar \t_2} \cont{\s} - 6 \BG_{\bar \t_1 \bar \t_2 -+} + \tfrac{1}{2} \e_{\bar \t_1 \bar \t_2}{}^\s \BP_{\bar k \s} \,, \cr
  \G^{\bar \t_1 \cdots \bar \t_3 +} &:& - 4 \BG_{k \bar \t_1 \cdots \bar \t_3} + \tfrac{1}{24} \e_{\bar \t_1 \cdots \bar \t_3} ( LP + 2 \BP_{k \bar k} - 2 \BP \cont{\s} - 2 \BP_{-+}) \,.
 \eea
The integrability conditions $\mathcal{I}_{k}$ are given by
 \bea
  \G^{\bar k} &:& - \tfrac{1}{2} \E_{k -} + 6 i \LF_{k -} \cont{\s} + 16 \e^{\s_1 \cdots \s_3} \BG_{\s_1 \cdots \s_3 -} \,, \cr
  \G^{\bar \t} &:& - 24 \e_{\bar \t}{}^{\s_1 \s_2} \BG_{k \s_1 \s_2 -} \,, \cr
  \G^{+} &:& \tfrac{1}{2} \E_{kk} + 8 \e^{\s_1 \cdots \s_3} \BG_{k \s_1 \cdots \s_3} \,, \cr
  \G^{\bar k \bar \t_1 \bar \t_2} &:& 3 i \LF_{k \bar \t_1 \bar \t_2 -} + 12 \e_{\bar \t_1 \bar \t_2}{}^{\s_1} \BG_{\s_1} \cont{\s_2} {}_- \,, \cr
  \G^{\bar k \bar \t +} &:& - \tfrac{1}{4} \E_{k \bar \t} + 3 i \LF_{k \bar \t} \cont{\s} - 3 i \LF_{k \bar \t -+} + 12 \e_{\bar \t}{}^{\s_1 \s_2} ( \BG_{\s_1 \s_2} \cont{\s_3} + \BG_{\s_1 \s_2 -+} ) \,, \cr
  \G^{\bar \t_1 \cdots \bar \t_3} &:& \tfrac{1}{12} \e_{\bar \t_1 \cdots \bar \t_3} (\LG_{k-} - 24 \BG_{k-} \cont{\s} ) \,, \cr
  \G^{\bar \t_1 \bar \t_2 +} &:& - \tfrac{1}{4} \e_{\bar \t_1 \bar \t_2}{}^{\s_1} ( \LG_{k \s_1} - 24 \BG_{k \s_1} \cont{\s_2} - 24 \BG_{k \s_1 -+} ) \,, \cr
  \G^{\bar k \bar \t_1 \cdots \bar \t_3 +} &:& \tfrac{1}{2} i \LF_{k \bar \t_1 \cdots \bar \t_3} - \tfrac{1}{24} \e_{\bar \t_1 \cdots \bar \t_3} ( \LG_{k \bar k} - 12 \BG \cont{\s_1} \cont{\s_2} - 24 \BG \cont{\s} {}_{-+} ) \,.
 \eea
Similarly, $\mathcal{I}_{\bar k}$ reads
 \bea
  \G^{\bar k} &:& - \tfrac{1}{2} \E_{\bar k -} + 6 i \LF_{\bar k -} \cont{\s} \,, \cr
  \G^{\bar \t} &:& 12 i \LF_{k \bar k \bar \t -} + 24 \e_{\bar \t}{}^{\s_1 \s_2} \BG_{\bar k \s_1 \s_2 -} \,, \cr
  \G^{+} &:& \tfrac{1}{2} \E_{k \bar k} + 6 i \LF_{k \bar k } \cont{\s} - 6 i \LF_{k \bar k -+} - 8 \e^{\s_1 \cdots \s_3} \BG_{\bar k \s_1 \cdots \s_3} \,, \cr
  \G^{\bar k \bar \t_1 \bar \t_2} &:& + 3 i \LF_{\bar k \bar \t_1 \bar \t_2 -} \,, \cr
  \G^{\bar k \bar \t +} &:& - \tfrac{1}{4} \E_{\bar k \bar \t} + 3 i \LF_{\bar k \bar \t} \cont{\s} - 3 i \LF_{\bar k \bar \t -+} \,, \cr
  \G^{\bar \t_1 \cdots \bar \t_3} &:& \tfrac{1}{12} \e_{\bar \t_1 \cdots \bar \t_3} (\LG_{\bar k -} + 24 \BG_{\bar k -} \cont{\s}) ,, \cr
  \G^{\bar \t_1 \bar \t_2 +} &:& 3 i \LF_{k \bar k \bar \t_1 \bar \t_2} - \tfrac{1}{4} \e_{\bar \t_1 \bar \t_2}{}^\s ( \LG_{\bar k \s} + 24 \BG_{\bar k \s} \cont{\s_2} + 24 \BG_{\bar k \s -+} ) \,, \cr
  \G^{\bar k \bar \t_1 \cdots \bar \t_3 +} &:& \tfrac{1}{2} i \LF_{\bar k \bar \t_1 \cdots \bar \t_3} \,.
 \eea
The integrability condition $\mathcal{I}_M$ with $M = \rho$ is given by
 \bea
  \G^{\bar k} &:& - \tfrac{1}{2} \E_{\rho -} - 6 i \LF_{k \bar k \rho -} + 6 i \LF_{\rho -} \cont{\s} \,, \cr
  \G^{\bar \t} &:& 12 i \LF_{k \rho \bar \t -} + 24 \e_{\bar \t}{}^{\s_1 \s_2} \BG_{\r \s_1 \s_2 -} \,, \cr
  \G^{+} &:& \tfrac{1}{2} \E_{\r k} + 6 i\LF_{k \r} \cont{\s} -6 i \LF_{k \rho -+} \,, \cr
  \G^{\bar k \bar \t_1 \bar \t_2} &:& 3 i \LF_{\r \bar \t_1 \bar \t_2 -} + 12 \e_{\bar \t_1 \bar \t_2}{}^{\s} \BG_{\bar k \r \s -} \,, \cr
  \G^{\bar k \bar \t +} &:& - \tfrac{1}{4} \E_{\r \bar \t} - 3 i \LF_{\r \bar \t k \bar k} + 3 i \LF_{\r \bar \t} \cont{\s} - 3 i \LF_{\r \bar \t -+} + 12 \e_{\bar \t}{}^{\s_1 \s_2} \BG_{\bar k \r \s_1 \s_2} \,, \cr
  \G^{\bar \t_1 \cdots \bar \t_3} &:& \tfrac{1}{12} \e_{\bar \t_1 \cdots \bar \t_3} ( \LG_{\r -} + 24 \BG_{\r} \cont{\s} {}_- - 24 \BG_{k \bar k \r -} ) \,, \cr
  \G^{\bar \t_1 \bar \t_2 +} &:& 3 i \LF_{k \r \bar \t_1 \bar \t_2} - \tfrac{1}{4} \e_{\bar \t_1 \bar \t_2}{}^{\s_1} (\LG_{\r \s_1} - 24 \BG_{\r \s_1 k \bar k} + 24 \BG_{\rho \s_1} \cont{\s_2} + 24 \BG_{\r \s_1 -+}) \,, \cr
  \G^{\bar k \bar \t_1 \cdots \bar \t_3 +} &:& \tfrac{1}{2} i \LF_{\r \bar \t_1 \cdots \bar \t_3} + \tfrac{1}{24} \e_{\bar \t_1 \cdots \bar \t_3} (\LG_{\bar k \r} + 24 \BG_{\bar k \r} \cont{\s} + 24 \BG_{\bar k \r -+}) \,.
 \eea
Furthermore, $\mathcal{I}_M$ with $M = \bar \rho$ is given by the expressions
 \bea
  \G^{\bar k} &:& - \tfrac{1}{2} \E_{\bar \r -} + 6 i \LF_{\bar \r - } \cont{\s} - 6 i \LF_{k \bar k \bar \r -} - 48 \e_{\bar \r}{}^{\s_1 \s_2} \BG_{\bar k \s_1 \s_2 -} \,, \cr
  \G^{\bar \t} &:& 12 i \LF_{k \bar \rho \bar \tau -} - 48 \e_{\bar \r \bar \t}{}^{\s_1} (\BG_{k \bar k \s_1 -} - \BG_{\s_1} \cont{\s_2} {}_-) - 24 \e_{\bar \t}{}^{\s_1 \s_2} \BG_{\bar \r \s_1 \s_2 -}  \,, \cr
  \G^{+} &:& \tfrac{1}{2} \E_{k \bar \rho} + 6 i \LF_{k \bar \r} \cont{\s} - 6 i \LF_{k \bar \r -+} - 24 \e_{\bar \r}{}^{\s_1 \s_2} ( \BG_{k \bar k \s_1 \s_2} - \BG_{\s_1 \s_2 -+}) \,, \cr
  \G^{\bar k \bar \t_1 \bar \t_2} &:& 3 i \LF_{\bar \r \bar \t_1 \bar \t_2 -} - 12  \e_{\bar \rho \bar \t_1 \bar \t_2} \BG_{\bar k - } \cont{\s} - 12 \e_{\bar \t_1 \bar \t_2}{}^{\s} \BG_{\bar k \bar \r \s -+} \,, \cr
  \G^{\bar k \bar \t +} &:& - \tfrac{1}{4} \E_{\bar \r \bar \t} - 3 i \LF_{k \bar k \bar \r \bar \t} + 3i \LF_{\bar \r \bar \t} \cont{\s} - 3 i \LF_{\bar \r \bar \t -+} - 12 \e_{\bar \t}{}^{\s_1 \s_2} \BG_{\bar k \bar \r \s_1 \s_2} + \cr
  && + 24 \e_{\bar \r \bar \t}{}^{\s_1} ( \BG_{ \bar k \s_1} \cont{\s_2} + \BG_{\bar k \s_1 -+} ) \,, \cr
  \G^{\bar \t_1 \cdots \bar \t_3} &:& \tfrac{1}{12} \e_{\bar \t_1 \cdots \bar \t_3} ( \LG_{\bar \r -} - 24 \BG_{\bar \r - } \cont{\s} + 24 \BG_{k \bar k \bar \r -} ) \,, \cr
  \G^{\bar \t_1 \bar \t_2 +} &:& 3 i \LF_{k \bar \r \bar \t_1 \bar \t_2} - \tfrac{1}{4} \e_{\bar \t_1 \bar \t_2}{}^{\s_1} (\LG_{\bar \r \s_1} + 24 \BG_{k \bar k \bar \r \s_1} - 24 \BG_{\bar \r \s_1} \cont{\s_2} - 24 \BG_{\bar \r \s_1 -+}) + \cr
  && + 3 \e_{\bar \r \bar \t_1 \bar \t_2} ( - 2 \BG_{k \bar k} \cont{\s} - 2 \BG_{k \bar k -+} + \BG \cont{\s_1} \cont{\s_2} + 2 \BG \cont{\s} {}_{-+}) \,, \cr
  \G^{\bar k \bar \t_1 \cdots \bar \t_3 +} &:& \tfrac{1}{2} i \LF_{\bar \r \bar \t_1 \cdots \bar \t_3} + \tfrac{1}{24} \e_{\bar \t_1 \cdots \bar \t_3} ( \LG_{\bar k \bar \r} - 24 \BG_{\bar k \bar \r} \cont{\s} - 24 \BG_{\bar k \bar \r -+}) \,.
 \eea
The integrability conditions $\mathcal{I}_-$ read
 \bea
  \G^{\bar k} &:& - \tfrac{1}{2} \E_{--} \,, \cr
  \G^{\bar \t} &:& 0 \,, \cr
  \G^{+} &:& \tfrac{1}{2} \E_{k-} +6 i \LF_{k-} \cont{\s} + 8 \e^{\s_1 \cdots \s_3} \BG_{\s_1 \cdots \s_3 -}\,, \cr
  \G^{\bar k \bar \t_1 \bar \t_2} &:& 0 \,, \cr
  \G^{\bar k \bar \t +} &:& - \tfrac{1}{4} \E_{\bar \t -} + 3 i \LF_{k \bar k \bar \t -} - 3 i \LF_{\bar \t -} \cont{\s} + 12 \e_{\bar \t}{}^{\s_1 \s_2} \BG_{\bar k \s_1 \s_2 -} \,, \cr
  \G^{\bar \t_1 \cdots \bar \t_3} &:& 0 \,, \cr
  \G^{\bar \t_1 \bar \t_2 +} &:& 3 i \LF_{k \bar \t_1 \bar \t_2 -} + \tfrac{1}{4} \e_{\bar \t_1 \bar \t_2}{}^{\s} (\LG_{\s -} - 24 \BG_{k \bar k \s -} + 24 \s_{\s -} \cont{\s_2}) \,, \cr
  \G^{\bar k \bar \t_1 \cdots \bar \t_3 +} &:& - \tfrac{1}{2} i \LF_{\bar \t_1 \cdots \bar \t_3 -} + \tfrac{1}{24} \e_{\bar \t_1 \cdots \bar \t_3} ( \LG_{\bar k -} + 24 \BG_{\bar k -} \cont{\s}) \,.
 \eea
Finally, for $\mathcal{I}_+$ we find
 \bea
  \G^{\bar k} &:& - \tfrac{1}{2} \E_{-+} + 6 i \LF_{k \bar k -+} - 6 i \LF \cont{\s}{}_{-+} + 16 \e^{\s_1 \cdots \s_3} \BG_{\bar k \s_1 \cdots \s_3} \,, \cr
  \G^{\bar \t} &:& 12 i \LF_{k \bar \t -+} + 24 \e_{\bar \t}{}^{\s_1 \s_2} (\BG_{k \bar k \s_1 \s_2} - \BG_{\s_1 \s_2} \cont{\s_3}) \,, \cr
  \G^{+} &:& \tfrac{1}{2} \E_{k +} + 6 i \LF_{k+} \cont{\s} - 8 \e^{\s_1 \cdots \s_3} \BG_{\s_1 \cdots \s_3 +} \,, \cr
  \G^{\bar k \bar \t_1 \bar \t_2} &:& - 3 i \LF_{\bar \t_1 \bar \t_2 -+} + 12 \e_{\bar \t_1 \bar \t_2}{}^\s \BG_{\bar k \s} \cont{\s_2}  \,, \cr
  \G^{\bar k \bar \t +} &:& - \tfrac{1}{4} \E_{\bar \t +} + 3 i \LF_{k \bar k \bar \t +} - 3 i \LF_{\bar \t +} \cont{\s} - 12 \e_{\bar \t}{}^{\s_1 \s_2} \BG_{\bar k \s_1 \s_2 +} \,, \cr
  \G^{\bar \t_1 \cdots \bar \t_3} &:& - \tfrac{1}{12} \e_{\bar \t_1 \cdots \bar \t_3} ( LG_{-+} - 24 \BG_{k \bar k} \cont{\s} + 12 \BG \cont{\s_1} \cont{\s_2} )\,, \cr
  \G^{\bar \t_1 \bar \t_2 +} &:& 3 i \LF_{k \bar \t_1 \bar \t_2 +} + \tfrac{1}{4} \e_{\bar \t_1 \bar \t_2}{}^\s ( \LG_{\s +} +24 \BG_{k \bar k \s +} - 24 \BG_{\s +} \cont{\s_2})  \,, \cr
  \G^{\bar k \bar \t_1 \cdots \bar \t_3 +} &:& - \tfrac{1}{2} i \LF_{\bar \t_1 \cdots \bar \t_3 +} + \tfrac{1}{24} \e_{\bar \t_1 \cdots \bar \t_3 } ( \LG_{\bar k +} - 24 \BG_{\bar k +} \cont{\s}) \,.
 \eea

\subsection{Integrability conditions on $e_{1234}$}

The expressions for the integrability condition $\mathcal{I}$ on the
basis element $e_{1234}$ read
 \bea
  1 &:& 4 \e^{\g_1 \cdots \g_4} \BG_{\g_1 \cdots \g_4} + \LP + 2 \BP \cont{\g} + 2 \BP_{-+} \,, \cr
  \Gamma^{\bar \b_1 \bar \b_2} &:& - \tfrac{1}{4} \e_{\bar \b_1 \bar \b_2}{}^{\g_1 \g_2} ( \LG_{\g_1 \g_2}
  -24 \BG_{\g_1 \g_2} \cont{\g_3} + 24 \BG_{\g_1 \g_2 -+}) + \BP_{\bar \b_1 \bar \b_2} \,, \cr
  \G^{\bar \b +} &:& 8 \e_{\bar \b}{}^{\g_1 \cdots \g_3} \BG_{\g_1 \cdots \g_3 +} + 2 \BP_{\bar \b +} \,, \cr
  \G^{\bar \b_1 \cdots \bar \b_4} &:& \tfrac{1}{96} \e_{\bar \b_1 \cdots \bar \b_4} (- \LG \cont{\g} + \LG_{-+}
  + 12 \BG \cont{\g_1} \cont{\g_2} - 24 \BG \cont{\g} {}_{-+} ) \,, \cr
  \G^{\bar \b_1 \cdots \bar \b_3 +} &:& - \tfrac{1}{12} \e_{\bar \b_1 \cdots \bar \b_3}{}^{\g_1} ( \LG_{\g_1 +}
  -24 \BG_{\g_1 +} \cont{\g_2}) \,.
 \eea
The integrability conditions $\mathcal{I}_{\a}$ are
 \bea
  \Gamma^{\bar \b} &:& - 2 i \e_{\bar \b}{}^{\g_1 \cdots \g_3} \LF_{\a \g_1 \cdots \g_3} - \LG_{\a \bar \b}
  - 24(\BG_{\a \bar \b} \cont{\g} + \BG_{\a \bar \b -+}) + \cr
  && + 12 g_{\a \bar \b}( \BG \cont{\g_1} \cont{\g_2}
  + 2 \BG \cont{\g} {}_{-+}) \,, \cr
  \G^+ &:& - \LG_{\a +} - 24 \BG_{\a +} \cont{\g} \,, \cr
  \G^{\bar \b_1 \cdots \bar \b_3} &:& - \tfrac{1}{12} \e_{\bar \b_1 \cdots \bar \b_3}{}^{\g_1}
  ( \tfrac{1}{2} \E_{\a \g_1} + 6 i \LF_{\a \g_1} \cont{\g_2} - 6 i \LF_{\a \g_1 -+})
  - 4 \BG_{\a \bar \b_1 \cdots \bar \b_3} + \cr
  && + 12 g_{\a [ \bar \b_1} (
  \BG_{\bar \b_2 \bar \b_3 ] } \cont{\g} + \BG_{\bar \b_2 \bar \b_3
  ] -+} ) \,,
  \cr
  \G^{\bar \b_1 \bar \b_2 +} &:& \tfrac{3}{2} i \e_{\bar \b_1 \bar \b_2}{}^{\g_1 \g_2} \LF_{\a \g_1 \g_2 +}
  - 12 \BG_{\a \bar \b_1 \bar \b_2 +} + 24 g_{\a [ \bar \b_1} \BG_{\bar \b_2 ] +} \cont{\g}
  \,, \cr
  \G^{\bar \b_1 \cdots \bar \b_4 +} &:& \tfrac{1}{96} \e_{\bar \b_1 \cdots \bar \b_4} ( \tfrac{1}{2} \E_{\a +}
  + 6 i \LF_{\a +} \cont{\g} ) + 4 g_{\a [ \bar \b_1} \BG_{\bar \b_2 \cdots \bar \b_4 ] +} \,.
 \eea
The integrability conditions $\mathcal{I}_{\bar \a}$ read
 \bea
 \Gamma^{\bar \b} &:& - 2 i \e_{\bar \b}{}^{\g_1 \cdots \g_3} \LF_{\bar \a \g_1 \cdots \g_3}
 -\LG_{\bar \a \bar \b} + 24 (\BG_{\bar \a \bar \b} \cont{\g} + \BG_{\bar \a \bar \b -+} ) \,, \cr
 \Gamma^+ &:& - \LG_{\bar \a +} + 24 \BG_{\bar \a +} \cont{\g} \,,
 \cr
 \Gamma^{\bar \b_1 \cdots \bar \b_3} &:& - \tfrac{1}{12} \e_{\bar \b_1 \cdots \bar \b_3}{}^{\g_1}
 ( \tfrac{1}{2} \E_{\bar \a \g_1} + 6 i (\LF_{\bar \a \g_1} \cont{\g_2} - \LF_{\bar \a \g_1 -+}))
 + 4 \BG_{\bar \a \bar \b_1 \cdots \bar \b_3} \,, \cr
 \G^{\bar \b_1 \bar \b_2 +} &:& \tfrac{3}{2} i \e_{\bar \b_1 \bar \b_2}{}^{\g_1 \g_2} \LF_{\bar \a \g_1 \g_2 +}
 + 12 \BG_{\bar \a \bar \b_1 \bar \b_2 +} \,, \cr
 \G^{\bar \b_1 \cdots \bar \b_4 +} &:& \tfrac{1}{96} \e_{\bar \b_1 \cdots \bar \b_4} ( \tfrac{1}{2} \E_{\bar \a +}
 + 6 i \LF_{\bar \a +} \cont{\g}) \,.
 \eea
Similarly, $\mathcal{I}_A$ with $A=-$ is given by
 \bea
 \Gamma^{\bar \b} &:& - 2 i \e_{\bar \b}{}^{\g_1 \cdots \g_3} \LF_{- \g_1 \cdots \g_3} + \LG_{\bar \b -} + 24 \BG_{\bar \b -} \cont{\g} \,, \cr
 \Gamma^+ &:& - \LG_{-+} + 12 \BG \cont{\g_1} \cont{\g_2} \,,
 \cr
 \Gamma^{\bar \b_1 \cdots \bar \b_3} &:& - \tfrac{1}{12} \e_{\bar \b_1 \cdots \bar \b_3}{}^{\g_1}
 ( \tfrac{1}{2} \E_{\g_1 -} - 6 i \LF_{\g_1 -} \cont{\g_2})+ 4 \BG_{\bar \b_1 \cdots \bar \b_3 -}    \,, \cr
 \G^{\bar \b_1 \bar \b_2 +} &:&  \tfrac{3}{2} i \e_{\bar \b_1 \bar \b_2}{}^{\g_1 \g_2} \LF_{\g_1 \g_2 -+}
 + 12 \BG_{\bar \b_1 \bar \b_2} \cont{\g} \,, \cr
 \G^{\bar \b_1 \cdots \bar \b_4 +} &:& \tfrac{1}{96} \e_{\bar \b_1 \cdots \bar \b_4} ( \tfrac{1}{2} \E_{-+} + 6 i \LF_{-+} \cont{\g})
 + \BG_{\bar \b_1 \cdots \bar \b_4} \,.
 \eea
Finally, the expressions for $\mathcal{I}_+$ read
 \bea
  \Gamma^{\bar \b} &:& 2 i \e_{\bar \b}{}^{\g_1 \cdots \g_3} \LF_{\g_1 \cdots \g_3 +} + \LG_{\bar \b +}
  - 24 \BG_{\bar \b +} \cont{\g} \,, \cr
  \G^+ &:& 0 \,, \cr
  \G^{\bar \b_1 \cdots \bar \b_3} &:& - \tfrac{1}{12} \e_{\bar \b_1 \cdots \bar \b_3}{}^{\g_1}
  (\tfrac{1}{2} \E_{\g_1 +} - 6 i \LF_{\g_1 +} \cont{\g_2}) - 4 \BG_{\bar \b_1 \cdots \bar \b_3 +} \,,
  \cr
  \G^{\bar \b_1 \bar \b_2 +} &:& 0 \,, \cr
  \G^{\bar \b_1 \cdots \bar \b_4 +} &:& \tfrac{1}{192} \e_{\bar \b_1 \cdots \bar \b_4} \E_{++} \,.
 \eea

\subsection{Integrability conditions on $e_{i_1 \cdots i_3 5}$}

The expressions for the integrability condition $\mathcal{I}$ on the
basis element $e_{i_1 \cdots i_3 5}$ read
 \bea
  1 &:& 16 \e^{\s_1 \cdots \s_3} \BG_{\s_1 \cdots \s_3 -} - 4 \BP_{k -} \,, \cr
  \G^{\bar k \bar \t} &:& 24 \e_{\bar \t}{}^{\s_1 \s_2} \BG_{\bar k \s_1 \s_2 -} + 2 \BP_{\bar \t -} \,, \cr
  \G^{\bar k +} &:& - 8 \e^{\s_1 \cdots \s_3} \BG_{\bar k \s_1 \cdots \s_3} + \tfrac{1}{2} \LP - \BP_{k \bar k} + \BP \cont{\s} - \BP_{-+} \,, \cr
  \G^{\bar \t_1 \bar \t_2} &:& - \tfrac{1}{2} \e_{\bar \t_1 \bar \t_2} {}^{\s_1} ( \LG_{\s_1 -} + 24 \BG_{k \bar k \s_1 -} - 24 \BG_{\s_1} \cont{\s_2} {}_- ) \,, \cr
  \G^{\bar \t +} &:& - \tfrac{1}{2} \e_{\bar \t}{}^{\s_1 \s_2} (\LG_{\s_1 \s_2} + 24 \BG_{k \bar k \s_1 \s_2} - 24 \BG_{\s_1 \s_2} \cont{\s_3} - 24 \BG_{\s_1 \s_2 -+} ) - 2 \BP_{k \bar \t} \,, \cr
  \G^{\bar k \bar \t_1 \cdots \bar \t_3} &:& - \tfrac{1}{12} \e_{\bar \t_1 \cdots \bar \t_3} (\LG_{\bar k -} - 24 \BG_{\bar k -} \cont{\s} ) \,, \cr
  \G^{\bar k \bar \t_1 \bar \t_2 +} &:& \tfrac{1}{4} \e_{\bar \t_1 \bar \t_2}{}^{\s_1} (\LG_{\bar k \s_1} - 24 \BG_{\bar k \s_1} \cont{\s_2} - 24 \BG_{\bar k \s_1 -+} ) + \tfrac{1}{2} \BP_{\bar \t_1 \bar \t_2} \,, \cr
  \G^{\bar \t_1 \cdots \bar \t_3 +} &:& \tfrac{1}{24} \e_{\bar \t_1 \cdots \bar \t_3} (\LG_{k \bar k} - \LG \cont{\s} - \LG_{-+} - 24 \BG_{k \bar k} \cont{\s} - 24 \BG_{k \bar k -+} + \cr
  && + 12 \BG \cont{\s_1} \cont{\s_2} + 24 \BG \cont{\s} {}_{-+}) \,.
 \eea
The integrability conditions $\mathcal{I}_{k}$ are given by
 \bea
  \G^{\bar k} &:& \LG_{k -} - 24 \BG_{k-} \cont{\s} \,, \cr
  \G^{\bar \t} &:& - 6 i \e_{\bar \t}{}^{\s_1 \s_2} \LF_{k \s_1 \s_2 -} \,, \cr
  \G^{+} &:& 2 i \e^{\s_1 \cdots \s_3} \LF_{k \s_1 \cdots \s_3} \,, \cr
  \G^{\bar k \bar \t_1 \bar \t_2} &:& 3 i \e_{\bar \t_1 \bar \t_2}{}^{\s} \LF_{k \bar k \s -} - 12 \BG_{k \bar \t_1 \bar \t_2 -} \,, \cr
  \G^{\bar k \bar \t +} &:& 3 i \e_{\bar \t}{}^{\s_1 \s_2} \LF_{k \bar k \s_1 \s_2} + \tfrac{1}{2} \LG_{k \bar \t} - 12 \BG_{k \bar \t} \cont{\s} + 12 \BG_{k \bar \t -+} \,, \cr
  \G^{\bar \t_1 \cdots \bar \t_3} &:& - \tfrac{1}{24} \e_{\bar \t_1 \cdots \bar \t_3} ( \E_{k-} + 12 i \LF_{k -} \cont{\s}) \,, \cr
  \G^{\bar \t_1 \bar \t_2 +} &:& \tfrac{1}{8} \e_{\bar \t_1 \bar \t_2}{}^{\s_1} ( \E_{k \s_1} + 12 i \LF_{k \s_1} \cont{\s_2} + 12 i \LF_{k \s_1 -+}) \,, \cr
  \G^{\bar k \bar \t_1 \cdots \bar \t_3 +} &:& \tfrac{1}{48} \e_{\bar \t_1 \cdots \bar \t_3} (\E_{k \bar k} + 12 i \LF_{k \bar k} \cont{\s} + 12 i \LF_{k \bar k -+}) - 2 \BG_{k \bar \t_1 \cdots \bar \t_3} \,.
 \eea
Similarly, $\mathcal{I}_{\bar k}$ reads
 \bea
  \G^{\bar k} &:& \LG_{\bar k -} + 24 \BG_{\bar k -} \cont{\s} \,, \cr
  \G^{\bar \t} &:& - 6 i \e_{\bar \t}{}^{\s_1 \s_2} \LF_{\bar k \s_1 \s_2 -} + 48 \BG_{\bar \t -} \cont{\s} \,, \cr
  \G^{+} &:& 2 i \e^{\s_1 \cdots \s_3} \LF_{\bar k \s_1 \cdots \s_3} + \LG_{k \bar k} + 12 \BG \cont{\s_1} \cont{\s_2} - 24 \BG \cont{\s} {}_{-+} \,, \cr
  \G^{\bar k \bar \t_1 \bar \t_2} &:& 12 \BG_{\bar k \bar \t_1 \bar \t_2 -} \,, \cr
  \G^{\bar k \bar \t +} &:& \tfrac{1}{2} \LG_{\bar k \bar \t} + 12 (\BG_{\bar k \bar \t} \cont{\s} - \BG_{\bar k \bar \t -+}) \,, \cr
  \G^{\bar \t_1 \cdots \bar \t_3} &:& - \tfrac{1}{24} \e_{\bar \t_1 \cdots \bar \t_3} (\E_{\bar k -} + 12 i \LF_{\bar k -} \cont{\s}) + 8 \BG_{\bar \t_1 \cdots \bar \t_3 -} \,, \cr
  \G^{\bar \t_1 \bar \t_2 +} &:& \tfrac{1}{8} \e_{\bar \t_1 \bar \t_2}{}^{\s_1} (\E_{\bar k \s_1} + 12 i \LF_{\bar k \s_1} \cont{\s_2} + 12 i \LF_{\bar k \s -+}) + 12 (\BG_{\bar \t_1 \bar \t_2} \cont{\s} - \BG_{\bar \t_1 \bar \t_2 -+}) \,, \cr
  \G^{\bar k \bar \t_1 \cdots \bar \t_3 +} &:& \tfrac{1}{48} \e_{\bar \t_1 \cdots \bar \t_3} \E_{\bar k \bar k} + 2 \BG_{\bar k \bar \t_1 \cdots \bar \t_3} \,.
 \eea
The integrability condition $\mathcal{I}_M$ with $M = \rho$ is given by
 \bea
  \G^{\bar k} &:& \LG_{\r -} - 24 \BG_{k \bar k \r -} + 24 \BG_{\r -} \cont{\s} \,, \cr
  \G^{\bar \t} &:& - 6 i \e_{\bar \t}{}^{\s_1 \s_2} \LF_{\r \s_1 \s_2 -} + 48 (\BG_{k \r \bar \t -} - g_{\r \bar \t} \BG_{k -} \cont{\s}) \,, \cr
  \G^{+} &:& \LG_{k \r} + 24 \BG_{k \r} \cont{\s} - 24 \BG_{k \r -+} \,, \cr
  \G^{\bar k \bar \t_1 \bar \t_2} &:& - 3 i \e_{\bar \t_1 \bar \t_2}{}^{\s} \LF_{\bar k \r \s -} + 24 g_{\r [ \bar \t_1} ( \BG_{\bar \t_2 ] k \bar k -} - \BG_{\bar \t_2 ] -} \cont{\s} ) + 12 \BG_{\rho \bar \tau_1 \bar \tau_2 -} \,, \cr
  \G^{\bar k \bar \t +} &:& - 3 i \e_{\bar \t}{}^{\s_1 \s_2} \LF_{\bar k \r \s_1 \s_2} + \tfrac{1}{2} \LG_{\r \bar \t} - 12 (\BG_{k \bar k  \r \bar \t} - \BG_{\r \bar \t} \cont{\s} + \BG_{ \r \bar \t -+}) + \cr
  && + 6 g_{\r \bar \t} ( 2 \BG_{k \bar k} \cont{\s} - 2 \BG_{k \bar k -+} - \BG \cont{\s_1} \cont{\s_2} + 2 \BG \cont{\s}{}_{-+} ) \,, \cr
  \G^{\bar \t_1 \cdots \bar \t_3} &:& - \tfrac{1}{24} \e_{\bar \t_1 \cdots \bar \t_3} ( \E_{\r -} - 12 i \LF_{k \bar k \r -} + 12 i \LF_{\r - } \cont{\s} ) - 24 g_{\r [ \bar \t_1} \BG_{\bar \t_2 \bar \t_3 ] k - } \,, \cr
  \G^{\bar \t_1 \bar \t_2 +} &:& \tfrac{1}{8} \e_{\bar \t_1 \bar \t_2}{}^{\s_1} ( \E_{\r \s_1} - 12 i \LF_{k \bar k \r \s_1} + 12 i \LF_{\r \s_1} \cont{\s_2} + 12 i \LF_{\r \s -+} )+ \cr
  && + 24 g_{\r [ \bar \t_1} ( \BG_{\bar \t_2 ] k} \cont{\s} - \BG_{\bar \t_2 ] k -+}) + 12 \BG_{k \rho \bar \tau_1 \bar \tau_2} \,, \cr
  \G^{\bar k \bar \t_1 \cdots \bar \t_3 +} &:& \tfrac{1}{48} \e_{\bar \t_1 \cdots \bar \t_3} ( \E_{\bar k \r} - 12 i \LF_{\bar k \r} \cont{\s} - 12 i \LF_{\bar \k \r -+}) + 6 g_{\r [ \bar \t_1} ( \BG_{\bar \t_2 \bar \t_3 ] k \bar k} + \BG_{\bar \t_2 \bar \t_3] -+} ) \,. ~~~~~~~~~~
 \eea
Furthermore, $\mathcal{I}_M$ with $M = \bar \rho$ is given by the expressions
 \bea
  \G^{\bar k} &:& \LG_{\bar \r -} + 24 \BG_{k \bar k \bar \r -} - 24 \BG_{\bar \r -} \cont{\s} \,, \cr
  \G^{\bar \t} &:& - 6 i \e_{\bar \t}{}^{\s_1 \s_2} \LF_{\bar \r \s_1 \s_2 -} - 48 \BG_{k \bar \r \bar \t -} \,, \cr
  \G^{+} &:& 2 i \e^{\s_1 \cdots \s_3} \LF_{\bar \r \s_1 \cdots \s_3} + \LG_{k \bar \r} - 24 \BG_{k \bar \r} \cont{\s} + 24 \BG_{k \bar \r -+}  \,, \cr
  \G^{\bar k \bar \t_1 \bar \t_2} &:& - 3 i \e_{\bar \t_1 \bar \t_2}{}^{\s} \LF_{\bar k \bar \r \s -} - 12 \BG_{\bar \r \bar \t_1 \bar \t_2 -}  \,, \cr
  \G^{\bar k \bar \t +} &:& - 3 i \e_{\bar \t}{}^{\s_1 \s_2} \LF_{\bar k \bar \r \s_1 \s_2} + \tfrac{1}{2} \LG_{\bar \r \bar \t} + 12 (\BG_{k \bar k \bar \r \bar \t} - \BG_{\bar \r \bar \t} \cont{\s} + \BG_{\bar \r \bar \t -+} ) \,, \cr
  \G^{\bar \t_1 \cdots \bar \t_3} &:& - \tfrac{1}{24} \e_{\bar \t_1 \cdots \bar \t_3} ( \E_{\bar \r -} - 12 i \LF_{k \bar k \bar \r -} + 12 i \LF_{\bar \r -} \cont{\s} ) \,, \cr
  \G^{\bar \t_1 \bar \t_2 +} &:& \tfrac{1}{8} \e_{\bar \t_2 \bar \t_2}{}^{\s_1} ( \E_{\bar \r \s_1} - 12 i \LF_{\bar \r \s_1 k \bar k} + 12 i \LF_{\bar \r \s_1} \cont{\s_2} + 12 i \LF_{\bar \r \s -+} ) - 12 \BG_{k \bar \r \bar \t_1 \bar \t_2}  \,, \cr
  \G^{\bar k \bar \t_1 \cdots \bar \t_3 +} &:& \tfrac{1}{48} \e_{\bar \t_1 \cdots \bar \t_3} (\E_{\bar k \bar \r} - 12 i \LF_{\bar k \bar \r} \cont{\s} - 12 i \LF_{\bar k \bar \r -+} )  \,.
 \eea
The integrability conditions $\mathcal{I}_-$ read
 \bea
  \G^{\bar k} &:& 0 \,, \cr
  \G^{\bar \t} &:& 0 \,, \cr
  \G^{+} &:& -2 i \e^{\s_1 \cdots \s_3} \LF_{\s_1 \cdots \s_3 -} + \LG_{k-} - 24 \BG_{k-} \cont{\s} \,, \cr
  \G^{\bar k \bar \t_1 \bar \t_2} &:& 0 \,, \cr
  \G^{\bar k \bar \t +} &:& - 3 i \e_{\bar \t}{}^{\s_1 \s_2} \LF_{\bar k \s_1 \s_2 -} - \tfrac{1}{2} \LG_{\bar \t -} - 12 ( \BG_{k \bar k \bar \t -} - \BG_{\bar \t -} \cont{\s}) \,, \cr
  \G^{\bar \t_1 \cdots \bar \t_3} &:& - \tfrac{1}{24} \e_{\bar \t_2 \cdots \bar \t_3} \E_{--} \,, \cr
  \G^{\bar \t_1 \bar \t_2 +} &:& \tfrac{1}{8} \e_{\bar \t_1 \bar \t_2}{}^{\s_1} ( \E_{\s_1 -} + 12 i \LF_{k \bar k \s_1 -} - 12 i \LF_{\s_1} \cont{\s_2}{}_- ) - 12 \BG_{k \bar \t_1 \bar \t_2 -} \,, \cr
  \G^{\bar k \bar \t_1 \cdots \bar \t_3 +} &:& \tfrac{1}{48} \e_{\bar \t_1 \cdots \bar \t_3} ( \E_{\bar k -} - 12 i \LF_{\bar k -} \cont{\s} ) + 2 \BG_{\bar \t_1 \cdots \bar \t_3 -} \,.
 \eea
Finally, for $\mathcal{I}_+$ we find
 \bea
  \G^{\bar k} &:& - \LG_{-+} + 24 \BG_{k \bar k} \cont{\s} - 12 \BG \cont{\s_1} \cont{\s_2} \,, \cr
  \G^{\bar \t} &:& 6 i \e_{\bar \t}{}^{\s_1 \s_2} \LF_{\s_1 \s_2 -+} + 48 \BG_{k \bar \t} \cont{\s} \,, \cr
  \G^{+} &:& - 2 i \e^{\s_1 \cdots \s_3} \LF_{\s_1 \cdots \s_3 +} + \LG_{k+} + 24 \BG_{k +} \cont{\s} \,, \cr
  \G^{\bar k \bar \t_1 \bar \t_2} &:& - 3 i \e_{\bar \t_1 \bar \t_2}{}^{\s} \LF_{\bar k \s -+} + 12 ( \BG_{k \bar k \bar \t_1 \bar \t_2} - \BG_{\bar \t_1 \bar \t_2} \cont{\s} ) \,, \cr
  \G^{\bar k \bar \t +} &:& - 3 i \e_{\bar \t}{}^{\s_1 \s_2} \LF_{\bar k \s_1 \s_2 +} - \tfrac{1}{2} \LG_{\bar \t +}  + 12 (\BG_{k \bar k \bar \t +} - \BG_{\bar \t +} \cont{\s})\,, \cr
  \G^{\bar \t_1 \cdots \bar \t_3} &:& - \tfrac{1}{24} \e_{\bar \t_1 \cdots \bar \t_3} ( \E_{-+} + 12 i \LF_{k \bar k -+} - 12 i \LF \cont{\s}{}_{-+} ) + 8 \BG_{k \bar \t_1 \cdots \bar \t_3} \,, \cr
  \G^{\bar \t_1 \bar \t_2 +} &:& \tfrac{1}{8} \e_{\bar \t_1 \bar \t_2}{}^{\s_1} (\E_{\s_1 +} + 12 i \LF_{k \bar k \s_1 +} - 12 i \LF_{\s_1} \cont{\s_2} {}_+) + 12 \BG_{k \bar \t_1 \bar \t_2 +} \,, \cr
  \G^{\bar k \bar \t_1 \cdots \bar \t_3 +} &:& \tfrac{1}{48} \e_{\bar \t_1 \dots \bar \t_3} ( \E_{\bar k +} -12 i \LF_{\bar k +} \cont{\s} ) - 2 \BG_{\bar \t_1 \cdots \bar \t_3 +} \,.
 \eea

\newsection{Generic half-maximal $SU(4)\ltimes \bR^8$-backgrounds}\la{conconap}

\subsection{The linear system}

We decompose the vector $SO(9,1)$ representation under $SU(4)$. This
is equivalent to decomposing the frame indices as $A=(+,-,\alpha, \bar\alpha)$.
Consequently, the fluxes and geometry decompose into  $SU(4)$
representations, i.e. $P_{A}$ decomposes as $P_{+}, P_{-}, P_{\a}$ and $P_{\bar\a}$
and similarly for the other fluxes and geometry\footnote{
If the fluxes are complex, like $P$ and $G$, then their various components
do not satisfy the `naive' complex conjugate relations, i.e. $(P_\a)^*\not= P_{\bar\a}$
and similarly for $G$.}.

Next, we construct the linear system associated with the algebraic and supercovariant
connection Killing spinor equations. In particular, the algebraic Killing spinor equations
give
\bea
(A_{11}+ i A_{12}) P_{\bar\a}+{1 \over 4}  G_{-+\bar{\alpha}}
+{1 \over 4} G_{\bar{\alpha}\b}{}^\b
\cr
+{1 \over 12} \epsilon_{\bar{\alpha}}{}^{\b_1 \b_2 \b_3} G_{\b_1\b_2\b_3} =0~,
\la{donea}
\eea
\bea
(A_{21}+ i A_{22}) P_{\bar\a}  +{i \over 4}  G_{-+\bar{\alpha}}
+{i \over 4} G_{\bar{\alpha}\b}{}^\b
\cr
-{i \over 12} \epsilon_{\bar{\alpha}}{}^{\b_1 \b_2 \b_3} G_{\b_1\b_2\b_3} =0~,
\la{doneb}
\eea
\bea
(A_{11}-iA_{12}) P_\alpha +{1 \over 4}  G_{-+\alpha}
-{1 \over 4} G_{\alpha \b}{}^\b
\cr
+{1 \over 12}  \epsilon_\alpha{}^{{\bar{\b_1}} {\bar{\b_2}} {\bar{\b_3}}}
G_{{\bar{\b_1}} {\bar{\b_2}} {\bar{\b_3}}} =0~,
\la{dtwoa}
\eea
\bea
(A_{21}-iA_{22}) P_\alpha   -{i \over 4} G_{-+\alpha}
+{i \over 4} G_{\alpha \b}{}^\b
\cr
+{i \over 12}  \epsilon_\alpha{}^{{\bar{\b_1}} {\bar{\b_2}} {\bar{\b_3}}}
G_{{\bar{\b_1}} {\bar{\b_2}} {\bar{\b_3}}} =0~,
\la{dtwob}
\eea
\bea
(A_{11}+iA_{12}) P_+ +{1 \over 4}  G_{+ \alpha}{}^\a =0~,
\la{dthreea}
\eea
\bea
(A_{21}+iA_{22})P_+ +{i \over 4}  G_{+ \alpha}{}^\a =0~,
\la{dthreeb}
\eea
\bea
(A_{11}-iA_{12})P_+ -{1 \over 4}  G_{+ \alpha}{}^\a =0~,
\la{dfoura}
\eea
\bea
(A_{21}-iA_{22})P_+ +{i \over 4}  G_{+ \alpha}{}^\a =0~,
\la{dfourb}
\eea
and
\be
 G_{+{\bar{\alpha}} {\bar{\beta}}}-{1 \over 2}
\epsilon_{{\bar{\alpha}}
{\bar{\beta}}}{}^{\g \d} G_{+ \g \d}=0~.
\la{dfivea}
\ee
\be
i G_{+{\bar{\alpha}} {\bar{\beta}}}+{i \over 2}
\epsilon_{{\bar{\alpha}}
{\bar{\beta}}}{}^{\g \d} G_{+ \g \d}=0~,
\la{dfiveb}
\ee
where we have set  $A=z^{-1}z^*$.

The Killing spinor equation associated
with the supercovariant derivative (\ref{kseqna}) also decomposes in $SU(4)$ representations.
In particular the conditions associated with ${\cal D}_\a$ are
\bea
(z^{-1}D_\a z)_{11}+i (z^{-1}D_\a z)_{12} +{1\over2}\Omega_{\a,\b}{}^\b+
{1\over2} \Omega_{\a,-+}
\cr
+i F_{\a-+\b}{}^\b
+{1\over4} (A_{11}+iA_{12}) G_{\a\b}{}^\b+
{1\over4} (A_{11}+iA_{12}) G_{\a-+}
=0~,
\la{aonea}
\eea
\bea
(z^{-1}D_\a z)_{21}+i (z^{-1}D_\a z)_{22}
+i[
 {1\over2}\Omega_{\a,\b}{}^\b
{1\over2} \Omega_{\a,-+}
\cr
+i F_{\a-+\b}{}^\b]
+ {1\over4} (A_{21}+iA_{22})[
G_{\a\b}{}^\b+
 G_{\a-+}]
=0~,
\la{aoneb}
\eea
\bea
 \Omega_{\a,\bar\b_1\bar\b_2}+i F_{\a\bar\b_1\bar\b_2\g}{}^\g
  +i F_{\a -+\bar\b_1\bar\b_2}+ {1\over2} (A_{11}+iA_{12}) G_{\a\bar\b_1\bar\b_2}
  +(B_{11}+iB_{12})\delta_{\a[\bar\b_1} P_{\bar\b_2]}
\cr
- {1\over2}
 [ \Omega_{\a,\g_1\g_2}+{1\over2} (A_{11}-iA_{12}) G_{\a\g_1\g_2}
 ] \epsilon^{\g_1\g_2}{}_{\bar\b_1\bar\b_2} =0~,
 \la{atwoa}
 \eea
 \bea
 i[\Omega_{\a,\bar\b_1\bar\b_2}+i F_{\a\bar\b_1\bar\b_2\g}{}^\g
  +i F_{\a -+\bar\b_1\bar\b_2}]+ {1\over2} (A_{21}+iA_{22}) G_{\a\bar\b_1\bar\b_2}
  +(B_{21}+iB_{22})\delta_{\a[\bar\b_1} P_{\bar\b_2]}
\cr
-
{1\over2} [-i \Omega_{\a,\g_1\g_2}+{1\over2} (A_{21}-iA_{22}) G_{\a\g_1\g_2}
 ] \epsilon^{\g_1\g_2}{}_{\bar\b_1\bar\b_2} =0~,
 \la{atwob}
 \eea
 \bea
 (z^{-1} D_\a z)_{11}-i (z^{-1} D_\a z)_{12}+{1\over2} (B_{11}-i B_{12}) P_\a
  -{1\over2} \Omega_{\a,\b}{}^\b+{1\over2} \Omega_{\a,-+}
\cr
-{1\over4} (A_{11}-iA_{12}) [G_{\a\b}{}^\b-G_{\a-+}]
+{i\over12} F_{\a\bar\b_1\bar\b_2\bar\b_3\bar\b_4}
 \epsilon^{\bar\b_1\bar\b_2\bar\b_3\bar\b_4}=0~,
  \la{athreea}
\eea
\bea
(z^{-1} D_\a z)_{21}-i (z^{-1} D_\a z)_{22}+{1\over2} (B_{21}-i B_{22}) P_\a
-i[ -{1\over2} \Omega_{\a,\b}{}^\b+{1\over2} \Omega_{\a,-+}]
\cr
-{1\over4} (A_{21}-iA_{22}) [G_{\a\b}{}^\b-G_{\a-+}]
-{1\over12} F_{\a\bar\b_1\bar\b_2\bar\b_3\bar\b_4}
 \epsilon^{\bar\b_1\bar\b_2\bar\b_3\bar\b_4}
=0~,
  \la{athreeb}
\eea
\bea
{1\over2} \Omega_{\a,+\bar\b}-{1\over 4} (B_{11}+iB_{12}) \delta_{\a\bar\b} P_++{1\over4} (A_{11}+iA_{12}) G_{\a+\bar\b}+
{i\over2} F_{\a+\bar\b\g}{}^\g
=0~,
\la{afoura}
\eea
\bea
[{i\over2} \Omega_{\a,+\bar\b}-{1\over 4} (B_{21}+iB_{22}) \delta_{\a\bar\b} P_++{1\over4} (A_{21}+iA_{22}) G_{\a+\bar\b}
-{1\over2} F_{\a+\bar\b\g}{}^\g]
=0~,
\la{afourb}
\eea
\bea
{i\over12} F_{\a+\bar\b_1\bar\b_2\bar\b_3}
+{1\over12}
 [{1\over2} \Omega_{\a,+\g}+{1\over4} (A_{11}-iA_{12})G_{\a+\g}] \epsilon^\g{}_{\bar\b_1\bar\b_2\bar\b_3} =0~,
\la{afivea}
\eea
\bea
-{1\over12} F_{\a+\bar\b_1\bar\b_2\bar\b_3}
+{1\over12}
 [-{i\over2} \Omega_{\a,+\g}+{1\over4} (A_{21}-i A_{22}) G_{\a+\g}] \epsilon^\g{}_{\bar\b_1\bar\b_2\bar\b_3} =0~,
\la{afiveb}
\eea
where $B=A^2$.

The conditions associated with  ${\cal D}_{\bar\a}$ are
\bea
(z^{-1}D_{\bar\a} z)_{11}+i (z^{-1}D_{\bar\a} z)_{12}+{1\over2} (B_{11}+i B_{12}) P_{\bar\a}
+{1\over2} \Omega_{\bar\a,\b}{}^\b+{1\over2} \Omega_{\bar\a, -+}
\cr
+{i\over12}  F_{\bar\a\g_1\g_2\g_2\g_4} \epsilon^{\g_1\g_2\g_3\g_4}
+{1\over4} (A_{11}+iA_{12}) [G_{\bar\a\b}{}^\b+ G_{\bar\a -+}]
=0~,
\la{abonea}
\eea
\bea
(z^{-1}D_{\bar\a} z)_{21}+i (z^{-1}D_{\bar\a} z)_{22}+{1\over2} (B_{21}+i B_{22}) P_{\bar\a}+
i[{1\over2} \Omega_{\bar\a,\b}{}^\b+{1\over2} \Omega_{\bar\a, -+}
\cr
-{i\over12}  F_{\bar\a\g_1\g_2\g_2\g_4} \epsilon^{\g_1\g_2\g_3\g_4}]
+{1\over4} (A_{21}+iA_{22}) [G_{\bar\a\b}{}^\b+G_{\bar\a -+}]
=0~,
\la{aboneb}
\eea
\bea
\Omega_{\bar\a, \bar\b_1\bar\b_2}+{1\over2} (A_{11}+i A_{12}) G_{\bar\a\bar\b_1\bar\b_2}
\cr
-{1\over2}[ \Omega_{\bar\a,\g_1\g_2}
-i F_{\bar\a\g_1\g_2\d}{}^\d+i F_{\bar\a-+\g_1\g_2}
\cr
+{1\over2} (A_{11}-i A_{12}) G_{\bar\a\g_1\g_2}+(B_{11}-i B_{12}) \delta_{\bar\a [\g_1} P_{\g_2]}]
\epsilon^{\g_1\g_2}{}_{\bar\b_1\bar\b_2}=0~,
\la{abtwoa}
\eea
\bea
i\Omega_{\bar\a, \bar\b_1\bar\b_2}+ {1\over2} (A_{21}+i A_{22}) G_{\bar\a \bar\b_1\bar\b_2}
\cr
+{i\over2}[\Omega_{\bar\a,\g_1\g_2}
-i F_{\bar\a\g_1\g_2\d}{}^\d+i F_{\bar\a-+\g_1\g_2}]
\epsilon^{\g_1\g_2}{}_{\bar\b_1\bar\b_2}
\cr
-{1\over 2} [{1\over2}(A_{21}-iA_{22}) G_{\bar\a\g_1\g_2}+
(B_{21}-iB_{22}) \delta_{\bar\a[\g_1} P_{\g_2]}]\epsilon^{\g_1\g_2}{}_{\bar\b_1\bar\b_2}=0~,
\la{abtwob}
\eea
\bea
(z^{-1}D_{\bar\a} z)_{11}-i(z^{-1}D_{\bar\a} z)_{12}
-{1\over2} \Omega_{\bar\a,\g}{}^\g+{1\over2} \Omega_{\bar\a,-+}
-i F_{\bar\a-+\g}{}^\g
\cr
-{1\over4} (A_{11}-i A_{12}) G_{\bar\a\g}{}^\g+{1\over4} (A_{11}-i A_{12})G_{\bar\a-+}=0~,
\la{abthreea}
\eea
\bea
(z^{-1}D_{\bar\a} z)_{21}-i(z^{-1}D_{\bar\a} z)_{22}-i
[-{1\over2} \Omega_{\bar\a,\g}{}^\g+{1\over2} \Omega_{\bar\a,-+}
-i F_{\bar\a-+\g}{}^\g]
\cr
-{1\over4} (A_{21}-iA_{22}) G_{\bar\a\g}{}^\g+{1\over4} (A_{21}-iA_{22})G_{\bar\a-+}=0~,
\la{abthreeb}
\eea
\bea
{1\over2} \Omega_{\bar\a,+\bar\b}+{1\over4} (A_{11}+i A_{12}) G_{\bar\a+\bar\b}
-{i\over6}  F_{\bar\a+\g_1\g_2\g_3} \epsilon^{\g_1\g_2\g_3}{}_{\bar\b}=0~,
\la{abfoura}
\eea
\bea
{i\over2} \Omega_{\bar\a,+\bar\b}+{1\over4} (A_{21}+i A_{22}) G_{\bar\a+\bar\b}
-{1\over6}  F_{\bar\a+\g_1\g_2\g_3} \epsilon^{\g_1\g_2\g_3}{}_{\bar\b}=0~,
\la{abfourb}
\eea
\bea
[{1\over2} \Omega_{\bar\a,+\g}
-{i\over2} F_{\bar\a+\g\d}{}^\d] \epsilon^\g{}_{\bar\b_1\bar\b_2\bar\b_3}
\cr
+{1\over4} (A_{11}-iA_{12})
G_{\bar\a+\g}\epsilon^\g{}_{\bar\b_1\bar\b_2\bar\b_3}-{1\over4} (B_{11}-iB_{12}) P_+\epsilon_{\bar\a\bar\b_1\bar\b_2\bar\b_3}=0~,
\la{abfivea}
\eea
\bea
 -i[{1\over2} \Omega_{\bar\a,+\g}
-{i\over2} F_{\bar\a+\g\d}{}^\d] \epsilon^\g{}_{\bar\b_1\bar\b_2\bar\b_3}
\cr
+{1\over4} (A_{21}-iA_{22}) G_{\bar\a+\g}\epsilon^\g{}_{\bar\b_1\bar\b_2\bar\b_3}
-{1\over4} (B_{21}-iB_{22}) P_+ \epsilon_{\bar\a\bar\b_1\bar\b_2\bar\b_3}=0~.
\la{abfiveb}
\eea

The conditions associated with  ${\cal D}_-$ are
\bea
(z^{-1}D_-z)_{11}+i (z^{-1}D_-z)_{12}
+{1\over2} \Omega_{-,\g}{}^\g+{1\over2} \Omega_{-,-+}
+{i\over4} F_{-\g}{}^\g{}_\d{}^\d
\cr
+{i\over12}  F_{-\g_1\g_2\g_3\g_4} \epsilon^{\g_1\g_2\g_3\g_4}
+{1\over4} (A_{11}+iA_{12})G_{-\g}{}^\g
=0~,
\la{monea}
\eea
\bea
(z^{-1}D_-z)_{21}+i (z^{-1}D_-z)_{22}+i
[{1\over2} \Omega_{-,\g}{}^\g+{1\over2} \Omega_{-,-+}
+{i\over4} F_{-\g}{}^\g{}_\d{}^\d]
\cr
+{1\over4} (A_{21}+i A_{22}) G_{-\g}{}^\g
+{1\over12}  F_{-\g_1\g_2\g_3\g_4} \epsilon^{\g_1\g_2\g_3\g_4}=0~,
\la{moneb}
\eea
\bea
\Omega_{-,\bar\b_1\bar\b_2}+i F_{-\bar\b_1\bar\b_2\g}{}^\g+{1\over2}(A_{11}+iA_{12})G_{-\bar\b_1\bar\b_2}
\cr
-{1\over2} [ \Omega_{-,\g_1\g_2}
-i F_{-\g_1\g_2\d}{}^\d] \epsilon^{\g_1\g_2}{}_{\bar\b_1\bar\b_2}
-{1\over4} (A_{11}-iA_{12}) G_{-\g_1\g_2} \epsilon^{\g_1\g_2}{}_{\bar\b_1\bar\b_2}
=0~,
\la{mtwoa}
\eea
\bea
i[\Omega_{-,\bar\b_1\bar\b_2}+i F_{-\bar\b_1\bar\b_2\g}{}^\g]+{1\over2} (A_{21}+i A_{22}) G_{-\bar\b_1\bar\b_2}
\cr
+{i\over2} [ \Omega_{-,\g_1\g_2}
-i F_{-\g_1\g_2\d}{}^\d] \epsilon^{\g_1\g_2}{}_{\bar\b_1\bar\b_2}-{1\over4} (A_{21}-i A_{22}) G_{-\g_1\g_2}
\epsilon^{\g_1\g_2}{}_{\bar\b_1\bar\b_2}
=0~,
\la{mtwob}
\eea
\bea
(z^{-1}D_- z)_{11}-i (z^{-1}D_- z)_{12}-{1\over2} \Omega_{-,\g}{}^\g+{1\over2}\Omega_{-,-+}
+{i\over4} F_{-\g}{}^\g{}_\d{}^\d -{1\over4} (A_{11}-iA_{12}) G_{-\g}{}^\g
\cr
+{i\over 12}  F_{-\bar\b_1\bar\b_2\bar\b_3\bar\b_4}
\epsilon^{\bar\b_1\bar\b_2\bar\b_3\bar\b_4}=0,~~~
\la{mthreea}
\eea
\bea
(z^{-1}D_- z)_{21}-i (z^{-1}D_- z)_{22}
-i [-{1\over2} \Omega_{-,\g}{}^\g+{1\over2}\Omega_{-,-+}
+{i\over4} F_{-\g}{}^\g{}_\d{}^\d]-{1\over4}  (A_{21}-iA_{22})G_{-\g}{}^\g
\cr
-{1\over 12}  F_{-\bar\b_1\bar\b_2\bar\b_3\bar\b_4}
\epsilon^{\bar\b_1\bar\b_2\bar\b_3\bar\b_4}=0,~~~
\la{mthreeb}
\eea
\bea
{1\over2} \Omega_{-,+\bar\b}+{i\over2} F_{-+\bar\b\g}{}^\g+{1\over4} (A_{11}+iA_{12})G_{-+\bar\b}
+{1\over4} (B_{11}+iB_{12}) P_{\bar\b}
\cr
-{i\over6}  F_{-+\g_1\g_2\g_3} \epsilon^{\g_1\g_2\g_3}{}_{\bar\b}=0~,
\la{mfoura}
\eea
\bea
i[{1\over2} \Omega_{-,+\bar\b}+{i\over2} F_{-+\bar\b\g}{}^\g]+{1\over4} (A_{21}+iA_{22})G_{-+\bar\b}
+{1\over4} (B_{21}+iB_{22})P_{\bar\b}
\cr
-{1\over6}  F_{-+\g_1\g_2\g_3} \epsilon^{\g_1\g_2\g_3}{}_{\bar\b}=0~,
\la{mfourb}
\eea
\bea
i F_{-+\bar\b_1\bar\b_2\bar\b_3}
+
[{1\over2} \Omega_{-,+\g}-{i\over2}
F_{-+\g\d}{}^\d] \epsilon^\g{}_{\bar\b_1\bar\b_2\bar\b_3}
\cr
+{1\over4} [(A_{11}-iA_{12}) G_{-+\g}+
(B_{11}-iB_{12}) P_\g]\epsilon^\g{}_{\bar\b_1\bar\b_2\bar\b_3}
=0~.
\la{mfivea}
\eea
\bea
- F_{-+\bar\b_1\bar\b_2\bar\b_3}
-i
[{1\over2} \Omega_{-,+\g}-{i\over2}
F_{-+\g\d}{}^\d] \epsilon^\g{}_{\bar\b_1\bar\b_2\bar\b_3}
\cr
+{1\over4} [(A_{21}-iA_{22})G_{-+\g}+(B_{21}-iB_{22}) P_\g] \epsilon^\g{}_{\bar\b_1\bar\b_2\bar\b_3}
=0~.
\la{mfiveb}
\eea

The conditions associated  with ${\cal D}_+$  are
\bea
(z^{-1}D_+ z)_{11}+i (z^{-1}D_+ z)_{12}+{1\over2} (B_{11}+iB_{12}) P_+ +{1\over2} \Omega_{+,\g}{}^\g
\cr
+{1\over2} \Omega_{+,-+}+ {1\over4} (A_{11}+iA_{12})  G_{+\g}{}^\g
=0~,
\la{ponea}
\eea
\bea
(z^{-1}D_+ z)_{21}+i (z^{-1}D_+ z)_{22}+{1\over2} (B_{21}+iB_{22}) P_+ +i
[{1\over2} \Omega_{+,\g}{}^\g
\cr
+{1\over2} \Omega_{+,-+}]
+{1\over4} (A_{21}+iA_{22}) G_{+\g}{}^\g=0~,
\la{poneb}
\eea
\bea
\Omega_{+,\bar\b_1\bar\b_2}+{1\over2} (A_{11}+iA_{12}) G_{+\bar\b_1\bar\b_2}
\cr
-{1\over2} \Omega_{+,\g_1\g_2}
 \epsilon^{\g_1\g_2}{}_{\bar\b_1\bar\b_2}-{1\over4} (A_{11}-iA_{12})
G_{+\g_1\g_2} \epsilon^{\g_1\g_2}{}_{\bar\b_1\bar\b_2}
=0~,
\la{ptwoa}
\eea
\bea
i\Omega_{+,\bar\b_1\bar\b_2}+{1\over2} (A_{21}+iA_{22}) G_{+\bar\b_1\bar\b_2}
\cr
+{i\over2} \Omega_{+,\g_1\g_2}
 \epsilon^{\g_1\g_2}{}_{\bar\b_1\bar\b_2}-{1\over4} (A_{21}-i A_{22}) G_{+\g_1\g_2}
\epsilon^{\g_1\g_2}{}_{\bar\b_1\bar\b_2}
=0~,
\la{ptwob}
\eea
\bea
(z^{-1}D_+ z)_{11}-i(z^{-1}D_+ z)_{12}+{1\over2} (B_{11}-iB_{12}) P_+
-{1\over2} \Omega_{+,\g}{}^\g
\cr
+{1\over2} \Omega_{+,-+}
 -{1\over4} (A_{11}-iA_{12}) G_{+\g}{}^\g
=0~,
 \la{pthreea}
 \eea
 \bea
 (z^{-1}D_+ z)_{21}-i(z^{-1}D_+ z)_{22}+{1\over2} (B_{21}-iB_{22}) P_+
-i[-{1\over2} \Omega_{+,\g}{}^\g
\cr
+{1\over2} \Omega_{+,-+}
]
-{1\over4} (A_{21}-i A_{22}) G_{+\g}{}^\g
=0~,
 \la{pthreeb}
 \eea
 and
 \bea
 \Omega_{+,+\a}=\Omega_{+,+\bar\a}=0~.
 \la{pfour1}
 \eea
As we have already mentioned, all the equations that arise from the
Killing spinor equations are linear in the fluxes, geometry and the first derivatives
of the functions  $z$ that determine the Killing spinors.
The system may appear involved but it can be solved. It also simplifies in some special
cases, like for example whenever $z$ is a real matrix. In this case $A=B=1$ and so the terms in the linear
system above that contain
the fluxes and geometry do not depend on the functions $z$.

\subsection{The solution to the linear system}

We shall first solve the last six equations of the linear system associated with the algebraic
Killing spinor equation. Equations ({\ref{dthreea}})-({\ref{dfourb}}) imply that
\be
G_{+\a}{}^\a=P_+=0~,
\la{sone}
\ee
and ({\ref{dfivea}}), ({\ref{dfiveb}}) imply that
\be
G_{+ \alpha \beta}=G_{+ \bar{\alpha} \bar{\beta}}=0 \ .
\la{stwo}
\ee

Next consider the equations ({\ref{afivea}}), ({\ref{afiveb}}), ({\ref{abfoura}}),
({\ref{abfourb}}). These imply

\be
\Omega_{\alpha, \beta +}=0, \quad F_{+ \alpha {\bar{\beta}}_1 {\bar{\beta}}_2 {\bar{\beta}}_3}=0
\la{sthree}
\ee

and ({\ref{ptwoa}}), ({\ref{ptwob}}) and (\ref{pfour1}) imply that

\be
\Omega_{+, \alpha \beta}=0 \,, \quad  \Omega_{+,+\a}=0 \ .
\la{sfour}
\ee

The remaining components of the ${\cal D}_+$ equations ({\ref{ponea}}), ({\ref{poneb}}),
({\ref{pthreea}}), ({\ref{pthreeb}}) imply that

\bea
\label{pluscomp}
(z^{-1}D_+z)_{11} &=& (z^{-1}D_+z)_{22} = -{1 \over 2} \Omega_{+,-+}
\cr
(z^{-1}\partial_+z)_{12} &=& - (z^{-1}\partial_+z)_{21} = {i \over 2} \Omega_{+, \alpha}{}^\alpha \ .
\eea

The equations ({\ref{monea}}), ({\ref{moneb}}), ({\ref{mthreea}}), ({\ref{mthreeb}})
constrain $z^{-1} D_-z$ via

\bea
\label{minuscomp}
(z^{-1}D_-z)_{11} &=& -{1 \over 2} \Omega_{-,-+}-{i \over 4}F_{- \alpha}{}^\alpha{}_\beta{}^\beta
-{i \over 24} (F_{- \alpha_1 \alpha_2 \alpha_3 \alpha_4} \epsilon^{\alpha_1 \alpha_2 \alpha_3 \alpha_4}
\cr
&+& F_{- {\bar{\alpha}}_1 {\bar{\alpha}}_2 {\bar{\alpha}}_3 {\bar{\alpha}}_4}
\epsilon^{{\bar{\alpha}}_1 {\bar{\alpha}}_2 {\bar{\alpha}}_3 {\bar{\alpha}}_4})
-{i \over 4}A_{12}G_{- \alpha}{}^\alpha
\cr
(z^{-1}D_-z)_{22} &=& -{1 \over 2} \Omega_{-,-+}-{i \over 4}F_{- \alpha}{}^\alpha{}_\beta{}^\beta
+{i \over 24} (F_{- \alpha_1 \alpha_2 \alpha_3 \alpha_4} \epsilon^{\alpha_1 \alpha_2 \alpha_3 \alpha_4}
\cr
&+& F_{- {\bar{\alpha}}_1 {\bar{\alpha}}_2 {\bar{\alpha}}_3 {\bar{\alpha}}_4}
\epsilon^{{\bar{\alpha}}_1 {\bar{\alpha}}_2 {\bar{\alpha}}_3 {\bar{\alpha}}_4})
+{i \over 4}A_{21}G_{- \alpha}{}^\alpha
\cr
(z^{-1}\partial_- z)_{12}&=&{i \over 2} \Omega_{-,\alpha}{}^\alpha -{1 \over 24}
((F_{- \alpha_1 \alpha_2 \alpha_3 \alpha_4} \epsilon^{\alpha_1 \alpha_2 \alpha_3 \alpha_4}
- F_{- {\bar{\alpha}}_1 {\bar{\alpha}}_2 {\bar{\alpha}}_3 {\bar{\alpha}}_4}
\epsilon^{{\bar{\alpha}}_1 {\bar{\alpha}}_2 {\bar{\alpha}}_3 {\bar{\alpha}}_4})
\cr
&+&{i \over 4}A_{11}G_{- \alpha}{}^\alpha
\cr
(z^{-1}\partial_- z)_{21}&=&-{i \over 2} \Omega_{-,\alpha}{}^\alpha -{1 \over 24}
((F_{- \alpha_1 \alpha_2 \alpha_3 \alpha_4} \epsilon^{\alpha_1 \alpha_2 \alpha_3 \alpha_4}
- F_{- {\bar{\alpha}}_1 {\bar{\alpha}}_2 {\bar{\alpha}}_3 {\bar{\alpha}}_4}
\epsilon^{{\bar{\alpha}}_1 {\bar{\alpha}}_2 {\bar{\alpha}}_3 {\bar{\alpha}}_4})
\cr
&-&{i \over 4}A_{22}G_{- \alpha}{}^\alpha \ .
\eea

And from ({\ref{mtwoa}}) and ({\ref{mtwob}}), we find

\bea
G_{-{\bar{\alpha}}_1 {\bar{\alpha}}_2} &=& -{(A_{11}+A_{22}
+i(A_{21}-A_{12})) \over A_{11}A_{22}-A_{12}A_{21}} (\Omega_{-,{\bar{\alpha}}_1 {\bar{\alpha}}_2}+iF_{-{\bar{\alpha}}_1
{\bar{\alpha}}_2 \gamma}{}^\gamma)
\cr
&+&{(A_{22}-A_{11}+i(A_{12}+A_{21})) \over 2(A_{11}A_{22}-A_{12}A_{21})} (\Omega_{-,\beta_1 \beta_2}
-iF_{-\beta_1 \beta_2 \gamma}{}^\gamma) \epsilon_{{\bar{\alpha}}_1 {\bar{\alpha}}_2}{}^{\beta_1 \beta_2}
\cr
G_{- \alpha_1 \alpha_2} &=& -{(A_{11}+A_{22}-i(A_{21}-A_{12})) \over A_{11}A_{22}-A_{12}A_{21}}
(\Omega_{-,\alpha_1 \alpha_2}-iF_{-\alpha_1 \alpha_2 \gamma}{}^\gamma)
\cr
&-&{(A_{11}-A_{22}+i(A_{12}+A_{21})) \over 2(A_{11}A_{22}-A_{12}A_{21})}(\Omega_{-,{\bar{\beta}}_2
{\bar{\beta}}_2}+iF_{-{\bar{\beta}}_1 {\bar{\beta}}_2\g}{}^\g) \epsilon_{\alpha_1 \alpha_2}{}^{{\bar{\beta}}_1
{\bar{\beta}}_2} \ .
\la{sfive}
\eea

In order to proceed we shall consider two separate cases,
according as $(A_{11}-A_{22})^2 + (A_{12}+A_{21})^2$ vanishes or not.
Observe that $(A_{11}-A_{22})^2 + (A_{12}+A_{21})^2=0$ is equivalent to
$A_{11}=A_{22}$ and $A_{12}+A_{21}=0$.

First we shall assume that $(A_{11}-A_{22})^2 + (A_{12}+A_{21})^2 \neq 0$. Then
({\ref{afoura}}), ({\ref{afourb}}), ({\ref{abfivea}}),
({\ref{abfiveb}}) imply that

\be
G_{+ \alpha {\bar{\beta}}} = \Omega_{\alpha, {\bar{\beta}} +} = 0
\la{ssix}
\ee
and
\be
F_{+ \alpha_1 \alpha_2 {\bar{\beta}}_1 {\bar{\beta}}_2} =0 \ .
\la{sseven}
\ee

Next consider the equations which have one free holomorphic index,
excluding for the moment those equations involving $z^{-1}Dz$; these are
({\ref{dtwoa}}), ({\ref{dtwob}}), the trace of the duals of ({\ref{atwoa}}), ({\ref{atwob}})
and the duals of ({\ref{mfivea}}), ({\ref{mfiveb}}).
These fix $P_\alpha$, $\xonet$, $\xtwot$, $\xthreet$, $\xfourt$, $\xfivet$ in terms of the
components of the spin connection $\yonet$, $\ytwot$, $\ythreet$ and $A_{ij}$.

The corresponding equations with one free antiholomorphic constraint are
({\ref{donea}}), ({\ref{dtwob}}), the traces of ({\ref{atwoa}}), ({\ref{atwob}})
and ({\ref{mfoura}}), ({\ref{mfourb}}). These fix $P_{\bar{\alpha}}$,
$\xone$, $\xtwo$, $\xthree$, $\xfour$, $\xfive$ in terms of the
components of the spin connection $\yone$, $\ytwo$, $\ythree$.

By comparing the complex expressions for $\xfourt$, $\xfivet$
from the former equations
with the complex conjugates of the expressions $\xfour$, $\xfive$
from the latter, we find the following geometric constraints

\bea
\yone &=& -{(6 A_{11}A_{22}-4 A_{12}A_{21}-A_{12}{}^2 -A_{21}{}^2) \over
(A_{11}{}^2+A_{22}{}^2+4 A_{11}A_{22}-2A_{12}A_{21})} \ythree
\cr
\ytwo &=& -2i{(A_{12}-A_{21})(A_{11}-A_{22}+i(A_{12}+A_{21})) \over
(A_{11}{}^2+A_{22}{}^2+4 A_{11}A_{22}-2A_{12}A_{21})} \ythree \ .
\la{seight}
\eea

Using these constraints, we obtain the following simplifications

\bea
P_{\bar{\alpha}} &=& {4 \over (A_{11}{}^2+A_{22}{}^2+4 A_{11}A_{22}-2A_{12}A_{21})} \ythree
\cr
\xone &=& -8{(A_{11}+A_{22}) \over (A_{11}{}^2+A_{22}{}^2+4 A_{11}A_{22}-2A_{12}A_{21})} \ythree
\cr
\xtwo &=& -8i{(A_{12}-A_{21}) \over (A_{11}{}^2+A_{22}{}^2+4 A_{11}A_{22}-2A_{12}A_{21})} \ythree
\cr
\xthree &=& -24{(A_{11}-A_{22}+i(A_{12}+A_{21})) \over (A_{11}{}^2+A_{22}{}^2+4 A_{11}A_{22}-2A_{12}A_{21})}
\ythree
\la{snine}
\eea

and

\bea
P_\alpha &=& {4 \over (A_{11}{}^2+A_{22}{}^2+4 A_{11}A_{22}-2A_{12}A_{21})} \ythreet
\cr
\xonet &=& -8{(A_{11}+A_{22}) \over (A_{11}{}^2+A_{22}{}^2+4 A_{11}A_{22}-2A_{12}A_{21})} \ythreet
\cr
\xtwot &=& -8i{(A_{12}-A_{21}) \over (A_{11}{}^2+A_{22}{}^2+4 A_{11}A_{22}-2A_{12}A_{21})} \ythreet
\cr
\xthreet &=& -24{(A_{11}-A_{22}-i(A_{12}+A_{21})) \over (A_{11}{}^2+A_{22}{}^2+4 A_{11}A_{22}-2A_{12}A_{21})}
\la{sten}
\eea

and
\bea
\xfour &=& -3i {(A_{11}+A_{22})(A_{11}-A_{22}+i(A_{12}+A_{21})) \over
(A_{11}{}^2+A_{22}{}^2+4 A_{11}A_{22}-2A_{12}A_{21})} \ythree
\cr
\xfive &=& {(A_{11}+A_{22})(A_{12}-A_{21})\over (A_{11}{}^2+A_{22}{}^2+4 A_{11}A_{22}-2A_{12}A_{21})} \ .
\ythree
\la{seleven}
\eea

Substituting these constraints back into ({\ref{abonea}}), ({\ref{aboneb}}), ({\ref{abthreea}})
and ({\ref{abthreeb}}) we find the following constraints on $z^{-1} D_{\bar{\alpha}} z$:

\bea
\label{genericcon1}
(z^{-1} D_{\bar{\alpha}} z)_{11} &=& -{1 \over 2} \yfour
\cr
&-&{1 \over 2} {(A_{22}{}^2-3A_{11}{}^2+4A_{12}{}^2-4 A_{11}A_{22}-2A_{12}A_{21})
\over  ((A_{11}+A_{22})^2+2(A_{11}A_{22}-A_{12}A_{21}))} \ythree
\cr
(z^{-1} D_{\bar{\alpha}} z)_{22} &=&  -{1 \over 2} \yfour
\cr
&+&{1 \over 2} {(3 A_{22}{}^2-A_{11}{}^2-4A_{21}{}^2+4A_{11}A_{22}+2A_{12}A_{21})
\over  ((A_{11}+A_{22})^2+2(A_{11}A_{22}-A_{12}A_{21}))} \ythree
\cr
(z^{-1} \partial_{\bar{\alpha}} z)_{12} &=& {i \over 2} \yfive
\cr
&+&{i \over 2} ((A_{11}+A_{22})^2+2(A_{11}A_{22}-A_{12}A_{21}))^{-1}
(A_{11}{}^2+A_{22}{}^2-2iA_{12}A_{22}
\cr
&-&2iA_{21}A_{22}-6iA_{11}A_{12}
+2iA_{11}A_{21}+2A_{12}A_{21})\ythree
\cr
(z^{-1} \partial_{\bar{\alpha}} z)_{21} &=& -{i \over 2} \yfive
\cr
&-&{i \over 2} ((A_{11}+A_{22})^2+2(A_{11}A_{22}-A_{12}A_{21}))^{-1}
(A_{11}{}^2+A_{22}{}^2+6iA_{21}A_{22}
\cr
&-&2iA_{12}A_{22}
+2iA_{11}A_{12}+2iA_{11}A_{21}+2A_{12}A_{21})  \ythree \ .
\eea

And from  ({\ref{aonea}}), ({\ref{aoneb}}), ({\ref{athreea}})
and ({\ref{athreeb}}) we find the following constraints on $z^{-1} D_\alpha z$:

\bea
\label{genericcon2}
(z^{-1}D_\alpha z)_{11} &=& -{1 \over 2} \yfourt
\cr
&+&{1 \over 2} {(3A_{11}{}^2 -A_{22}{}^2 -4A_{12}{}^2+4A_{11}A_{22}+2A_{12}A_{21})
\over ((A_{11}+A_{22})^2+2(A_{11}A_{22}-A_{12}A_{21}))} \ythreet
\cr
(z^{-1}D_\alpha z)_{22} &=& -{1 \over 2} \yfourt
\cr
&-&{1 \over 2}{(A_{11}{}^2-3A_{22}{}^2 +4A_{21}{}^2-4A_{11}A_{22}-2A_{12}A_{21})
\over ((A_{11}+A_{22})^2+2(A_{11}A_{22}-A_{12}A_{21}))} \ythreet
\cr
(z^{-1} \partial_\alpha z)_{12} &=& {i \over 2} \yfivet
\cr
&-&{i \over 2}  ((A_{11}+A_{22})^2+2(A_{11}A_{22}-A_{12}A_{21}))^{-1}
(A_{11}{}^2+A_{22}{}^2-2iA_{11}A_{21}
\cr
&+&6i A_{11}A_{12}+2iA_{21}A_{22}+2iA_{12}A_{22}+2A_{12}A_{21})  \ythreet
\cr
(z^{-1} \partial_\alpha z)_{21} &=& -{i \over 2} \yfivet
\cr
&+&{i \over 2}  ((A_{11}+A_{22})^2+2(A_{11}A_{22}-A_{12}A_{21}))^{-1}
(A_{11}{}^2+A_{22}{}^2-2iA_{11}A_{21}
\cr
&-&2i A_{11}A_{12}+2iA_{12}A_{22}-6iA_{21}A_{22}+2A_{12}A_{21})  \ythreet \ .
\eea

Note that by means of an appropriate $U(1)$ transformation, one can work in a gauge
for which $\det z = \det z^*$, so that $\det A=1$. Then
({\ref{genericcon1}}), ({\ref{genericcon2}}) and ({\ref{pluscomp}})
imply that

\be
Q_+ = Q_\alpha =0
\la{stwelve}
\ee
though in general ({\ref{minuscomp}}) does not constrain $Q_-$ to vanish.

Finally, we consider the equations ({\ref{atwoa}}), ({\ref{atwob}}), ({\ref{abtwoa}}), ({\ref{abtwob}}).
The constraints obtained from taking traces of these equations have already been obtained;
the remaining constraints consist of the fixing of two components of the $G$-flux via

\bea
G_{\alpha {\bar{\beta}}_1 {\bar{\beta}}_2} &=& {2 \over A_{11}-A_{22}+i(A_{12}+A_{21})}
\Omega_{\alpha, \gamma_1 \gamma_2} \epsilon_{ {\bar{\beta}}_1 {\bar{\beta}}_2}{}^{\gamma_1 \gamma_2}
\cr
&-& {8i (A_{21}-A_{12}) \over (A_{11}+A_{22})^2+2(A_{11}A_{22}-A_{12}A_{21})} \Omega_{-,+[{\bar{\beta}}_1}
\delta_{{\bar{\beta}}_2] \alpha}
\cr
G_{{\bar{\alpha}} \gamma_1 \gamma_2} &=& {2 \over A_{11}-A_{22}-i(A_{12}+A_{21})} \Omega_{{\bar{\alpha}},
{\bar{\beta}}_1 {\bar{\beta}}_2} \epsilon^{{\bar{\beta}}_1 {\bar{\beta}}_2}{}_{\gamma_1 \gamma_2}
\cr
&-&{8i (A_{12}-A_{21}) \over (A_{11}+A_{22})^2+2(A_{11}A_{22}-A_{12}A_{21})} \Omega_{-,+[\gamma_1}
\delta_{\gamma_2] {\bar{\alpha}}}
\la{sthirteen}
\eea

together with the fixing of a component of the $F$-flux

\bea
F_{-+ \alpha  {\bar{\beta}}_1 {\bar{\beta}}_2} &=& (A_{11}+A_{22})
\big( {i \over 4(A_{11}-A_{22}+i(A_{12}+A_{21}))} \Omega_{\alpha, \gamma_1 \gamma_2}
\epsilon^{\gamma_1 \gamma_2}{}_{ {\bar{\beta}}_1 {\bar{\beta}}_2}
\cr
&-&{(A_{12}-A_{21}) \over  (A_{11}+A_{22})^2+2(A_{11}A_{22}-A_{12}A_{21})} \Omega_{-,+[{\bar{\beta}}_1}
\delta_{{\bar{\beta}}_2]\alpha} \big)
\la{sfourteen}
\eea

and a geometric constraint

\bea
\Omega_{\alpha, {\bar{\beta}}_1 {\bar{\beta}}_2} +{i (A_{12}-A_{21}) \over
2(A_{11}-A_{22}+i(A_{12}+A_{21}))}  \Omega_{\alpha, \gamma_1 \gamma_2}
\epsilon^{\gamma_1 \gamma_2}{}_{ {\bar{\beta}}_1 {\bar{\beta}}_2}
\cr
- {4(A_{11}A_{22}-A_{12}A_{21}) \over  (A_{11}+A_{22})^2+2(A_{11}A_{22}-A_{12}A_{21})} \Omega_{-,+[{\bar{\beta}}_1}
\delta_{{\bar{\beta}}_2]\alpha}=0 \ .
\la{sfifteen}
\eea

Next we consider the special case when $(A_{11}-A_{22})^2 + (A_{12}+A_{21})^2 =0$. Then
({\ref{afoura}}), ({\ref{afourb}}), ({\ref{abfivea}}),
({\ref{abfiveb}}) imply that

\bea
\label{special1}
G_{+ \alpha {\bar{\beta}}} &=& {1 \over A_{11}}(\Omega_{\alpha,+ {\bar{\beta}}}
- \Omega_{{\bar{\beta}},+ \alpha})
\eea
\bea
\label{special2}
F_{+ \alpha {\bar{\beta}} \gamma}{}^\gamma &=& -{i \over 2 A_{11}}
((A_{11}-iA_{12})\Omega_{\alpha,+ {\bar{\beta}}}+(A_{11}+iA_{12})
\Omega_{{\bar{\beta}},+ \alpha}) \ .
\eea

On comparing ({\ref{special2}}) with its complex conjugate, the geometric constraint
\be
\Omega_{\alpha,+ {\bar{\beta}}}+ \Omega_{{\bar{\beta}}, + \alpha}=0
\ee
is obtained. Substituting this back into ({\ref{special1}}) and ({\ref{special2}})
we obtain some simplification:

\bea
G_{+ \alpha {\bar{\beta}}} &=& {2 \over A_{11}}\Omega_{\alpha,+ {\bar{\beta}}}
\cr
F_{+ \alpha {\bar{\beta}} \gamma}{}^\gamma &=& -{A_{12} \over A_{11}} \Omega_{\alpha,+ {\bar{\beta}}} \ .
\eea

Next consider the equations which have one free holomorphic index,
excluding for the moment those equations involving $z^{-1}Dz$; these are
({\ref{dtwoa}}), ({\ref{dtwob}}), the trace of the duals of ({\ref{atwoa}}), ({\ref{atwob}})
and the duals of ({\ref{mfivea}}), ({\ref{mfiveb}}).
Unlike the generic case, these do not
fix $P_\alpha$, $\xonet$, $\xtwot$, $\xthreet$, $\xfourt$, $\xfivet$ uniquely in terms of the
components of the spin connection $\yonet$, $\ytwot$, $\ythreet$ and $A_{ij}$.
Rather, $\xtwo$, $\xfive$ and $\yone$ are arbitrary with

\bea
P_{\bar{\alpha}} &=& -{1 \over 3(A_{11}+iA_{12})} \xtwo -{2 \over 3 (A_{11}+iA_{12})^2}
(\yone-i\xfive)
\cr
\xone &=& {1 \over 3} \xtwo +{8 \over 3(A_{11}+iA_{12})}(\yone-i\xfive)
\cr
G_{\alpha_1 \alpha_2 \alpha_3} &=&0
\cr
F_{-+ \alpha_1 \alpha_2 \alpha_3}&=&0
\cr
\Omega_{[\alpha_1, \alpha_2 \alpha_3]} &=&0
\cr
\ythree &=& - \yone \ .
\eea

The corresponding equations with one free antiholomorphic constraint are
({\ref{donea}}), ({\ref{dtwob}}), the traces of ({\ref{atwoa}}), ({\ref{atwob}})
and ({\ref{mfoura}}), ({\ref{mfourb}}). Taking $\xtwot$, $\xfivet$ and $\yonet$
to be arbitrary we find the additional constraints

\bea
P_\alpha &=& {1 \over 3(A_{11}-iA_{12})} \xtwot -{2 \over 3(A_{11}-iA_{12})^2}(\yonet
+i\xfivet)
\cr
\xonet &=& -{1 \over 3} \xtwot +{8 \over 3 (A_{11}-iA_{12})}(\yonet+i\xfivet)
\cr
G_{{\bar{\alpha}}_1 {\bar{\alpha}}_2 {\bar{\alpha}}_3} &=&0 \ .
\la{spfour}
\eea

Substituting these constraints back into ({\ref{abonea}}), ({\ref{aboneb}}), ({\ref{abthreea}})
and ({\ref{abthreeb}}) we find the following constraints on $z^{-1} D_{\bar{\alpha}} z$:

\bea
\label{specialder1}
(z^{-1}D_{\bar{\alpha}}z)_{11}&=& (z^{-1}D_{\bar{\alpha}}z)_{22} =
-{i \over 6}A_{12} \xtwo -{1 \over 2} \yfour
\cr
&+&{1 \over 6 (A_{11}+iA_{12})}(-3A_{11}+iA_{12}) \yone
\cr
&+&{i \over 3 (A_{11}+iA_{12})}(3A_{11}+iA_{12}) \xfive
\cr
(z^{-1} \partial_{\bar{\alpha}} z)_{12}&=&-(z^{-1} \partial_{\bar{\alpha}} z)_{21}
= {i \over 2} \yfive +{i \over 6} A_{11} \xtwo
\cr
&+&{i \over 6(A_{11}+iA_{12})}
(-A_{11}+3iA_{12}) \yone
\cr
&-&{2 A_{11} \over 3(A_{11}+iA_{12})} \xfive \ .
\eea

And from  ({\ref{aonea}}), ({\ref{aoneb}}), ({\ref{athreea}})
and ({\ref{athreeb}}) we find the following constraints on $z^{-1} D_\alpha z$:

\bea
\label{specialder2}
(z^{-1}D_\alpha z)_{11} &=& (z^{-1}D_\alpha z)_{22}=-{1 \over 2} \yfourt
-{i \over 6} A_{12} \xtwot
\cr
&-&{1 \over 6(A_{11}-iA_{12})}(3A_{11}+iA_{12}) \yonet
\cr
&+&{i \over 3 (A_{11}-iA_{12})}(-3A_{11}+iA_{12}) \xfivet
\cr
(z^{-1} \partial_\alpha z)_{12}&=&-(z^{-1} \partial_\alpha z)_{21}
={i \over 2} \yfivet  +{i \over 6}A_{11} \xtwot
\cr
&+&{i \over 6(A_{11}-iA_{12})}(A_{11}+3iA_{12}) \yonet
\cr
&-&{2 A_{11} \over 3(A_{11}-iA_{12})} \xfivet \ .
\eea

Finally, we consider the equations ({\ref{atwoa}}), ({\ref{atwob}}), ({\ref{abtwoa}}), ({\ref{abtwob}}).
These fix two components of the $G$-flux via

\bea
G_{\alpha {\bar{\beta}}_1 {\bar{\beta}}_2} &=& -{2 \over A_{11}+iA_{12}}
(\Omega_{\alpha, {\bar{\beta}}_1 {\bar{\beta}}_2}+2iF_{-+ \alpha {\bar{\beta}}_1 {\bar{\beta}}_2})
\cr
&-& {1 \over A_{11}+iA_{12}}\delta_{\alpha [{\bar{\beta}}_1}
\big( -{2 \over 3} (A_{11}+iA_{12}) G_{{\bar{\beta}}_2] \gamma}{}^\gamma
-{4 \over 3} \Omega_{|\gamma|,}{}^\gamma{}_{{\bar{\beta}}_2]}
-{8i \over 3} F_{{\bar{\beta}}_2] -+\gamma}{}^\gamma \big)
\cr
G_{{\bar{\alpha}} \gamma_1 \gamma_2} &=& -{2 \over A_{11}-iA_{12}}
(\Omega_{{\bar{\alpha}}, \gamma_1 \gamma_2}+2iF_{-+{\bar{\alpha}} \gamma_1 \gamma_2})
\cr
&-&{1 \over A_{11}-iA_{12}} \delta_{{\bar{\alpha}}[\gamma_1}
\big( {2 \over 3}(A_{11}-iA_{12}) G_{\gamma_2]\beta}{}^\beta-{4 \over 3}
\Omega_{|{\bar{\beta}}|,}{}^{\bar{\beta}}{}_{\gamma_2]}
+{8i \over 3}F_{\gamma_2]-+\beta}{}^\beta \big)
\eea

together with the geometric constraint

\be
\Omega_{\alpha_1,\alpha_2 \alpha_3}=0 \ .
\la{spfive}
\ee

 As the first order equations in spacetime derivatives of $z$ are non-linear, we shall not
 investigate the general case here. Instead we shall focus on two examples that illustrate some of the properties
 of this non-linear system.

\subsection{Special cases} \la{coconapsp}

We shall first consider the case that $z$ is diagonal with complex entries.
We have seen that it can be arranged such that the Killing spinors can be written as
$\epsilon_1=\rho e^{i\varphi}\eta_1$ and $\epsilon_1=\rho^{-1} e^{-i\varphi}\eta_2$.
Using this,
the equations ({\ref{genericcon1}}) and ({\ref{genericcon2}}) imply that

\bea
\Omega_{\alpha, \beta}{}^\beta &=& {\cos 4 \varphi \over 2 + \cos 4 \varphi} \Omega_{-,+\alpha}
\cr
Q_\alpha &=&0
\cr
\Omega_{\alpha,-+}+\Omega_{-,\alpha +}&=&0
\eea

and

\bea
\partial_\alpha \rho &=&0
\cr
\partial_\alpha \varphi +{\sin 4 \varphi \over 2+ \cos 4 \varphi} \Omega_{-,+\alpha} &=&0 \ .
\eea

{}From ({\ref{minuscomp}}) we obtain

\bea
Q_- &=& {1 \over 2}F_{- \alpha}{}^\alpha{}_\beta{}^\beta
\cr
\Omega_{-,\alpha}{}^\alpha &=& -{1 \over 2} \cos 2 \varphi G_{-\alpha}{}^\alpha
\cr
F_{-\alpha_1 \alpha_2 \alpha_3 \alpha_4} \epsilon^{\alpha_1 \alpha_2 \alpha_3 \alpha_4}
-F_{- {\bar{\alpha}}_1 {\bar{\alpha}}_2 {\bar{\alpha}}_3 {\bar{\alpha}}_4}
\epsilon^{{\bar{\alpha}}_1 {\bar{\alpha}}_2 {\bar{\alpha}}_3 {\bar{\alpha}}_4}
&=& 6 \sin 2 \varphi G_{- \alpha}{}^\alpha
\cr
\Omega_{-,-+} &=&0
\eea

and

\bea
\partial_- \varphi &=& -{1 \over 24}(F_{-\alpha_1 \alpha_2 \alpha_3 \alpha_4} \epsilon^{\alpha_1 \alpha_2 \alpha_3 \alpha_4}
-F_{- {\bar{\alpha}}_1 {\bar{\alpha}}_2 {\bar{\alpha}}_3 {\bar{\alpha}}_4}
\epsilon^{{\bar{\alpha}}_1 {\bar{\alpha}}_2 {\bar{\alpha}}_3 {\bar{\alpha}}_4})
\cr
\partial_- \rho &=&0 \ .
\eea

{}From ({\ref{pluscomp}}) we obtain

\bea
\Omega_{+,\alpha}{}^\alpha &=&0
\cr
Q_+ &=&0
\cr
\Omega_{+,-+} &=&0
\eea

and

\bea
\partial_+ \varphi &=&0
\cr
\partial_+ \rho &=&0 \ .
\eea

{}For the other example we take $z$ to be real and so  $A$ is the identity matrix.
Then from ({\ref{specialder1}}) we find

\bea
Q_\alpha &=& 2 F_{-+ \alpha \beta}{}^\beta
\cr
F_{-+ \alpha \beta}{}^\beta &=& {i \over 8}(G_{\alpha \beta}{}^\beta
+(G_{\bar{\alpha} \beta}{}^\beta)^*)
\eea

and

\bea
(z^{-1} \partial_\alpha z)_{11}&=&(z^{-1} \partial_\alpha z)_{22}=-{1 \over 2}
\Omega_{\alpha,-+}-{1 \over 2} \Omega_{{\bar{\beta}},}{}^{\bar{\beta}}{}_\alpha
\cr
(z^{-1} \partial_\alpha z)_{12}&=&-(z^{-1} \partial_\alpha z)_{21}={i \over 2}
\Omega_{\alpha,\beta}{}^\beta+{i \over 6}  \Omega_{{\bar{\beta}},}{}^{\bar{\beta}}{}_\alpha + {i \over 12}(G_{\alpha \beta}{}^\beta
-(G_{\bar{\alpha} \beta}{}^\beta)^*) \ .
\eea

{}From ({\ref{minuscomp}}) we obtain

\bea
Q_- &=& {1 \over 2} F_{- \alpha}{}^\alpha{}_\beta{}^\beta
\cr
F_{- \alpha_1 \alpha_2 \alpha_3 \alpha_4} &=&0
\cr
(G_{- \alpha}{}^\alpha)^* + G_{- \alpha}{}^\alpha &=&0
\eea

and

\bea
(z^{-1} \partial_- z)_{11}&=&(z^{-1} \partial_- z)_{22}=-{1 \over 2}
\Omega_{-,-+}
\cr
(z^{-1} \partial_- z)_{12}&=&-(z^{-1} \partial_- z)_{21}={i \over 2} \Omega_{-,\alpha}{}^\alpha +{i \over 4}G_{- \alpha}{}^\alpha \ .
\eea

{}From ({\ref{pluscomp}}) we obtain

\be
Q_+ =0
\ee

and

\bea
(z^{-1} \partial_+ z)_{11}&=&(z^{-1} \partial_+ z)_{22}=-{1 \over 2}
\Omega_{+,-+}
\cr
(z^{-1} \partial_+ z)_{12}&=&-(z^{-1} \partial_+ z)_{21}={i \over 2} \Omega_{+,\alpha}{}^\alpha \ .
\eea

\newsection{The linear system of degenerate half-maximal $SU(4)\ltimes\bR^8$-backgrounds}\la{degencaseap}

It is straightforward to construct the linear system for the degenerate half-maximal $SU(4)\ltimes\bR^8$ case
using the linear system of \cite{ugjggpa} for one $SU(4)\ltimes\bR^8$-invariant spinor and those of section \ref{degencase}.
In particular, we have that the algebraic Killing spinor equations give
\bea
(f+g_2 -i g_1) P_{\bar{\alpha}}&=&0~,
\cr
 {1 \over 4} (f-g_2 +i g_1) G_{-+\bar{\alpha}}
+{1 \over 4} (f-g_2+ig_1)G_{\bar{\alpha}\b}{}^\b
\cr
+{1 \over 12}(f+g_2+ig_1) \epsilon_{\bar{\alpha}}{}^{\b_1 \b_2 \b_3} G_{\b_1\b_2\b_3} &=&0~,
\la{done}
\eea
\bea
(f-g_2 -i g_1) P_\alpha&=&0~,
\cr
 {1 \over 4} (f+g_2+ig_1) G_{-+\alpha}
-{1 \over 4} (f+g_2+ig_1)G_{\alpha \b}{}^\b
\cr
+{1 \over 12} (f-g_2+ig_1) \epsilon_\alpha{}^{{\bar{\b_1}} {\bar{\b_2}} {\bar{\b_3}}}
G_{{\bar{\b_1}} {\bar{\b_2}} {\bar{\b_3}}} &=&0~,
\la{dtwo}
\eea
\bea
(f+g_2 -i g_1)P_+ &=&0~,
\cr
{1 \over 4} (f-g_2 +i g_1) G_{+ \alpha}{}^\a&=&0~,
\la{dthree}
\eea
\bea
(f-g_2 -i g_1)P_+&=&0~,
\cr
 -{1 \over 4} (f+g_2 +i g_1) G_{+ \alpha}{}^\a &=&0~,
\la{dfour}
\eea
and
\be
(f-g_2+ig_1) G_{+{\bar{\alpha}} {\bar{\beta}}}-{1 \over 2} (f+g_2+ig_1)
\epsilon_{{\bar{\alpha}}
{\bar{\beta}}}{}^{\g \d} G_{+ \g \d}=0~.
\la{dfive}
\ee

The conditions arising from the ${\cal D}_\a$ component of the supercovariant derivative are
\bea
[D_\a +(w-w^*)^{-1} \partial_\a w+{1\over2}\Omega_{\a,\b}{}^\b+
{1\over2} \Omega_{\a,-+}
+{i\over4} F_{\a\b}{}^\b{}_{\g}{}^\g~~~~~~~~
\cr
+{i\over2} F_{\a-+\b}{}^\b](f-g_2+ig_1) =0~,
\la{aonex}
\eea
\bea
-(w-w^*)^{-1} \partial_\a w (f-g_2+ig_1)
+(f+g_2-ig_1) [{1\over4} G_{\a\b}{}^\b
+{1\over4} G_{-+\a}]=0~,
\la{aoney}
\eea
\bea
 (f-g_2+ig_1) [\Omega_{\a,\bar\b_1\bar\b_2}+i F_{\a\bar\b_1\bar\b_2\g}{}^\g
  +i F_{\a -+\bar\b_1\bar\b_2}]
 \cr
-
(f+g_2+ig_1) [{1\over2} \Omega_{\a,\g_1\g_2}-{i\over2}
F_{\a\g_1\g_2\d}{}^\d+{i\over2} F_{\a-+\g_1\g_2}
 ] \epsilon^{\g_1\g_2}{}_{\bar\b_1\bar\b_2}=0~,
 \la{atwox}
 \eea
 \bea
(f+g_2-ig_1) [{1\over2} G_{\a\bar\b_1\bar\b_2}
-{1\over4} g_{\a[\bar\b_1} G_{\bar\b_2]\g}{}^\g
-{1\over4} g_{\a[\bar\b_1} G_{\bar\b_2]-+}]
\cr
 -{1\over8}(f-g_2-ig_1) G_{\a\g_1\g_2} \epsilon^{\g_1\g_2}{}_{\bar\b_1\bar\b_2}=0~,
 \la{atwoy}
 \eea
 \bea
[ D_\a +(w-w^*)^{-1} \partial_\a w -{1\over2} \Omega_{\a,\b}{}^\b+{1\over2} \Omega_{\a,-+}
+{i\over4} F_{\a\b}{}^\b{}_{\g}{}^\g~~~~~~~~~~~~~~~
\cr
-{i\over2}
F_{\a-+\g}{}^\g](f+g_2+ig_1)
+{i\over12} F_{\a\bar\b_1\bar\b_2\bar\b_3\bar\b_4}
 \epsilon^{\bar\b_1\bar\b_2\bar\b_3\bar\b_4}
(f-g_2+ig_1)=0~,
  \la{athreex}
\eea
\bea
-(w-w^*)^{-1} \partial_\a w(f+g_2+ig_1)
+ [-{1\over8} G_{\a\g}{}^\g+{1\over8} G_{-+\a}] (f-g_2-ig_1)
\cr
-{1\over24}
 \epsilon_{\a}{}^{\bar\b_1\bar\b_2\bar\b_3}
  G_{\bar\b_1\bar\b_2\bar\b_3}(f+g_2-ig_1)=0~,
  \la{athreey}
\eea
\bea
[{1\over2} \Omega_{\a,+\bar\b}+{i\over2} F_{\a+\bar\b\g}{}^\g] (f-g_2+i g_1)
-{i\over 6} F_{\a+\b_1\b_2\b_3} \epsilon^{\b_1\b_2\b_3}{}_{\bar\b}
(f+g_2+ig_1)=0~,
\la{afourx}
\eea
\bea
[{1\over16}g_{\a\bar\b} G_{+\g}{}^\g-
{1\over4} G_{+\a\bar\b}](f+g_2-i g_1)
=0~,
\la{afoury}
\eea
\bea
{i\over12} F_{\a+\bar\b_1\bar\b_2\bar\b_3} (f-g_2+ig_1)
+{1\over12} (f+g_2+ig_1)
 [{1\over2} \Omega_{\a,+\g}
-{i\over2} F_{\a+\g\d}{}^\d] \epsilon^\g{}_{\bar\b_1\bar\b_2\bar\b_3}=0~,
\la{afivex}
\eea
\bea
{1\over32} g_{\a[\bar\b_1} G_{\bar\b_2\bar\b_3]+}
(f+g_2-ig_1) -
{1\over 96}  (f-g_2-ig_1) G_{+\a\g} \epsilon^\g{}_{\bar\b_1\bar\b_2\bar\b_3}=0~.
\la{afivey}
\eea

The conditions arising from the   ${\cal D}_{\bar\a}$ component of the supercovariant derivative are
\bea
[D_{\bar\a}+(w-w^*)^{-1} \partial_{\bar\a}w+{1\over2} \Omega_{\bar\a,\b}{}^\b+{1\over2} \Omega_{\bar\a, -+}
+{i\over4} F_{\bar\a\b}{}^\b{}_\g{}^\g~~~~~~~~~~~~~
\cr
+{i\over2} F_{\bar\a-+\b}{}^\b](f-g_2+ig_1)
+{i\over12} (f+g_2+ig_1) F_{\bar\a\g_1\g_2\g_2\g_4} \epsilon^{\g_1\g_2\g_3\g_4}
=0~,
\la{abonex}
\eea
\bea
-(w-w^*)^{-1} \partial_{\bar\a} w (f-g_2+ig_1)
+{1\over8} [ G_{\bar\a\b}{}^\b+ G_{\bar\a-+}] (f+g_2-ig_1)
\cr
-{1\over24}
(f-g_2-ig_1)  \epsilon_{\bar\a}{}^{\g_1\g_2\g_3}
G_{\g_1\g_2\g_3}=0~,
\la{aboney}
\eea
\bea
[\Omega_{\bar\a, \bar\b_1\bar\b_2}+i F_{\bar\a\bar\b_1\bar\b_2\g}{}^\g
+i F_{\bar\a-+\bar\b_1\bar\b_2}] (f-g_2+ig_1)
\cr
-(f+g_2+ig_1)[{1\over2}\Omega_{\bar\a,\g_1\g_2}
-{i\over2} F_{\bar\a\g_1\g_2\d}{}^\d+{i\over2} F_{\bar\a-+\g_1\g_2}]
\epsilon^{\g_1\g_2}{}_{\bar\b_1\bar\b_2}=0~,
\la{abtwox}
\eea
\bea
{1\over4} G_{\bar\a\bar\b_1\bar\b_2} (f+g_2-ig_1)
-(f-g_2-ig_1)[{1\over8} g_{\bar\a\g_1} G_{\g_2\d}{}^\d
\cr
+{1\over4}G_{\bar\a\g_1\g_2}-{1\over8} g_{\bar\a\g_1} G_{\g_2-+}]
\epsilon^{\g_1\g_2}{}_{\bar\b_1\bar\b_2}=0~,
\la{abtwoy}
\eea
\bea
[D_{\bar\a}+(w-w^*)^{-1} \partial_{\bar\a} w-{1\over2} \Omega_{\bar\a,\g}{}^\g+{1\over2} \Omega_{\bar\a,-+}
\cr
+{i\over4} F_{\bar\a\g}{}^\g{}_\d{}^\d
-{i\over2} F_{\bar\a-+\g}{}^\g](f+g_2+ig_1)=0~,
\la{abthreex}
\eea
\bea
-(w-w^*)^{-1} \partial_{\bar\a} w(f+g_2+ig_1)
+[-{1\over4} G_{\bar\a\g}{}^\g+{1\over4} G_{\bar\a-+}] (f-g_2-ig_1)=0~,
\la{abthreey}
\eea
\bea
[{1\over2} \Omega_{\bar\a,+\bar\b}+{i\over2} F_{\bar\a+\bar\b\g}{}^\g] (f-g_2+ig_1)
\cr
-{i\over6} (f+g_2+ig_1) F_{\bar\a+\g_1\g_2\g_3} \epsilon^{\g_1\g_2\g_3}{}_{\bar\b}=0~,
\la{abfourx}
\eea
\bea
-{1\over8} G_{+\bar\a\bar\b} (f+g_2-ig_1)
-{1\over16} (f-g_2-ig_1) G_{\g_1\g_2+} \epsilon^{\g_1\g_2}{}_{\bar\a\bar\b}=0~,
\la{abfoury}
\eea
\bea
i F_{\bar\a+\bar\b_1\bar\b_2\bar\b_3} (f-g_2+ig_1)+ (f+g_2+ig_1)
[{1\over2} \Omega_{\bar\a,+\g}
-{i\over2} F_{\bar\a+\g\d}{}^\d] \epsilon^\g{}_{\bar\b_1\bar\b_2\bar\b_3}=0~,
\la{abfivex}
\eea
\bea
(f-g_2-ig_1)[-{1\over16} g_{\bar\a\g} G_{+\d}{}^\d
+{1\over4} G_{\bar\a+\g}] \epsilon^\g{}_{\bar\b_1\bar\b_2\bar\b_3}=0~.
\la{abfivey}
\eea

The conditions arising from the   ${\cal D}_{-}$ component of the supercovariant derivative are
\bea
[D_-+(w-w^*)^{-1} \partial_- w+{1\over2} \Omega_{-,\g}{}^\g+{1\over2} \Omega_{-,-+}
+{i\over4} F_{-\g}{}^\g{}_\d{}^\d](f-g_2+ig_1)
\cr
+{i\over12} (f+g_2+ig_1) F_{-\g_1\g_2\g_3\g_4} \epsilon^{\g_1\g_2\g_3\g_4}=0~,
\la{monex}
\eea
\bea
-(w-w^*)^{-1} \partial_- w(f-g_2+ig_1)
+{1\over4} G_{-\g}{}^\g (f+g_2-ig_1)=0~,
\la{money}
\eea
\bea
[\Omega_{-,\bar\b_1\bar\b_2}+i F_{-\bar\b_1\bar\b_2\g}{}^\g]
(f-g_2+ig_1)
\cr
-(f+g_2+ig_1) [{1\over2} \Omega_{-,\g_1\g_2}
-{i\over2} F_{-\g_1\g_2\d}{}^\d] \epsilon^{\g_1\g_2}{}_{\bar\b_1\bar\b_2}=0~,
\la{mtwox}
\eea
\bea
{1\over2} G_{-\bar\b_1\bar\b_2}
(f+g_2-ig_1)
-{1\over4} (f-g_2-ig_1) G_{-\g_1\g_2} \epsilon^{\g_1\g_2}{}_{\bar\b_1\bar\b_2}=0~,
\la{mtwoy}
\eea
\bea
[D_-+(w-w^*)^{-1} \partial_- w-{1\over2} \Omega_{-,\g}{}^\g+{1\over2}\Omega_{-,-+}
+{i\over4} F_{-\g}{}^\g{}_\d{}^\d](f+g_2+ig_1)
\cr
+{i\over 12} (f-g_2+ig_1) F_{-\bar\b_1\bar\b_2\bar\b_3\bar\b_4}
\epsilon^{\bar\b_1\bar\b_2\bar\b_3\bar\b_4}=0~,
\la{mthreex}
\eea
\bea
-(w-w^*)^{-1} \partial_- w(f+g_2+ig_1)
-{1\over4} (f-g_2-ig_1) G_{-\g}{}^\g
=0~,
\la{mthreey}
\eea
\bea
[{1\over2} \Omega_{-,+\bar\b}+{i\over2} F_{-+\bar\b\g}{}^\g] (f-g_2+ig_1)
-{i\over6} (f+g_2+i g_1) F_{-+\g_1\g_2\g_3} \epsilon^{\g_1\g_2\g_3}{}_{\bar\b}
=0~,
\la{mfourx}
\eea
\bea
[-{1\over16} G_{\bar\b\g}{}^\g+{3\over16} G_{-+\bar\b}]
(f+g_2-ig_1)
+{1\over 48} (f-g_2-i g_1) G_{\g_1\g_2\g_3} \epsilon^{\g_1\g_2\g_3}{}_{\bar\b}=0~,
\la{mfoury}
\eea
\bea
i F_{-+\bar\b_1\bar\b_2\bar\b_3} (f-g_2+ig_1)
+
[{1\over2} \Omega_{-,+\g}-{i\over2}
F_{-+\g\d}{}^\d] \epsilon^\g{}_{\bar\b_1\bar\b_2\bar\b_3}
(f+g_2+i g_1)=0~.
\la{mfivex}
\eea
\bea
-{1\over8} G_{\bar\b_1\bar\b_2\bar\b_3} (f+g_2-ig_1)
+ [{1\over16} G_{\g\d}{}^\d +{3\over16} G_{-+\g}]
(f-g_2-i g_1) \epsilon^\g{}_{\bar\b_1\bar\b_2\bar\b_3}=0~.
\la{mfivey}
\eea

The conditions arising from the   ${\cal D}_{+}$ component of the supercovariant derivative are
\bea
[D_++(w-w^*)^{-1} \partial_+ w+{1\over2} \Omega_{+,\g}{}^\g
+{1\over2} \Omega_{+,-+}+{i\over4} F_{+\g}{}^\g{}_\d{}^\d](f-g_2+ig_1)
=0~,
\la{ponex}
\eea
\bea
-(w-w^*)^{-1} \partial_+ w(f-g_2+ig_1)
+{1\over8} G_{+\g}{}^\g (f+g_2-ig_1)
=0~,
\la{poney}
\eea
\bea
[\Omega_{+,\bar\b_1\bar\b_2}+i F_{+\bar\b_1\bar\b_2\g}{}^\g]
 (f-g_2+ig_1)
\cr
-[{1\over2} \Omega_{+,\g_1\g_2}
-{i\over2} F_{+\g_1\g_2\d}{}^\d] (f+g_2+ig_1) \epsilon^{\g_1\g_2}{}_{\bar\b_1\bar\b_2}
=0~,
\la{ptwox}
\eea
\bea
{1\over4} (f+g_2-ig_1) G_{+\bar\b_1\bar\b_2}
-{1\over8} (f-g_2-ig_1) G_{+\g_1\g_2} \epsilon^{\g_1\g_2}{}_{\bar\b_1\bar\b_2}=0~,
\la{ptwoy}
\eea
\bea
[D_++(w-w^*)^{-1} \partial_+ w-{1\over2} \Omega_{+,\g}{}^\g+{1\over2} \Omega_{+,-+}
+{i\over4} F_{+\g}{}^\g{}_\d{}^\d] (f+g_2+ig_1)=0~,
 \la{pthreex}
 \eea
 \bea
-(w-w^*)^{-1} \partial_+ w (f+g_2+ig_1)
-{1\over8} G_{+\g}{}^\g (f-g_2-ig_1)
=0~,
 \la{pthreey}
 \eea
 and
 \bea
 \Omega_{+,+\a}=\Omega_{+,+\bar\a}=0~.
 \la{pfour}
 \eea
The solution of the above linear system and the conditions on the geometry and the fluxes are summarized
in section \ref{degencase}.


\begin{thebibliography}{00}
\addcontentsline{toc}{section}{References} \frenchspacing \small
\addtolength{\itemsep}{-4pt}

\bibitem{gibbons}
A.~Dabholkar, G.~W.~Gibbons, J.~A.~Harvey and F.~Ruiz Ruiz,
``Superstrings And Solitons,''
Nucl.\ Phys.\ B {\bf 340} (1990) 33.

\bibitem{duff}
M.~J.~Duff and J.~X.~Lu,
``The Selfdual type IIB superthreebrane,''
Phys.\ Lett.\ B {\bf 273} (1991) 409.

``Elementary Five-Brane Solutions Of D = 10 Supergravity,''
Nucl.\ Phys.\ B {\bf 354} (1991) 141.

\bibitem{strominger}
G.~T.~Horowitz and A.~Strominger,
``Black strings and P-branes,''
Nucl.\ Phys.\ B {\bf 360} (1991) 197.


\bibitem{maldacena}
O.~Aharony, S.~S.~Gubser, J.~M.~Maldacena, H.~Ooguri and Y.~Oz,
``Large N field theories, string theory and gravity,''
Phys.\ Rept.\  {\bf 323}, 183 (2000)
[arXiv:hep-th/9905111].

\bibitem{schwarz}
J.~H.~Schwarz, ``Covariant Field Equations Of Chiral N=2 D = 10
Supergravity,'' Nucl.\ Phys.\ B {\bf 226} (1983) 269.

\bibitem{klebanov}
I.~R.~Klebanov and M.~J.~Strassler,
``Supergravity and a confining gauge theory: Duality cascades and
chiSB-resolution of naked singularities,''
JHEP {\bf 0008}, 052 (2000)
[arXiv:hep-th/0007191].

\bibitem{chamseddine}
A.~H.~Chamseddine and M.~S.~Volkov,
``Non-Abelian solitons in N = 4 gauged supergravity and leading order  string
theory,''
Phys.\ Rev.\ D {\bf 57} (1998) 6242
[arXiv:hep-th/9711181].

\bibitem{nunez}
J.~M.~Maldacena and C.~Nunez,
``Towards the large N limit of pure N = 1 super Yang Mills,''
Phys.\ Rev.\ Lett.\  {\bf 86}, 588 (2001)
[arXiv:hep-th/0008001].

\bi{georgea}
M.~Blau, J.~Figueroa-O'Farrill, C.~Hull and G.~Papadopoulos,
``A new maximally supersymmetric background of IIB superstring theory,''
JHEP {\bf 0201} (2002) 047
[arXiv:hep-th/0110242].

\bi{jfgpa}
 J.~Figueroa-O'Farrill and
G.~Papadopoulos,
``Maximally supersymmetric solutions of ten-dimensional
and eleven-dimensional supergravities,''
JHEP {\bf 0303} (2003) 048:
[arXiv:hep-th/0211089].

``Pluecker-type relations for orthogonal planes,''
[arXiv:math.ag/0211170].


\bi{georgeb}
M.~Blau, J.~Figueroa-O'Farrill, C.~Hull and G.~Papadopoulos,
``Penrose limits and maximal supersymmetry,''
Class.\ Quant.\ Grav.\  {\bf 19} (2002) L87
[arXiv:hep-th/0201081].

\bibitem{ugjggpa}
  U.~Gran, J.~Gutowski and G.~Papadopoulos,
  ``The spinorial geometry of supersymmetric IIB backgrounds,''
  Class.\ Quant.\ Grav.\  {\bf 22} (2005) 2453
  [arXiv:hep-th/0501177].



\bibitem{ugjggpb}
U.~Gran, J.~Gutowski and G.~Papadopoulos,
``The G(2) spinorial geometry of supersymmetric IIB backgrounds,''
arXiv:hep-th/0505074.



\bibitem{uggp}
  J.~Gillard, U.~Gran and G.~Papadopoulos,
  ``The spinorial geometry of supersymmetric backgrounds,''
  Class.\ Quant.\ Grav.\  {\bf 22} (2005) 1033
  [arXiv:hep-th/0410155].



\bibitem{ugdrgp}
  U.~Gran, G.~Papadopoulos and D.~Roest,
  ``Systematics of M-theory spinorial geometry,''
  Class.\ Quant.\ Grav.\  {\bf 22} (2005) 2701
  [arXiv:hep-th/0503046].

\bibitem{jose}
J.~M.~Figueroa-O'Farrill,
``Breaking the M-waves,''
Class.\ Quant.\ Grav.\  {\bf 17}, 2925 (2000)
[arXiv:hep-th/9904124].

\bibitem{west}
J.~H.~Schwarz and P.~C.~West,
``Symmetries And Transformations Of Chiral N=2 D = 10 Supergravity,''
Phys.\ Lett.\ B {\bf 126} (1983) 301.



\bibitem{howe}
P.~S.~Howe and P.~C.~West, ``The Complete N=2, D = 10
Supergravity,'' Nucl.\ Phys.\ B {\bf 238} (1984) 181.


\bibitem{eric}
E.~A.~Bergshoeff, M.~de Roo, S.~F.~Kerstan and F.~Riccioni,
``IIB supergravity revisited,''
arXiv:hep-th/0506013.


\bibitem{tsimpis}
G.~Papadopoulos and D.~Tsimpis, ``The holonomy of IIB
supercovariant
connection,'' Class.\ Quant.\ Grav.\  {\bf 20} (2003) L253
[arXiv:hep-th/0307127].

\bibitem{tomas}
J.~Bellorin and T.~Ortin,
``A note on simple applications of the Killing spinor identities,''
Phys.\ Lett.\ B {\bf 616} (2005) 118
[arXiv:hep-th/0501246].


\bibitem{gamma}
  U.~Gran,
  ``GAMMA: A Mathematica package for performing Gamma-matrix algebra and  Fierz
  transformations in arbitrary dimensions,''
  arXiv:hep-th/0105086.

  \bi{grayhervella}
A. Gray and L.M. Hervella, ``The sixteen classes of
 almost Hermitian manifolds and their linear invariants'',
 Ann.\ Mat.\ Pura\ e \ Appl.\ {\bf 282} (1980) 1.


 \bibitem{hull}
C.~M.~Hull,
``Exact Pp Wave Solutions Of 11-Dimensional Supergravity,''
Phys.\ Lett.\ B {\bf 139} (1984) 39.

  \bibitem{mateos}
  R.~Emparan, D.~Mateos and P.~K.~Townsend,
  ``Supergravity supertubes,''
  JHEP {\bf 0107} (2001) 011, arXiv:hep-th/0106012.




\bibitem{pope}
M.~Cvetic, G.~W.~Gibbons, H.~Lu and C.~N.~Pope,
``Ricci-flat metrics, harmonic forms and brane resolutions,''
Commun.\ Math.\ Phys.\  {\bf 232} (2003) 457
[arXiv:hep-th/0012011].


\bibitem{chen}
C.~M.~Chen and J.~F.~Vazquez-Poritz,
``Resolving the M2-brane,''
arXiv:hep-th/0403109.




\bibitem{wang} McKenzie Y. Wang,
 ``Parallel spinors and parallel forms'', Ann. Global Anal Geom.
Vol 7, No 1 (1989), 59.




\bibitem{lawson} H. Blaine Lawson and Marie-Louise Michelsohn, ``Spin geometry,'' Princeton
University Press (1989).


\bi{harvey} F. R. Harvey, ``Spinors and Calibrations,'' Academic
Press, London (1990).

\end{thebibliography}
\end{document}